\documentclass[pdflatex,sn-mathphys-num]{sn-jnl}
 
\usepackage{graphicx}%
\usepackage{multirow}%
\usepackage{amsmath,amssymb,amsfonts}%
\usepackage{amsthm}%
\usepackage{mathrsfs}%
\usepackage[title]{appendix}%
\usepackage{xcolor}%
\usepackage{textcomp}%
\usepackage{manyfoot}%
\usepackage{booktabs}%
\usepackage{algorithm}%
\usepackage{algorithmicx}%
\usepackage{algpseudocode}%
\usepackage{listings}%
 
\usepackage{url}
\usepackage{float}
\usepackage{subcaption}
\usepackage{siunitx}
\usepackage{makecell}
\usepackage{array}
\usepackage{glossaries}
 
\glsdisablehyper
\newacronym{pinn}{PINN}{Physics-Informed Neural Network}
\newacronym{piml}{PIML}{Physics-Informed Machine Learning}
\newacronym{sir}{SIR}{Susceptible-Infectious-Recovered}
\newacronym{seir}{SEIR}{Susceptible-Exposed-Infectious-Recovered}
\newacronym{seird}{SEIRD}{Susceptible-Exposed-Infectious-Recovered-Dead}
\newacronym{pde}{PDE}{Partial Differential Equation}
\newacronym{ode}{ODE}{Ordinary Differential Equation}
\newacronym{mcmc}{MCMC}{Markov Chain Monte Carlo}
\newacronym{abc}{ABC}{Approximate Bayesian Computation}
\newacronym{abcsmc}{ABC-SMC}{Approximate Bayesian Computation Sequential Monte Carlo}
\newacronym{ili}{ILI}{Influenza-Like-Illness}
\newacronym{ari}{ARI}{Acute Respiratory Infection}
\newacronym{is}{IS}{Interval Score}
\newacronym{wis}{WIS}{Weighted Interval Score}
\newacronym{ae}{AE}{Absolute Error}
\newacronym{mae}{MAE}{Mean Absolute Error}
\newacronym{mse}{MSE}{Mean Squared Error}
\newacronym{rmse}{RMSE}{Root Mean Squared Error}
\newacronym{wmape}{WMAPE}{Weighted Mean Absolute Percentage Error}
 
\theoremstyle{thmstyleone}%

\theoremstyle{thmstyletwo}%

\theoremstyle{thmstylethree}%

\begin{document}

\title[Article Title]{A systematic comparison of physics-informed neural networks and mechanistic models for epidemic modeling and forecasting}

\author*[1,2]{\fnm{Martina} \sur{Rama}}\email{mrama@fbk.eu}

\author[3]{\fnm{Yuhan} \sur{Li}}

\author[2]{\fnm{Bruno} \sur{Lepri}}

\author[3]{\fnm{Nicola} \sur{Perra}}

\author[4]{\fnm{Michele} \sur{Tizzoni}}

\affil[1]{\orgdiv{Department of Information Engineering and Computer Science}, \orgname{University of Trento}, \city{Trento}, \country{Italy}}

\affil*[2]{\orgdiv{Mobile and Social Computing (MobS) Lab}, \orgname{Fondazione Bruno Kessler (FBK)}, \city{Trento}, \country{Italy}}

\affil[3]{\orgdiv{School of Mathematical Sciences}, \orgname{Queen Mary University}, \city{London}, \country{United Kingdom}}

\affil[4]{\orgdiv{Department of Sociology and Social Research}, \orgname{University of Trento}, \city{Trento}, \country{Italy}}

\abstract{Physics-Informed Neural Networks (PINNs) have recently been applied to epidemic modeling and forecasting with promising results. These methods integrate mechanistic knowledge with data-driven learning. Yet, their comparative performance against well-established, and simpler epidemic models is underexplored. Here, we conduct a systematic comparison of a PINN-based framework and a stochastic mechanistic model. Both approaches share the same age-structured SEIRD core and rolling-window forecasting pipeline, in a deliberately minimal setting. 
We evaluate their retrospective fits and probabilistic forecasts in two synthetic scenarios and the first COVID-19 wave in London. Interestingly, while PINNs underperform in synthetic scenarios, they achieve better performance on real data. Furthermore, PINNs provide narrower uncertainty intervals and sharper estimation of parameters in synthetic scenarios. Our findings confirm PINNs as a promising tool in epidemiology. At the same time, the results reaffirm the value of simpler approaches such as traditional compartmental models, and promote additional benchmarking studies.\\[0.7ex]
}

\maketitle

\section{Introduction}

Accurate epidemic forecasting is fundamental for public health preparedness, as it informs resource allocation, the timing of both pharmaceutical and non-pharmaceutical interventions, and strategic decision-making~\cite{marathe2013computational,desai2019real,lauer2021infectious}. 
Indeed, during the COVID-19 pandemic, forecasts of cases, hospitalizations, and deaths guided policy decisions amid substantial uncertainty, rapidly evolving transmission dynamics influenced by human behavior, limited and noisy surveillance data~\cite{pollett2021recommended, wolffram2023collaborative}. 
Collaborative forecasting efforts, such as the US and European COVID-19 Forecast Hubs and analogous initiatives for seasonal influenza~\cite{reich2019accuracy, howerton2023evaluation,sherratt2023predictive,lopez2024challenges, fox2024optimizing, fiandrino2025collaborative}, have become central tools to guide disease response, underscoring the value of collaboration across diverse modeling approaches. By collecting the forecasts of many independent teams in a common format and evaluating them with shared metrics, these hubs provide public health authorities with timely, comparable, and rigorously validated forecasts.

Traditional mechanistic compartmental models, such as \gls{sir}, \gls{seir}, and their extensions, describe disease transmission through systems of differential equations, dividing the population into disease-state compartments, with individuals moving between them according to specific transition rates~\cite{kermack1927contribution,anderson1991infectious,keeling2008modeling}. 
While these models provide an explainable and reliable framework for exploring scenarios, evaluating the effect of interventions to support policy decisions, and making predictions, their effectiveness relies on strong modeling assumptions, accurate parameter estimation, and sufficient data availability, which can limit their flexibility. 
This is particularly limiting in real-time forecasts of complex health emergencies such as the COVID-19 pandemic, when rapidly capturing multi-source data-driven patterns is crucial. 

In contrast, the advent of big data has introduced computational challenges for traditional methods, but has also enabled the application of modern AI techniques --- including recurrent and convolutional neural network architectures --- to epidemic forecasting~\cite{zeroual2020deep,lim2021time,nabi2021forecasting,lee2021predictive}. These methods can learn temporal patterns directly from diverse datasets, often achieving strong short-term predictive accuracy without explicit mechanistic assumptions. This, however, typically comes at the cost of a much lower interpretability~\cite{rodriguez2024machine, ye2025integrating, ray2025flusion}. 

To bridge the gap between the interpretability of mechanistic models and the flexibility of data-driven approaches, hybrid frameworks have emerged. These frameworks preserve knowledge of disease transmission mechanisms while also extracting relevant epidemiological features from data using flexible machine learning models~\cite{karniadakis2021physics, o2022semi, ye2025integrating}. Among these, we find \glspl{pinn}~\cite{raissi2019physics,cuomo2022scientific}. These methods embed the physical knowledge of a system, written in terms of \glspl{ode} (\glspl{pde} in general), into neural networks through a proper regularization mechanism, driving the network to approximate the system while simultaneously fitting the data. 

\glspl{pinn} have recently attracted increasing attention in epidemiology~\cite{shaier2021data,bertaglia2022asymptotic,ning2023epi,qian2025physics,rama2026forecasting}. By incorporating the governing differential equations of compartmental models directly into the neural network's loss function, they can learn the disease-spreading process and estimate the parameters, even with limited or noisy data.
At the same time, \gls{pinn}-based frameworks present open challenges. Examples include the reliance on multiple neural networks and on retraining when applied to new outbreaks or updated observations. These constraints increase the computational cost, delay the forecasts' availability, and, in several existing implementations, limit the treatment of forecast uncertainty~\cite{kharazmi2021identifiability, berkhahn2022physics, rodriguez2023einns, millevoi2024physics, madden2024deep}.

Despite the growing adoption of \gls{pinn}-based approaches in epidemiological modeling, a systematic evaluation of their forecasting performance in real-world scenarios relative to well-established mechanistic approaches remains largely lacking. Several recent studies have benchmarked epidemic forecasting models against one another, comparing deep learning architectures~\cite{zeroual2020deep,nabi2021forecasting}, foundation models~\cite{kalahasti2026foundation}, or broader collections of statistical, mechanistic, and machine learning approaches~\cite{wynants2020prediction, sherratt2025influence, wang2026naive}. However, \gls{pinn}-based models have typically been tested in isolation, against other deep learning models, or benchmarked against simplified or limited baselines~\cite{berkhahn2022physics,bertaglia2022asymptotic,millevoi2024physics,qian2025physics}, rather than against consolidated approaches under shared, comparable settings. 
This leaves open the question of whether, and under which conditions, \gls{pinn}-based methods offer genuine advantages over state-of-the-art methodologies in terms of parameter estimation, forecast accuracy, and uncertainty quantification. 

In this work, we address this gap by systematically comparing a \gls{pinn}-based forecasting framework with a consolidated mechanistic modeling approach, on synthetic scenarios and real epidemic data capturing the first COVID-19 wave in London. 
Rather than building two forecasting systems independently optimized for operational use, we compare the two modeling paradigms under minimal and shared settings that intentionally exclude any time-varying epidemiological contexts, such as interventions or mobility patterns. 
This design allows us to investigate how each framework behaves when relying solely on observed incidence data. 

\begin{figure}[tbp]
    \centering
    \includegraphics[width=\textwidth]{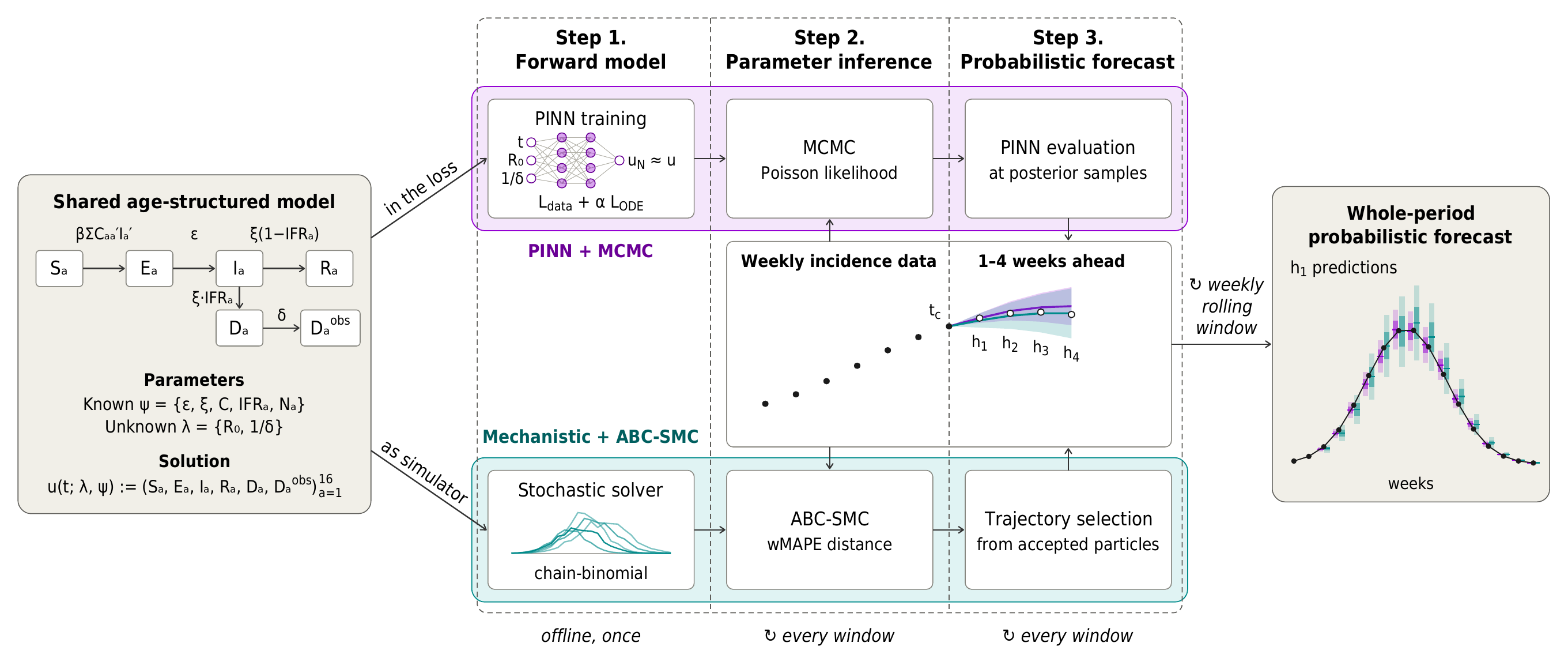}
    \caption{{\bf Overview of the two forecasting frameworks.}
    The PINN-based approach is shown in violet and the mechanistic approach in teal. Both rely on the same age-structured Susceptible-Exposed-Infectious-Recovered-Dead (SEIRD) model ($a = 1, \dots, 16$; left panel), with known parameters $\psi = \{\varepsilon, \xi, C, \text{IFR}_a, N_a\}$, unknown parameters $\lambda = \{R_0, 1/\delta\}$, and solution $u(t;\lambda,\psi)$.
    \textit{Step~1 (offline, once).} The PINN $u_N(t, R_0, 1/\delta)$ is trained on a parameter grid, with the ODE system embedded in the loss through $L_{\text{ODE}}$, to approximate $u$. The mechanistic approach builds a pool of stochastic chain-binomial simulations, one for each parameter set.
    \textit{Step~2 (every window).} The unknown parameters are inferred from the weekly incidence data in the look-back window, via MCMC with a Poisson likelihood for the PINN and via likelihood-free ABC-SMC with a Weighted Mean Absolute Percentage Error (WMAPE) distance for the mechanistic model. For the London COVID-19 data, the observed quantity is weekly deaths and both $R_0$ and $1/\delta$ are estimated; for the synthetic scenarios, the observed quantity is weekly new infections and only $R_0$ is estimated.
    \textit{Step~3 (every window).} Probabilistic forecasts at horizons $h_1$--$h_4$ (1--4 weeks ahead) are obtained by evaluating the PINN at the posterior samples with their uncertainty and by selecting the pool trajectories associated with the accepted particles for the mechanistic-based framework.
    The central panel illustrates a single window: the dashed line separating Steps~2 and~3 passes through the last available observation at time $t_c$, from which the forecasts start; filled circles are the observed data used in Step~2, open circles the future observations against which the forecasts are evaluated.
    Steps~2--3 are repeated every week in a rolling-window manner. The right panel shows, as an example, the resulting 1-week-ahead forecasts ($h_1$) over the whole period (median, 50\% and 90\% prediction intervals), against the observed data (black); the same is done for horizons $h_2$--$h_4$.}
\label{fig:frameworks}
\end{figure}

Building on our previous work~\cite{rama2026forecasting}, which demonstrated the potential of \gls{pinn}-based forecasting for seasonal influenza, we optimize and extend the framework to deal with COVID-19-like dynamics, and compare its forecasting performance against a mechanistic model calibrated via \gls{abcsmc}.  
In addition to sharing the same forecasting pipeline, both frameworks are based on an age-structured compartmental model, as schematically shown in Fig~\ref{fig:frameworks}.
Age structure is essential to describe diseases like COVID-19 where both contacts and mortality vary strongly across age groups~\cite{prem2017projecting, verity2020estimates, ram2021modified}, however, only a few \gls{pinn}-based studies account for it~\cite{zhang2025dynamics, dimarco2025augmented, kong2025age}. 

The comparison is performed in two case studies. First, we consider two synthetic scenarios, representing different severity levels of a COVID-19-like epidemic wave, where the basic epidemiological parameters and the weekly incidences are known. This allows us to evaluate not only the forecasts, but also how well each framework fits the data retrospectively and recovers the parameters used to generate them. Second, we apply the same pipeline to the real, more challenging, case of the first COVID-19 wave in London. 
In all scenarios, we evaluate the probabilistic forecasts from $1$ to $4$ weeks ahead with the same metrics, both by horizon and aggregated, normalizing the error-based scores by those of a baseline model, a simple data-driven benchmark, which does not rely on any epidemiological knowledge.

Our comparison shows that neither approach consistently outperforms the other: the advantage of \gls{pinn}-based methods depends on the setting. In synthetic scenarios, both approaches recover $R_0$ accurately and outperform the baseline, with the mechanistic framework being slightly more accurate, and the \gls{pinn}-based one providing overconfident forecasts. 
Owing to their admittedly simplified structure, in the first COVID-19 wave in London, both approaches underperform the baseline, yet the \gls{pinn}-based framework achieves lower forecasting errors than the mechanistic approach.

These findings highlight the importance of systematic comparisons under shared modeling assumptions to evaluate the strengths and limitations of hybrid approaches relative to mechanistic ones in epidemic forecasting.

\section{Methods}
\label{sec: methods}
\subsection{Epidemiological data}
\label{sec: epidemiological_data}

\paragraph{Synthetic data.}
We generate two synthetic scenarios mimicking a COVID-19-like epidemic wave. Both scenarios are obtained by integrating an age-structured \gls{seir} model, derived from the \glspl{ode} system~\ref{eq: ODEs system}. To match the real COVID-19 scenario discussed below, we divide the population into $16$ age classes, and use the contact matrix $C$ for the UK. For more details on the parameter values used to generate the data, refer to Section~\ref{sec: common_structural_setting}. For both synthetic scenarios, we set the initial number of infected individuals equal to $I_0 = 1\,000$, distributed across age classes proportionally to their population size and split between the exposed and infectious compartments according to the mean latency and infectious periods (see Eqs.~(\ref{eq: E0}) and~(\ref{eq: I0}) in Appendix \ref{sec: methods_details} in \nameref{S1_Appendix}). All remaining individuals are set to be susceptible. The two scenarios differ only in the basic reproduction number, $R_0=1.5$ and $R_0=2.0$, simulating either a slower epidemic with a later and lower peak and a faster, larger wave with an earlier peak, respectively. 
The system is integrated deterministically via \textit{odeint} (from \texttt{SciPy} library~\cite{Virtanen2020}) over $34$ weeks for $R_0=1.5$ and $24$ weeks for $R_0=2.0$. A shorter window is sufficient in the latter since the epidemic ends earlier. In this way, the time window of both cases contains the entire epidemic wave, with an initial growth phase and a final decay phase of approximately comparable duration.

From the deterministic solution we then compute the weekly incidence of infections, namely $\mu_w$ for each week $w$, aggregated across age classes from the susceptible compartments. To mimic the observational noise of real surveillance data, we draw the observed counts (independently for each week) from a Negative Binomial distribution with mean $\mu_w$ and variance $\mu_w+\mu_w^2/r$, i.e. $y_w\sim\mathrm{NB}(\mu_w,r)$, following the parametrization commonly used for COVID-19 count data~\cite{flaxman2020estimating,congdon2021mid,kim2023evaluation}. 
The dispersion parameter $r$ controls the noise level: smaller $r$ values increase the overdispersion, while the Poisson limit is recovered for $r\to\infty$. Here, we consider a baseline noise level, $r=100$, while Appendix \ref{sec: appendix_noise_synthetic} in \nameref{S1_Appendix} reports the same analysis on data generated with a stronger noise level, $r=50$. We thus obtain four datasets ($2$ values of $R_0$ $\times$ $2$ noise levels), each consisting of a single aggregated noisy incidence, the only information used for parameter inference and model calibration (Section~\ref{sec: PINN_inference&forecasting} and Section~\ref{sec: calibration&forecast_mech}). The noise-free trajectories are retained as ground truth. In particular, we look at the age-specific components in a sanity check test reported in Appendix \ref{sec: appendix_sanity_synthetic} in \nameref{S1_Appendix}. 

\paragraph{COVID-19 London data.}

We use COVID-19 mortality data for London, provided by Public Health England (PHE)~\cite{glacovid2020}, consisting of both daily death counts and weekly death incidence data stratified by $19$ age groups. We consider the period from mid-February to early August 2020, specifically from 2020-02-16 to 2020-07-19, chosen to fully contain London's first COVID-19 wave, including its initial growth phase, its mortality peak (around early April 2020), and its subsequent declining phase. In our analysis, we specifically use the weekly death incidence.  
Furthermore, we analyze the observed weekly deaths aggregated across age classes to obtain the reference incidence used for parameter inference and model calibration (see Appendix \ref{sec: methods_details} in \nameref{S1_Appendix}), and for the forecast evaluation (Section~\ref{sec: metrics}).

\subsection{Forecasting setting}
\label{sec: forecasting_setting}

To reproduce a real-time forecasting scenario, we adopt a rolling-window strategy, schematically illustrated in Fig~\ref{fig:frameworks}. Each forecasting round corresponds to a fixed current week. The data available up to that week form the look-back observation window, on which each framework is first calibrated (\textit{Step 2.}): parameters are inferred via \gls{mcmc} for the \gls{pinn}-based model (Section~\ref{sec: PINN_inference&forecasting}), and via \gls{abcsmc} for the mechanistic model (Section~\ref{sec: calibration&forecast_mech}). The estimated parameters are then used to produce probabilistic forecasts of the following weeks (\textit{Step 3.}). As a new week of data becomes available, the window moves forward and the whole procedure is repeated: the models are re-calibrated on the updated past data and a new forecast is produced. In this way, each round relies only on the information available up to the current week. This procedure is repeated for every week of the study period, from the first forecasting round to the last one. 

For each round, we analyze how the performance of both frameworks is affected by the forecast horizon, ranging from $1$ to $4$ weeks ahead. By default, we use all past data available up to the current week as the look-back observation window. Note that during the final weeks of the real COVID-19 case-study period, when the requested forecast horizon would extend beyond the last week considered (i.e., 2020-07-19), we restrict the evaluation to the forecast up to the last week considered, shortening the maximum forecast horizon accordingly.

As a general rule, for the observations included in the look-back window, we further implement a weighting strategy that down-weights data points in proportion to their temporal distance from the current forecasting week, so that more recent weeks are given greater reliability. Specifically, for an observation window of length $M$, the $i$-th observation (with $i=1$ denoting the most distant, and $i=M$ the most recent) is assigned the harmonic weight
\begin{equation}
w_i = \frac{1}{1+M-i}, \qquad i = 1, \dots, M,
\label{eq: weights}
\end{equation}
normalized by the sum of the weights over the window. 

In addition, in Appendix \ref{sec: appendix_COVID_n5} in \nameref{S1_Appendix} we perform a sensitivity analysis for COVID-19 forecasting by shortening the length of the look-back window to $M = 5$ observations as in~\cite{rama2026forecasting}, without weighting due to the shorter length window. 

\subsection{Common structural settings}
\label{sec: common_structural_setting}

Although the two frameworks compared in this work differ substantially in their nature, they follow the same three-step pipeline, illustrated in Fig~\ref{fig:frameworks}: (\textit{Step 1.}) a forward epidemiology-based model, (\textit{Step 2.}) the inference of its unknown parameters from the look-back window of observed data, and (\textit{Step 3.}) the probabilistic forecast of the following weeks based on the estimated parameters. Steps 2 and 3 are repeated at every round of the rolling-window strategy described in Section~\ref{sec: forecasting_setting}, to produce forecasts which cover the whole period of interest. What differs is how each step is implemented: a \gls{pinn} surrogate of the \glspl{ode} system (\ref{eq: ODEs system}), presented later in this Section, followed by \gls{mcmc} inference (shown in the violet box in Fig~\ref{fig:frameworks}, and described in Section~\ref{sec: PINN_model}), and a stochastic mechanistic model of the same system with \gls{abcsmc} calibration (shown in the teal box in Fig~\ref{fig:frameworks}, and described in Section~\ref{sec: mechanistic_model}).

Furthermore, the two frameworks share the epidemiological structural core, represented in the first left block in Fig~\ref{fig:frameworks}. Consistently with our aim of investigating the baseline behavior of each framework, the shared setting is intentionally minimal (e.g., without any intervention or additional data such as human mobility), leaving both models to rely solely on the observed incidence data to produce their forecasts.
Both frameworks rely on the same compartmental structure, together with a set of common modeling assumptions. This structure is more articulated than the classical \gls{sir} model~\cite{kermack1927contribution}, following instead a \gls{seird} formulation. We consider an age-structured model and introduce an \textit{Exposed} (\textit{E}) compartment, taking into account the presence of a latency period before infectiousness, together with a \textit{Deaths} (\textit{D}) compartment paired with a corresponding observed-deaths quantity (\textit{$D^{obs}$}). This pair reflects both the age-sensitivity of mortality and the delay between an individual's death and its official notification. 

The deterministic formulation of this structure, learned by the \gls{pinn}-based framework (as described in Section~\ref{sec: PINN_modelapproximation}), is the following system of \glspl{ode}, while the mechanistic framework simulates it as a stochastic process (Section~\ref{sec: mechanistic_model}):
 
\begin{equation}
\begin{cases}
\frac{ds_a}{dt} &= -s_a \beta\sum_{a'=1}^{16}  \, C_{aa'} i_{a'} \\
\frac{de_a}{dt} &= s_a \beta\sum_{a'=1}^{16}  \, C_{aa'} i_{a'} - \varepsilon e_a \\
\frac{di_a}{dt} &= \varepsilon e_a - \xi i_a \\
\frac{dr_a}{dt} &= \xi (1 - \text{IFR}_a)\, i_a \\
\frac{dd_a}{dt} &= \xi \cdot \text{IFR}_a \cdot i_a - \delta d_a \\
\frac{dd^{\text{obs}}_a}{dt} &= \delta d_a
\label{eq: ODEs system}
\end{cases}
\end{equation}
where $s_a, e_a, i_a, r_a, d_a, d^{\text{obs}}_a : [t_0, T] \to [0,1]$ denote, respectively, the fraction of susceptible, exposed, infectious, recovered, deceased, and officially notified deceased individuals in each age class $a$. We consider $k_{\text{age}} = 16$ age classes, indexed by $a, a' = 1, \dots, 16$, with a five-year interval except for the last group, which is $75+$. Each compartment is normalized with respect to the total population of the corresponding age class in London, $N_a$, provided by official demographic data~\cite{nomis2020}, so that, for every age class $a$ and $\forall t \in [t_0, T]$,
\begin{equation}
s_a(t) + e_a(t) + i_a(t) + r_a(t) + d_a(t) + d^{\text{obs}}_a(t) = 1.
\label{eq: pop_conservation}
\end{equation}
The total population size is $N_{\text{tot}} = \sum_{a=1}^{16} N_a$.

Among the parameters of the \glspl{ode} system~(\ref{eq: ODEs system}), the latency rate $\varepsilon\in \mathbb{R}_{\geq 0}$ (i.e., the rate at which exposed individuals become infectious), the removal rate $\xi\in \mathbb{R}_{\geq 0}$ from the infectious compartment, the Infection Fatality Ratio $\text{IFR}_a$, and the contact matrix $C = (C_{aa'})_{a,a'=1}^{16}$ are fixed and shared by the two frameworks.
Since we consider a COVID-19-like disease spreading in London and the first wave of the pandemic in London for both the synthetic and the real COVID-19 scenarios, respectively, the fixed parameters are taken from the literature. In particular, we use the age-specific $\text{IFR}_a$ from~\cite{verity2020estimates} and the age-stratified contact matrix $C$ from~\cite{prem2017projecting}. According to the literature~\cite{kissler2020projecting,ma2020epidemiological,backer2020incubation,zhao2021estimating,xin2022estimating}, we set $\varepsilon = 1/3.7 \approx 0.27027$ day$^{-1}$ and $\xi = 1/2.5 = 0.4$ day$^{-1}$, to have a mean latency period of $3.7$ day and a mean infectious period of $2.5$ day, respectively. The remaining parameters: the disease transmission rate $\beta\in \mathbb{R}_{\geq 0}$ and, in the real COVID-19 scenario, the death notification rate $\delta\in \mathbb{R}_{\geq 0}$, are treated as unknown. 

For each week, both approaches estimate the basic reproduction number, $R_0$, that characterizes the epidemic threshold of the \glspl{ode} system~(\ref{eq: ODEs system}), from which $\beta$ can be obtained, together with the time-dependent effective reproduction number $R_t$. More details on the mathematical formulation and interpretation of both quantities can be found in the Appendix \ref{sec: methods_details} in \nameref{S1_Appendix}.   
The mean reporting delay $1/\delta$ between an individual's death and its official notification is also estimated in the real scenario.

Finally, both approaches share the same bounds during the calibration/inference phase (\textit{Step 2.} in Fig~\ref{fig:frameworks}), namely $R_0 \in [1.3, 3]$ and, for the real scenario, $1/\delta \in [7, 35]$ days.

In the synthetic scenarios, described in~\ref{sec: epidemiological_data}, the observed quantity is the weekly incidence of new infections (i.e., newly exposed individuals) rather than the reported deaths and it depends only on $s$, $i$ and $\beta$, unaffected by $\delta$ and $\text{IFR}_a$. The observation process (mortality and reporting delay) that characterizes the real data is removed, while the epidemic dynamics are left unchanged. Accordingly, the reporting delay is not estimated, and each framework fixes it to a constant value. 
Since the incidence of newly exposed individuals does not depend on $\delta$, this choice has no effect on the observed quantity, and the same \gls{pinn} and the same stochastic model can be used without modifications.

\subsection{Physics-informed neural network based model}
\label{sec: PINN_model}

This section presents the first of the two frameworks compared in this work, namely the \gls{pinn}-based framework coupled with \gls{mcmc} inference, shown in violet in Fig~\ref{fig:frameworks}. Following the three-step pipeline introduced in Section~\ref{sec: common_structural_setting}, the framework consists of:
\begin{itemize}
\item[] \textit{Step 1.} a \gls{pinn} trained once, on synthetic data, to approximate the solution of the compartmental \glspl{ode} system~(\ref{eq: ODEs system}) over a wide range of epidemic parameters, and then reused without retraining in all the scenarios and at every round of the rolling-window strategy. 
\item[] \textit{Step 2.} the inference, via \gls{mcmc}, of the unknown parameters from the look-back window of observed data, using the pre-trained network as forward model.
\item[] \textit{Step 3.} the probabilistic forecast of the following weeks, obtained by evaluating the network at the estimated parameters with their uncertainty.
\end{itemize}

\subsubsection{Compartmental model approximation via PINN}
\label{sec: PINN_modelapproximation}

The \gls{pinn} framework presented here is based on the hybrid framework introduced in our previous work, SIR-INN~\cite{rama2026forecasting}, and adapted to learn the compartmental epidemiological \glspl{ode} model~(\ref{eq: ODEs system}) introduced in Section~\ref{sec: common_structural_setting}.
This system can be written in the compact form
\begin{equation}
\frac{du(t;\lambda)}{dt} + F\big[u(t;\lambda); \lambda, \psi\big] = 0, \qquad t \in [t_0, T],
\label{eq: compact ODEs system}
\end{equation}
where $u(t;\lambda) := \big(s_a(t), e_a(t), i_a(t), r_a(t), d_a(t), d^{\text{obs}}_a(t)\big)_{a=1}^{16} \in [0,1]^{96}$ is the solution map, and $F$ is the parametric non-linear differential operator associated to the \glspl{ode} system~(\ref{eq: ODEs system})~\cite{Folland1995,Evans2022}. Here, $\lambda := \{R_0, 1/\delta\} \in \mathbb{R}^2_{>0}$ denotes the set of free parameters, and $\psi := \{\varepsilon, \xi, C, \text{IFR}_a, N_a\}$ denotes the set of parameters fixed a priori and shared with the mechanistic model, as illustrated in the first box of Fig~\ref{fig:frameworks}. Since $\psi$ is fixed, we omit its dependence in the notation of the neural network and in the following equations, in the same way we omit the dependence of the network on its own parameters (i.e., weights and biases), learned via neural network training~\cite{lecun2015deep,goodfellow2016deep}.
 
Given this formulation, the objective of the \gls{pinn} approach is to find a neural network $u_N$ that approximates the solution $u(t;\lambda)$ of the \gls{ode} system~(\ref{eq: ODEs system}), for fixed parameters. To allow the network to efficiently learn across a variety of epidemic scenarios, and consistently with our previous SIR-INN framework, we extend the temporal domain to a temporal-epidemic domain, which considers the parameters $\lambda$ as additional variables besides time. Hence, we construct the \gls{pinn} as a function
\begin{equation}
u_N: \mathcal{T} \times \mathcal{R} \times \mathcal{D} \to [0,1]^{96}, \qquad u_N(t, R_0, 1/\delta),
\end{equation}
where $\mathcal{T} = [t_{\min}, t_{\max}]$, $\mathcal{R} = [R_{0,\min}, R_{0,\max}]$, and $\mathcal{D} = [(1/\delta)_{\min}, (1/\delta)_{\max}]$. For the specific parameter ranges and the exact values of the parameters considered fixed, we refer the reader to Section~\ref{sec: common_structural_setting} and to Appendix \ref{sec: methods_details} in \nameref{S1_Appendix}.

Despite the essential role of age structure in describing COVID-19, where both contacts and mortality vary strongly across age groups~\cite{prem2017projecting, verity2020estimates, ram2021modified}, it remains rarely considered in \gls{pinn}-based studies~\cite{zhang2025dynamics, dimarco2025augmented, kong2025age}.
Motivated by this concern, we embed the age structure in the \gls{pinn}, by directly approximating the age-structured system of \glspl{ode}~(\ref{eq: ODEs system}). As a consequence, the network outputs each compartment of each age class, that is, $16\times6=96$ components. The age-specific dynamics are therefore learned explicitly for all the variables, and the aggregated quantities are learned accordingly.

The network's parameters are optimized during training by minimizing a loss function that, consistently with the \gls{pinn} approach~\cite{raissi2019physics}, combines a data-dependent term and a model-dependent term, allowing the neural network to simultaneously learn the underlying model and fit the data. Hence, we can write the loss function as
\begin{equation}
\mathcal{L} = \mathcal{L}_{data} + \alpha\mathcal{L}_{ODE},
\label{eq: loss_pinn}
\end{equation}
where $\alpha \in \mathbb{R}_{\geq 0}$ is a hyperparameter that balances the relative contribution of the two loss terms, weighting how strongly the physical constraint has to be relative to the data-fitting term.  
The term $\mathcal{L}_{data}$ is the standard \gls{mse} between the network output and a set of training data $(X_{tr}, Y_{tr})$, namely $\mathcal{L}_{data} = \text{MSE}(u_N, X_{tr}, Y_{tr})$. The term $\mathcal{L}_{ODE}$ instead penalizes deviations from the governing equations of the \glspl{ode} system~(\ref{eq: ODEs system}) depending on a so-called residual function $R(u,t;\lambda)$, Eq.~(\ref{eq: residual}) in Appendix \ref{sec: methods_details} in \nameref{S1_Appendix},
and it is expressed as the mean squared value of $R$ evaluated over a chosen set of collocation points, namely $\mathcal{L}_{ODE} = \frac{1}{|X_{coll}|} \sum_{(t,\lambda) \in X_{coll}} R\big(u_N, t; \lambda\big)^2$. 

As in the SIR-INN framework~\cite{rama2026forecasting}, we leverage a single feedforward neural network trained once on synthetic data generated according to the temporal-epidemic domain introduced above. The same trained network is then used in all the scenarios considered in this work, both real and synthetic, and at every round of the rolling-window strategy, without any retraining. This stage is schematically reported in \textit{Step 1.} of the violet box in Fig~\ref{fig:frameworks}. All the details about the neural network architecture and its training are described in Appendix \ref{sec: methods_details} in \nameref{S1_Appendix}.

\subsubsection{Parameter inference and forecasting}
\label{sec: PINN_inference&forecasting}

Once the neural network $u_N$ is trained, we implement an \gls{mcmc} method~\cite{neal1993probabilistic, richey2010evolution} (\textit{Step 12.} of the violet box in Fig~\ref{fig:frameworks}) to infer, from a limited window of data (e.g., weekly deaths in the real COVID-19 scenario), the epidemic features that characterize the disease, namely $\lambda = \{R_0, 1/\delta\}$. In the synthetic scenarios, even though the reporting delay $1/\delta$ is fixed to a constant value, whose choice is described in Appendix \ref{sec: methods_details} in \nameref{S1_Appendix}, it remains an input to the neural network, so that the same pre-trained \gls{pinn} can be used in both the synthetic and the real scenarios. 
 
The \gls{mcmc} method is implemented assuming as reference model our pre-trained \gls{pinn}, as prior distributions uniform priors for the parameters to be inferred, and as likelihood a negative Poisson log-likelihood, reflecting the count nature of the observed epidemiological data (i.e., weekly deaths or infections). Consistently with the mechanistic framework, the contribution of each weekly observation to the likelihood is weighted by the same weighting strategy, Eq.~(\ref{eq: weights}), so that more recent weeks have a greater influence.
 
For each rolling window of observations we run the \gls{mcmc} algorithm and, after selecting the distribution tails --- $1\,000$ samples from $10\,000$ iterations --- we obtain posterior samples of the inferred parameters, from which summary statistics and probabilistic forecasts can be derived.
In addition to these parameters, for each posterior set we derive a corresponding sample of the effective reproduction number $\hat R_t$.
 
Forecasting future epidemic states reduces to evaluating the pre-trained \gls{pinn} $u_N$ at the estimated parameters with their uncertainty, over the future time window (\textit{Step 3.} of the violet box in Fig~\ref{fig:frameworks}). 

Repeating parameter estimation and forecasting at each round of the rolling-window strategy provides probabilistic forecasts, in the form of prediction quantiles, which cover the entire period of interest.

For details on the parameter inference procedure, refer to Appendix \ref{sec: methods_details} in \nameref{S1_Appendix}.

\subsection{Mechanistic-based model}
\label{sec: mechanistic_model}

This section presents the second framework compared in this work, namely the stochastic mechanistic model coupled with \gls{abcsmc} calibration, shown in the teal box of Fig~\ref{fig:frameworks}. Following the three-step pipeline, it consists of:
\begin{itemize}
\item[] \textit{Step 1.} a stochastic simulation, for each parameter set, of the compartmental structure shared with the \gls{pinn}-based framework, which, unlike the neural network, requires no training.
\item[] \textit{Step 2.} the calibration of the unknown parameters on the look-back window of observed data via \gls{abcsmc}. 
\item[] \textit{Step 3.} the probabilistic forecast of the following weeks, obtained by extending beyond the look-back window the trajectories already simulated for the parameter sets accepted during the calibration step. 
\end{itemize}

\subsubsection{Compartmental model implementation}
\label{sec: mech_model_forward}

The mechanistic model relies on the same age-structured \gls{seird} structure, transition rates, and fixed parameters introduced in Section~\ref{sec: common_structural_setting}, but simulates it as a stochastic process in discrete time.  
Specifically, we use stochastic chain binomial processes denoted by $Bin(X, p)$, where $X$ is the number of individuals in a given compartment and $p$ is the probability of transitioning to another compartment within a time step $\Delta t$. 
This stage is schematically reported in \textit{Step 1.} of the teal box in Fig~\ref{fig:frameworks}. Further details on the stochastic formulation and implementation can be found in Appendix \ref{sec: methods_details} in \nameref{S1_Appendix}.

\subsubsection{Parameter inference and forecasting}
\label{sec: calibration&forecast_mech}

We calibrate the stochastic model using the \gls{abcsmc} algorithm~\cite{toni2009approximate}, following the procedure described in~\cite{Li2026}, to infer, from a limited window of data, the epidemic features that characterize the disease (\textit{Step 2.} of the teal box in Fig~\ref{fig:frameworks}). Equivalently to the \gls{pinn}-based framework, the parameters that are calibrated are the basic reproduction number $R_0$, from which $\beta$ is obtained, and, only in the real COVID-19 scenario, the death reporting delay $1/\delta$.

We assume uniform priors for all the unknown parameters, with the same bounds adopted for the \gls{pinn}-based framework where the parameters coincide (Section~\ref{sec: common_structural_setting}). Unlike the \gls{pinn}-based framework, which relies on a Poisson likelihood, the distance between simulated and observed data of the look-back window is measured through \gls{wmape}, with the same weighting criteria as Eq.~(\ref{eq: weights}), used in Step 2 of the \gls{pinn}-based framework. 
Details about the prior distributions, the exact form of the distance, and the algorithm settings are reported in Appendix \ref{sec: methods_details} in \nameref{S1_Appendix}.

For each rolling window of observations, we run the calibration via \gls{abcsmc} obtaining $1\,000$ posterior samples of the inferred parameters, from which summary statistics and probabilistic forecasts can be derived. In addition, as for the \gls{pinn}-based approach, for each posterior sample we derive the corresponding effective reproduction number $\hat R_t$.

Forecasts (\textit{Step 3.} of the teal box in Fig~\ref{fig:frameworks}) are obtained by extending, up to four weeks ahead of the look-back window, the trajectories already simulated for the parameter sets accepted during the \textit{Step 2.}. Since each trajectory is a single stochastic realization of the model, the forecast distribution reflects both the parameter uncertainty and the stochasticity of the process. 

As for the \gls{pinn}-based framework, iterating this two-step procedure over each rolling window of data provides probabilistic forecasts covering the entire period of interest.

\subsection{Forecasting evaluation}
\label{sec: metrics}
We adopt several metrics to evaluate the forecasting performance of both frameworks, partly following those adopted in \cite{fiandrino2025collaborative} and complementing them with additional measures. Together, these metrics allow us to investigate three distinct aspects of forecasting performance: the accuracy of the point forecasts (with \gls{mae}, and \gls{rmse}), the overall probabilistic accuracy of the forecasts, jointly capturing the interval- and point-forecast accuracy (with \gls{wis} and IS$_{80}$), and the calibration of the associated prediction intervals ($50\%$ and $90\%$ coverages). Depending on the analysis, each of these metrics can be computed on a single observation, aggregated across forecast horizons or rounds, or aggregated over the entire evaluation period. 

In contrast, \gls{wmape} is not used as a validation metric in this work, but rather as a selection criterion during the \gls{abcsmc} calibration process of the mechanistic framework.

In addition to the two frameworks, we also implemented a baseline model, used as a reference for the evaluation of the forecasts with the following metrics. It is a naive model that does not rely on any epidemiological knowledge and uses the last known value as forecasts across each horizon. The baseline is not part of the comparison between the two frameworks; for this reason, it is not shown in Fig~\ref{fig:frameworks}, and its detailed description is reported in Appendix \ref{sec: methods_details} in \nameref{S1_Appendix}.

\paragraph{Mean Absolute Error (MAE).}
The \gls{ae} quantifies the discrepancy between a probabilistic forecast $f$, with median $m$, and the corresponding observation $y$,
\begin{equation}
AE(f,y) = |m-y|.
\end{equation}
Given a sequence of observations $y=(y_1,\dots,y_M)$ and the associated forecasts $f=(f_1,\dots,f_M)$, with medians $m_1,\dots,m_M$, the \gls{mae} averages the \gls{ae} across time steps,
\begin{equation}
MAE(f,y) = \frac{1}{M}\sum_{i=1}^M AE(f_i,y_i).
\end{equation}

\paragraph{Weighted Mean Absolute Percentage Error (WMAPE).}
Given a temporal sequence of observations $y=(y_1,\dots,y_M)$ and the associated forecasts $f=(f_1,\dots,f_M)$, with medians $m_1,\dots,m_M$, the \gls{wmape} is a scale-independent metric, defined as
\begin{equation}
WMAPE(f,y) = \frac{\sum_{i=1}^M |m_i - y_i|}{\sum_{i=1}^M |y_i|}.
\end{equation}
Unlike the standard Mean Absolute Percentage Error (MAPE), which normalizes each individual error by the corresponding observation, \gls{wmape} normalizes the aggregated absolute error by the aggregated observed values, making it more robust when some observations are close to, or equal to, zero. 

\paragraph{Root Mean Squared Error (RMSE).}
Given a sequence of observations $y=(y_1,\dots,y_M)$ and the associated forecasts $f=(f_1,\dots,f_M)$, with medians $m_1,\dots,m_M$, the \gls{rmse} is defined as
\begin{equation}
RMSE(f,y) = \sqrt{\frac{1}{M}\sum_{i=1}^M (m_i - y_i)^2}.
\end{equation}
Unlike \gls{mae}, it penalizes larger deviations more heavily by squaring each individual error before averaging.

\paragraph{Interval Score (IS) and Weighted Interval Score (WIS).}
Denoting by $y$ the observed value and by $f$ a probabilistic forecast with median $m$, for a nominal level $\alpha \in [0,1)$ let $[\ell,u]$, $\ell,u \in \mathbb{R}$, denote the corresponding $(1-\alpha)\cdot100\%$ prediction interval of $f$. We first define the \gls{is} as
\begin{equation}
IS_\alpha(f,y) =
\begin{cases}
u - \ell + \dfrac{2}{\alpha}(\ell - y), & y < \ell, \\[4pt]
u - \ell, & \ell \le y \le u, \\[4pt]
u - \ell + \dfrac{2}{\alpha}(y - u), & u < y.
\end{cases}
\label{eq: is}
\end{equation}
It evaluates the probabilistic forecasting performance on the specific prediction interval of the $\alpha$ nominal level.

The Weighted Interval Score provides a proper approximation of the Continuous Ranked Probability Score (CRPS)~\cite{bracher2021evaluating}, jointly capturing the accuracy of the forecast median and the reliability of the prediction intervals in containing the observed value. It extends the \gls{is}, originally defined for a single prediction interval, to a weighted combination across multiple nominal coverage levels.  
The \gls{wis} is obtained by averaging the quantity of Eq.~(\ref{eq: is}) over $K \in \mathbb{N}$ nominal levels $\alpha_1,\dots,\alpha_K > 0$, with positive weights $w_0, w_1,\dots,w_K$:
\begin{equation}
WIS_{\alpha_1,\dots,\alpha_K}(f,y) = \frac{1}{K+\frac12}\left(w_0 \cdot |y-m| + \sum_{k=1}^K w_k \cdot IS_{\alpha_k}(f,y)\right).
\end{equation}
Following~\cite{howerton2023evaluation,fiandrino2025collaborative,wang2026naive}, we evaluate our forecasts with an \gls{is} of $\alpha=0.2$ (i.e., corresponding to the 80\% prediction interval), and for the \gls{wis}, we set $K=11$, $\alpha_k \in \{0.02, 0.05, 0.10, 0.20, 0.30, 0.40, 0.50, 0.60, 0.70, 0.80, 0.90\}$, $w_0 = 1/2$, and $w_k = \alpha_k/2$.

\paragraph{Coverage.}
Coverage evaluates the calibration of a probabilistic forecasting model, namely the reliability of its prediction intervals, measured as the empirical fraction of times the prediction interval actually contains the observed value. For a well-calibrated model, the empirical coverage is expected to closely approximate the nominal level of the interval considered (e.g., a 90\% prediction interval is expected to contain the observation approximately 90\% of the time). Considering the $(1-\alpha)\cdot100\%$ interval $[\ell,u]$ of forecast $f$, and a temporal sequence of observations $y=(y_1,\dots,y_M)$ with forecasts $f=(f_1,\dots,f_M)$, the coverage of level $\alpha \in [0,1)$ is defined as
\begin{equation}
C_\alpha(f,y) = \frac{1}{M}\left|\{i \in \{1,\dots,M\} : \ell_i \le y_i \le u_i\}\right|.
\end{equation}
We consider both the 50\% ($\alpha=0.5$) and 90\% ($\alpha=0.1$) coverage levels.

\section{Results}
\label{sec: results}

\subsection{Synthetic scenarios}
\label{sec: synth_results}

\subsubsection{Retrospective fitting}
\label{sec: synth_fitting&params_results}

Before comparing the forecasts, we investigate how well the two frameworks reproduce the data on which they are calibrated and how well they identify the parameter used to generate them. For each synthetic scenario, both frameworks are calibrated on the whole series of weekly incidence, aggregated across age classes ($34$ weeks for $R_0=1.5$ and $24$ weeks for $R_0=2.0$), and the incidence reconstructed by each framework is compared with the observed one (namely, retrospective fitting). In contrast with a real scenario in which the value of the basic reproduction number cannot be observed, in the case of synthetic data analyzed here, $R_0$ is known, and so we can also compare the posterior distributions of the estimates with it. 

\begin{figure}[tbp]
    \centering
    \resizebox{\textwidth}{!}{
        \includegraphics[height=4.7cm]{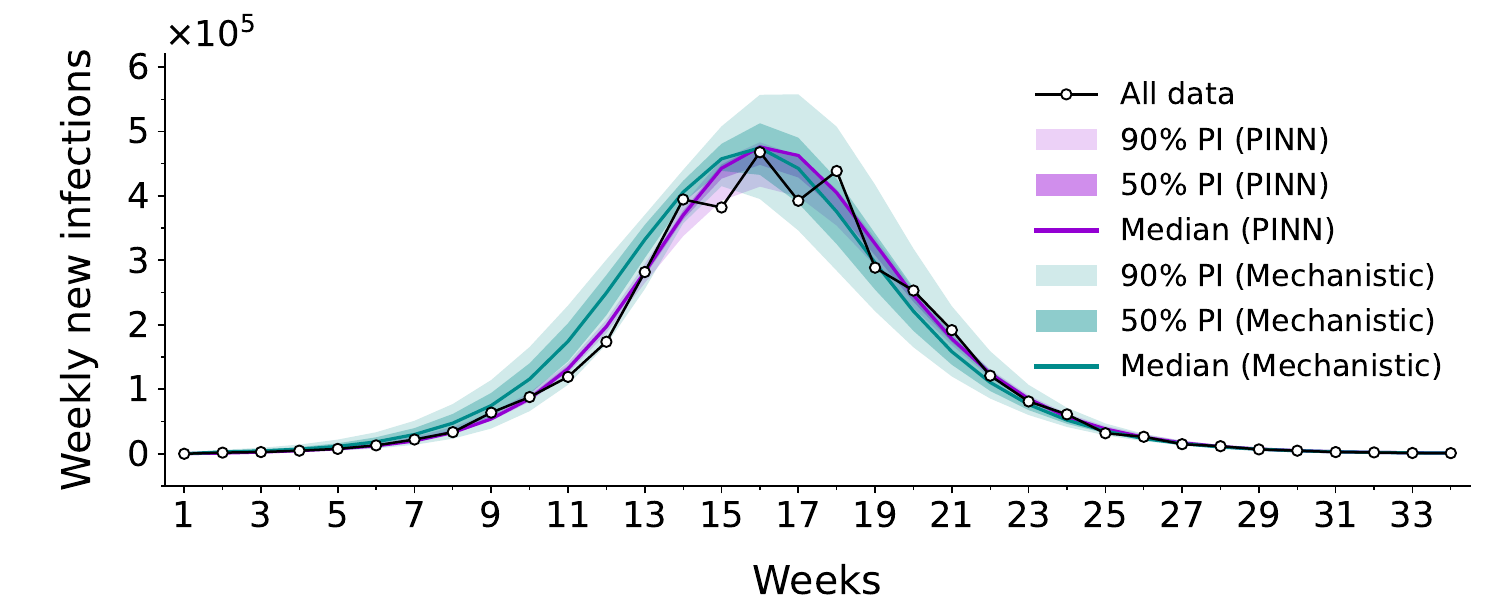}
        \hspace{0.3cm}%
        \includegraphics[height=4.7cm]{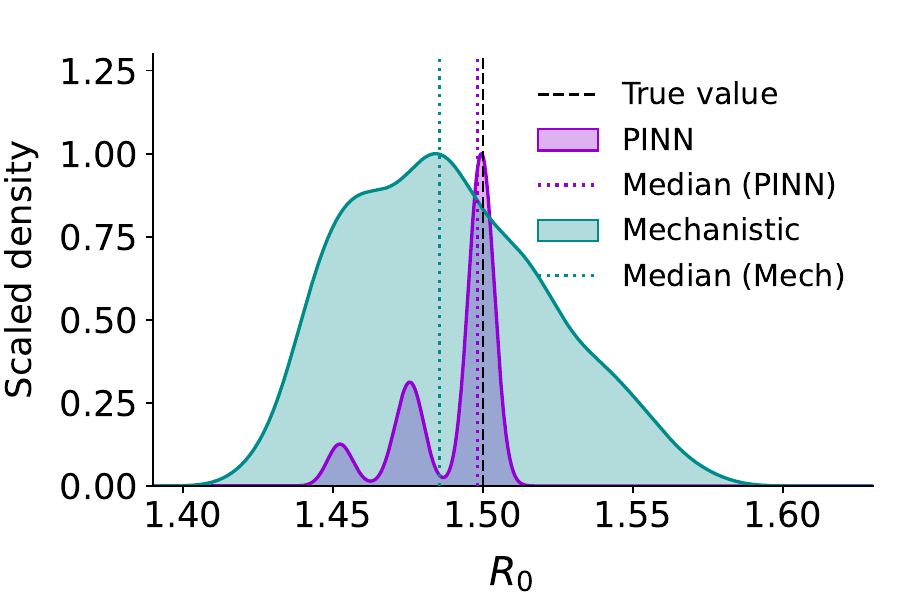}
        }

    \vspace{0.5em}

    \resizebox{\textwidth}{!}{
        \includegraphics[height=4.7cm]{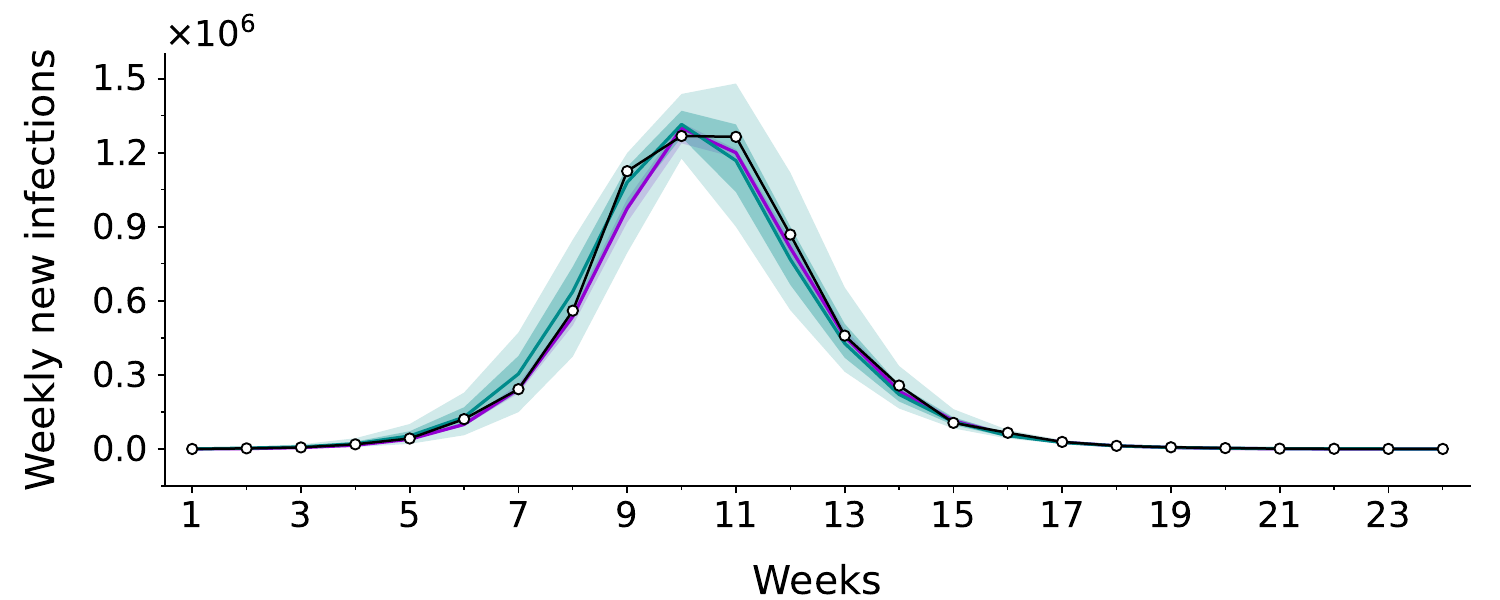}
        \hspace{0.3cm}
        \includegraphics[height=4.7cm]{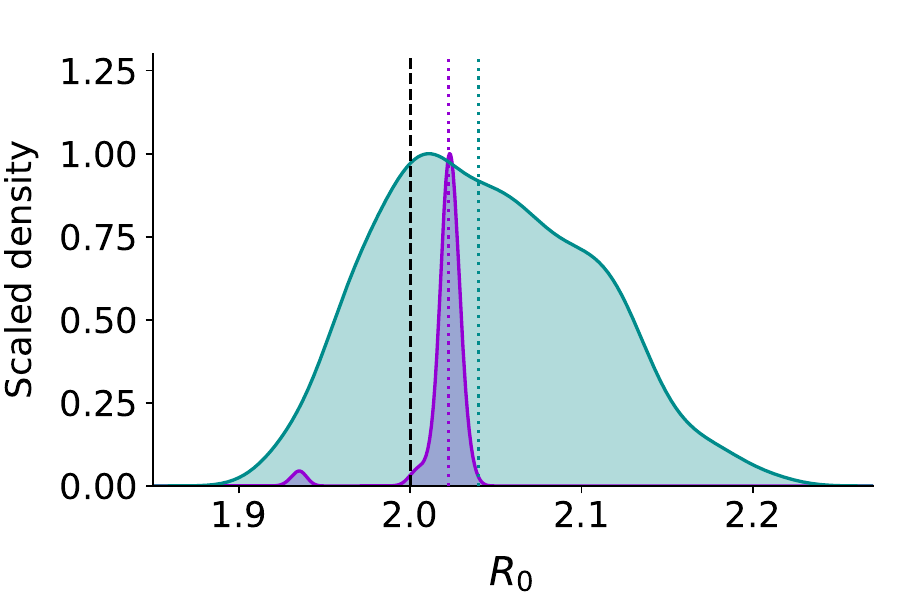}
        }
        \vspace{0.4em}
    \caption{\textbf{Retrospective fit and $R_0$ posteriors in the two synthetic scenarios.} Top row: first scenario ($R_0=1.5$), with data up to week $34$; bottom row: second scenario ($R_0=2.0$), with data up to week $24$. Left panels: retrospective fit of the weekly new infections obtained by the \gls{pinn}-based (violet) and mechanistic-based (teal) frameworks, calibrated on all the noisy observations up to the last week (white dots). Shaded areas show the $90\%$ (lighter) and $50\%$ (darker) prediction intervals, and lines the medians. Right panels: posterior distributions of $R_0$ estimated by the two frameworks at the same week, with densities scaled to a maximum of $1$. Dotted lines mark the posterior medians and the dashed black line the true value of $R_0$ used to generate the data.}
    \label{fig: synth_retro_R0}
\end{figure}

Figure~\ref{fig: synth_retro_R0} reports, for both scenarios ($R_0=1.5$ in the top row and $R_0=2.0$ in the bottom row), the retrospective fit of the aggregated incidence (left column) and the posterior distributions of $R_0$ (right column).
In the left column, the medians of both frameworks follow the epidemic curve from the growth phase to the tail in both scenarios, perfectly fitting the real incidence during the earliest weeks and in its last decline. Since the observation noise is not explicitly embedded into the models, they do not reproduce the week-to-week oscillations of the observed incidence, mostly concentrated around the peak. 
The main difference appears in the first scenario, where the mechanistic median overestimates the incidence in the growth phase by up to about $50\%$ and slightly underestimates it in the decline, while the \gls{pinn}-based median stays closer to the observations. In the second scenario, both medians are close to the observed curve over the whole period.

The two frameworks differ mainly in the width of their intervals. The intervals of the \gls{pinn}-based framework are narrow in both scenarios, almost invisible in the second one, and contain $59\%$ and $29\%$ of the observations at the $90\%$ level in the first and second scenario, respectively, against $97\%$ and $100\%$ for the mechanistic framework, whose intervals are wider, and widest around the peak, finally shrinking in the tail.

The right column shows the posterior distributions of $R_0$. The medians are close to the true value for both frameworks, with relative errors of $0.1\%$ (\gls{pinn}-based) and $1.0\%$ (mechanistic) in the first scenario, and $1.1\%$ and $2.0\%$ in the second. In the first scenario, the true value falls within the $90\%$ interval of both frameworks; in the second, it falls within that of the mechanistic one, but slightly below that of the \gls{pinn}-based one. The \gls{pinn}-based posteriors are much sharper than the mechanistic ones, by about $2$ and $7$ times in the first and second scenario, respectively.

Overall, the posterior median of the \gls{pinn}-based framework is closer to the true $R_0$, which is consistent with its retrospective fit following the epidemic curve more closely. However, its posteriors are also much narrower, which makes it overconfident about the reconstructed incidence. Similar observations can be derived from the estimate of the age-specific incidence, reported in Appendix \ref{sec: appendix_sanity_synthetic} in \nameref{S1_Appendix}.
For the results about the retrospective fitting and the basic reproduction number inference related to the synthetic scenarios generated from the same value of $R_0$ under a stronger level of noise, see Figure~\ref{fig: synth_retro_R0_noise} in Appendix \ref{sec: appendix_noise_synthetic} in \nameref{S1_Appendix}.

\subsubsection{Forecasting results}
\label{sec: synth_forecasting_results}

We evaluate the forecasting performance of the two frameworks on the two synthetic scenarios. Following the rolling-window strategy, at each round both frameworks produce probabilistic forecasts from $1$ to $4$ weeks ahead, using all the past data available up to the current week. The forecasts are evaluated on the same set of rounds at all the horizons, namely $27$ rounds for the first scenario ($34$ weeks in total) and $17$ rounds for the second one ($24$ weeks in total). The metrics (Section~\ref{sec: metrics}) are computed separately for each scenario, for each horizon and aggregated over all horizons and rounds. The error-based metrics (i.e., \gls{mae}, \gls{rmse}, \gls{wis}, and IS$_{80}$) are normalized by the score of the baseline, computed for these synthetic scenarios as explained in Appendix \ref{sec: methods_details} in \nameref{S1_Appendix}, so that values above $1$ indicate worse performance than the baseline.

\paragraph{Overall performance.}
Table~\ref{tab:results_pooled_synth1&2} reports the metrics aggregated over all horizons and rounds. In both scenarios, the two frameworks reduce the error-based scores by about $72$--$91\%$ with respect to the baseline, and the mechanistic framework obtains lower values than the \gls{pinn}-based one on all the error-based metrics, by about $14$--$22\%$ in the first scenario and $25$--$28\%$ in the second.
The two frameworks differ more in the coverage of their prediction intervals than in their accuracy. The intervals of the \gls{pinn}-based framework are too narrow: across the two scenarios, they contain the observations in $21$--$34\%$ of the cases for the $50\%$ interval and in $32$--$55\%$ for the $90\%$ interval. The mechanistic framework shows the opposite behavior, with intervals wider than needed: their coverage is $81$--$96\%$ for the $50\%$ interval, and $100\%$ for the $90\%$ interval in both scenarios.

\begin{table}[htbp]
\centering
\begin{tabular}{llcccccc}
\toprule
Scenario & Model & MAE & RMSE & WIS & IS$_{80}$ & Cov.\ 50\% & Cov.\ 90\% \\
\midrule
\multirow{3}{*}{\makecell[l]{Scenario 1\\ $R_0=1.5$}}
& \textsc{PINN} & 0.227 & 0.283 & 0.224 & 0.222 & \textbf{0.343} & 0.546 \\
& \textsc{Mechanistic} & \textbf{0.189} & \textbf{0.222} & \textbf{0.191} & \textbf{0.191} & 0.806 & \textbf{1.000} \\
& \textit{Baseline} & 1.000 & 1.000 & 1.000 & 1.000 & 0.333 & 0.583 \\
\midrule
\multirow{3}{*}{\makecell[l]{Scenario 2\\ $R_0=2.0$}}
& \textsc{PINN} & 0.127 & 0.162 & 0.126 & 0.137 & 0.206 & 0.324 \\
& \textsc{Mechanistic} & \textbf{0.094} & \textbf{0.119} & \textbf{0.094} & \textbf{0.099} & 0.956 & \textbf{1.000} \\
& \textit{Baseline} & 1.000 & 1.000 & 1.000 & 1.000 & \textbf{0.412} & 0.588 \\
\bottomrule
\end{tabular}
\caption{\textbf{Comparison of overall forecasting performance on the two synthetic scenarios.} Forecasts of weekly new infections, on the noisy synthetic data with $R_0=1.5$ and $R_0=2.0$, aggregated over all horizons (from 1 to 4) and rounds. MAE, RMSE, WIS, and IS$_{80}$ are normalized by the baseline score of the same scenario. Bold indicates the best performance among \textsc{PINN}, \textsc{Mechanistic}, and \textit{Baseline} within each scenario; for coverage, the value closest to the nominal level.}
\label{tab:results_pooled_synth1&2}
\end{table}

Taken together, both frameworks substantially over-perform the baseline. The mechanistic framework obtains the lowest errors in both point- and interval-forecast accuracy in the two scenarios. However, the two frameworks tend to mis-calibrate their prediction intervals, and in opposite directions, the \gls{pinn}-based one by under-covering and the mechanistic one by over-covering, so that the former is closer to the nominal level for the $50\%$ interval and the latter for the $90\%$ interval.

\paragraph{Performance by forecast horizon.} 
Table~\ref{tab:results_multihorizon_synth} reports the same metrics separately for each horizon and scenario. The mechanistic framework has lower errors than the \gls{pinn}-based one at almost all horizons and metrics, and the gap is largest at horizon 1, where its errors are lower by about $20$--$35\%$ in the first scenario and $31$--$51\%$ in the second. The exceptions are at horizon 4 in the first scenario, where the \gls{pinn}-based framework has slightly lower \gls{wis} and IS$_{80}$ (the mechanistic values are about $7\%$ and $19\%$ higher), and at horizon 3 in the second scenario, where the two frameworks are almost equivalent.
In both scenarios, the normalized scores of the two frameworks generally decrease with the horizon, with the exception of horizon 4 in the second scenario, where the \gls{pinn}-based scores increase for all the error-based metrics and the mechanistic ones for the \gls{rmse}. Since the scores are normalized by the baseline at the same horizon, this decrease indicates that the errors of the baseline grow faster with the horizon than those of the two frameworks, and not necessarily that the forecasts at longer horizons are more accurate in absolute terms.

As for the overall performance (Table~\ref{tab:results_pooled_synth1&2}), the two frameworks mis-calibrate their prediction intervals in opposite directions at every horizon. The intervals of the \gls{pinn}-based framework are too narrow, containing the observations in about $12$--$41\%$ of the cases for the $50\%$ interval and in $18$--$63\%$ for the $90\%$ interval, with lower values in the second scenario. The intervals of the mechanistic framework are instead too wide: the $90\%$ interval contains all the observations at every horizon in both scenarios, and the $50\%$ interval contains $78$--$100\%$ of them. As a result, the $50\%$ coverage closest to the nominal level is obtained by the baseline or by the \gls{pinn}-based framework, and the $90\%$ one by the mechanistic framework, at every horizon.

\begin{table}[htbp]
\centering
\footnotesize
\setlength{\tabcolsep}{2pt}   
\resizebox{\textwidth}{!}{%
\begin{tabular}{cl cccccc | cccccc}
\toprule
 & & \multicolumn{6}{c|}{Scenario 1: $R_0=1.5$} & \multicolumn{6}{c}{Scenario 2: $R_0=2.0$} \\
\cmidrule(lr){3-8} \cmidrule(lr){9-14}
Horizon & Model & MAE & RMSE & WIS & IS$_{80}$ & $C_{50\%}$ & $C_{90\%}$ & MAE & RMSE & WIS & IS$_{80}$ & $C_{50\%}$ & $C_{90\%}$ \\
\midrule
H1 & \textsc{PINN}        & 0.477 & 0.583 & 0.513 & 0.532 & 0.259 & 0.444 & 0.295 & 0.340 & 0.320 & 0.369 & 0.176 & 0.294 \\
   & \textsc{Mechanistic} & \textbf{0.382} & \textbf{0.453} & \textbf{0.359} & \textbf{0.346} & 0.778 & \textbf{1.000} & \textbf{0.204} & \textbf{0.228} & \textbf{0.186} & \textbf{0.181} & 0.941 & \textbf{1.000} \\
   & \textit{Baseline}    & 1.000 & 1.000 & 1.000 & 1.000 & \textbf{0.296} & 0.630 & 1.000 & 1.000 & 1.000 & 1.000 & \textbf{0.412} & 0.647 \\
\midrule
H2 & \textsc{PINN}        & 0.272 & 0.337 & 0.277 & 0.288 & 0.333 & 0.556 & 0.136 & 0.191 & 0.139 & 0.156 & 0.235 & 0.471 \\
   & \textsc{Mechanistic} & \textbf{0.257} & \textbf{0.303} & \textbf{0.240} & \textbf{0.237} & 0.778 & \textbf{1.000} & \textbf{0.104} & \textbf{0.115} & \textbf{0.100} & \textbf{0.112} & 1.000 & \textbf{1.000} \\
   & \textit{Baseline}    & 1.000 & 1.000 & 1.000 & 1.000 & \textbf{0.370} & 0.630 & 1.000 & 1.000 & 1.000 & 1.000 & \textbf{0.412} & 0.588 \\
\midrule
H3 & \textsc{PINN}        & 0.203 & 0.289 & 0.204 & 0.201 & \textbf{0.407} & 0.556 & 0.082 & 0.100 & \textbf{0.081} & \textbf{0.086} & 0.294 & 0.353 \\
   & \textsc{Mechanistic} & \textbf{0.155} & \textbf{0.202} & \textbf{0.164} & \textbf{0.165} & 0.778 & \textbf{1.000} & \textbf{0.074} & \textbf{0.088} & 0.081 & 0.088 & 0.941 & \textbf{1.000} \\
   & \textit{Baseline}    & 1.000 & 1.000 & 1.000 & 1.000 & 0.333 & 0.556 & 1.000 & 1.000 & 1.000 & 1.000 & \textbf{0.412} & 0.529 \\
\midrule
H4 & \textsc{PINN}        & 0.142 & 0.190 & \textbf{0.128} & \textbf{0.123} & \textbf{0.370} & 0.630 & 0.109 & 0.154 & 0.100 & 0.104 & 0.118 & 0.176 \\
   & \textsc{Mechanistic} & \textbf{0.118} & \textbf{0.151} & 0.137 & 0.146 & 0.889 & \textbf{1.000} & \textbf{0.074} & \textbf{0.122} & \textbf{0.076} & \textbf{0.077} & 0.941 & \textbf{1.000} \\
   & \textit{Baseline}    & 1.000 & 1.000 & 1.000 & 1.000 & 0.333 & 0.519 & 1.000 & 1.000 & 1.000 & 1.000 & \textbf{0.412} & 0.588 \\
\bottomrule
\end{tabular}%
}
\caption{\textbf{Comparison of horizon-specific forecasting performance on the two synthetic scenarios.} Forecasts of weekly new infections on the noisy synthetic data with $R_0=1.5$ and $R_0=2.0$, aggregated over all rounds (27 rounds in the first scenario and 17 in the second) for each horizon, from 1 to 4. MAE, RMSE, WIS, and IS$_{80}$ are normalized by the baseline score of the same scenario and horizon. Bold indicates the best performance among \textsc{PINN}, \textsc{Mechanistic}, and \textit{Baseline} within each scenario and horizon; for coverage, the value closest to the nominal level.}
\label{tab:results_multihorizon_synth}
\end{table}

\begin{figure}[htbp]
    \centering
    \resizebox{\textwidth}{!}{%
        \includegraphics[height=10cm]{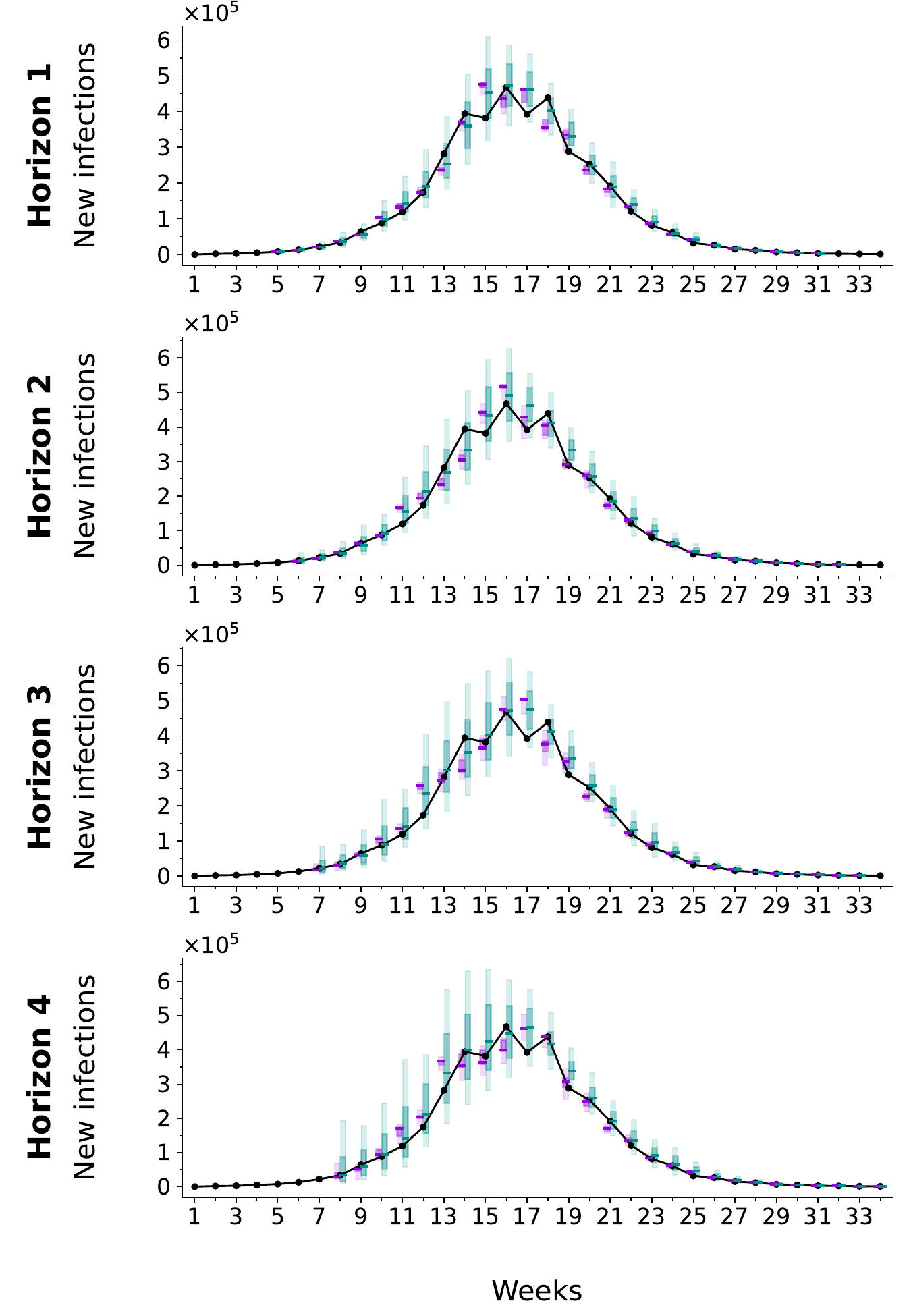}%
        \hspace{0.3cm}%
        \includegraphics[height=10cm]{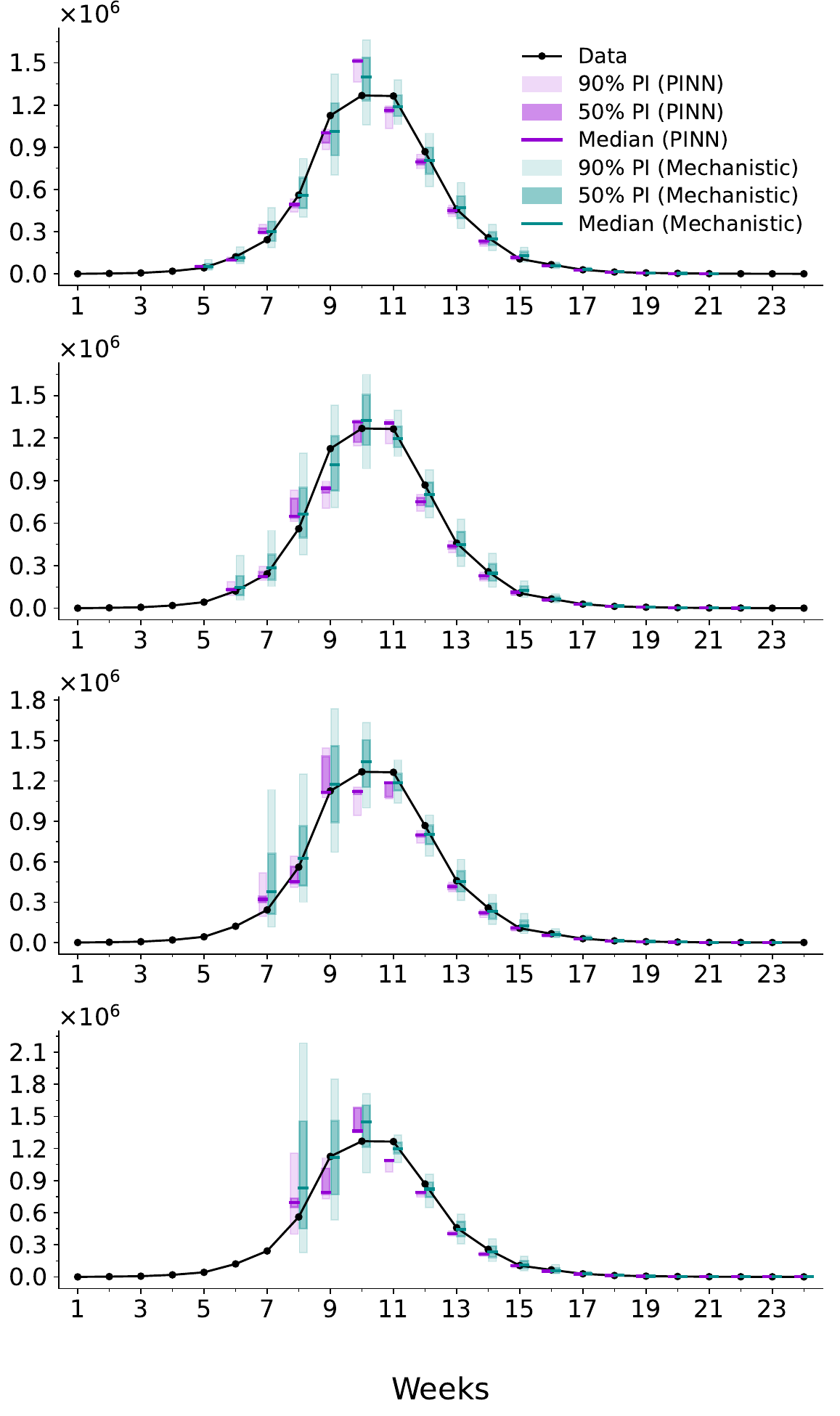}%
    }
   \caption{\textbf{Horizon-specific probabilistic forecasts on the two synthetic scenarios.} Forecasts from the \gls{pinn}-based (in violet) and mechanistic (in teal) frameworks of the weekly new infections on the noisy synthetic data with $R_0=1.5$ (on the left) and $R_0=2.0$ (on the right), at fixed horizons of 1 to 4 weeks ahead (panels from top to bottom). In each panel, every box refers to the week on the x-axis and shows the forecast made, for that week, as many weeks earlier as the horizon of the panel, using only the data available at that time. The lighter and darker boxes show the 90\% and 50\% prediction intervals, respectively, and the horizontal line inside each box shows the median. The observed weekly new infections are shown in black.}    \label{fig:comparison_horizons_synth}
\end{figure}

Figure~\ref{fig:comparison_horizons_synth} shows the probabilistic forecasts of the two frameworks at the four horizons, round by round, against the weekly new infections of the two scenarios, with $R_0=1.5$ in the left column and $R_0=2.0$ in the right column. Each row corresponds to a fixed horizon, so that the boxes at a given week summarize the forecast that each framework had produced for that week, 1 to 4 weeks earlier, without knowing the subsequent observations. At each week, the boxes of the two frameworks are placed side by side, in violet for the \gls{pinn}-based framework and in teal for the mechanistic one: the lighter box represents the $90\%$ prediction interval, the darker one the $50\%$ prediction interval, and the horizontal line inside the box marks the median. The incidence is shown in black over the whole simulated period. The two scenarios describe different epidemics, so the scales of the two columns differ: with $R_0=2.0$ the epidemic grows faster, reaches a higher peak, and ends earlier, so it is observed for $24$ weeks, against $34$ weeks for $R_0=1.5$. We consider only the rounds for which all four horizons fall within the simulated period, so that each panel contains the same number of box pairs, $27$ in the first scenario and $17$ in the second, shifted by one week from one horizon to the next.

In general, Figure~\ref{fig:comparison_horizons_synth} shows that the forecasts of both frameworks are accurate in both scenarios, as they follow the whole epidemic curve at all horizons, with the largest discrepancies and the largest uncertainty concentrated around the peak, as expected. Already the earliest forecasts, produced after observing only $4$ weeks of data, are close to the observed values, and their $90\%$ prediction intervals contain them at all horizons. As the growth accelerates, the medians of both frameworks alternate between over- and underestimation. 

Around the peak, the noise added to the data produces week-to-week oscillations of the observed incidence, which neither framework can anticipate.
Even so, the median forecasts of both frameworks for the peak week remain close to the observed peak: they deviate from it by at most about $15\%$ (first scenario) and $18\%$ (second scenario) for the \gls{pinn}-based framework, and by about $4\%$ and $14\%$ for the mechanistic one, and the deviation does not increase systematically with the horizon. The median forecasts of the mechanistic framework vary over a range about three times narrower across the horizons than those of the \gls{pinn}-based one, in both scenarios. Moreover, the observed peak falls within the $50\%$ interval of the mechanistic framework at all horizons in both scenarios, while for the \gls{pinn}-based framework this happens only once, in the second scenario at horizon 2.

After the peak, the forecasts of both frameworks are close to the observations at all horizons, with only small, scenario-dependent deviations.
Towards the tail, the boxes of both frameworks approach the observed incidence, which is almost zero, at all horizons.

The intervals of the \gls{pinn}-based framework remain narrow over the whole simulated period, so that around the peak, where the oscillations of the observed incidence are largest, the observations often fall outside both its $50\%$ and $90\%$ intervals. The intervals of the mechanistic framework are wider and contain all the observations, and their width depends on the phase of the epidemic and on the horizon: they are widest in the first rounds of the growth phase at the longest horizons, where the upper limit of the $90\%$ interval can reach about $4$--$6$ times the observed value, and they shrink progressively in the decline. This behavior is consistent with the coverages in Table~\ref{tab:results_multihorizon_synth}: below the nominal levels for the \gls{pinn}-based framework and above them for the mechanistic one.

\vspace{0.5cm}
In summary, in both synthetic scenarios the error-based metrics of both frameworks are well below those of the baseline, and the mechanistic framework obtains lower errors than the \gls{pinn}-based one for almost all the metrics and horizons. Furthermore, the \gls{pinn}-based framework tends to produce intervals that are too narrow, with coverages below the nominal levels, while the intervals predicted by the mechanistic framework are too wide, with coverages above them.
The same pattern can be observed in the forecasts of all the rounds, reported in Figure~\ref{fig:forecastig_comparison_synt1} and Figure~\ref{fig:forecastig_comparison_synt2} for the first and second scenario, respectively, in Appendix \ref{sec: appendix_full_synthetic} in \nameref{S1_Appendix}.

\begin{figure}[tbp]
    \centering
    \includegraphics[width=0.9\textwidth]{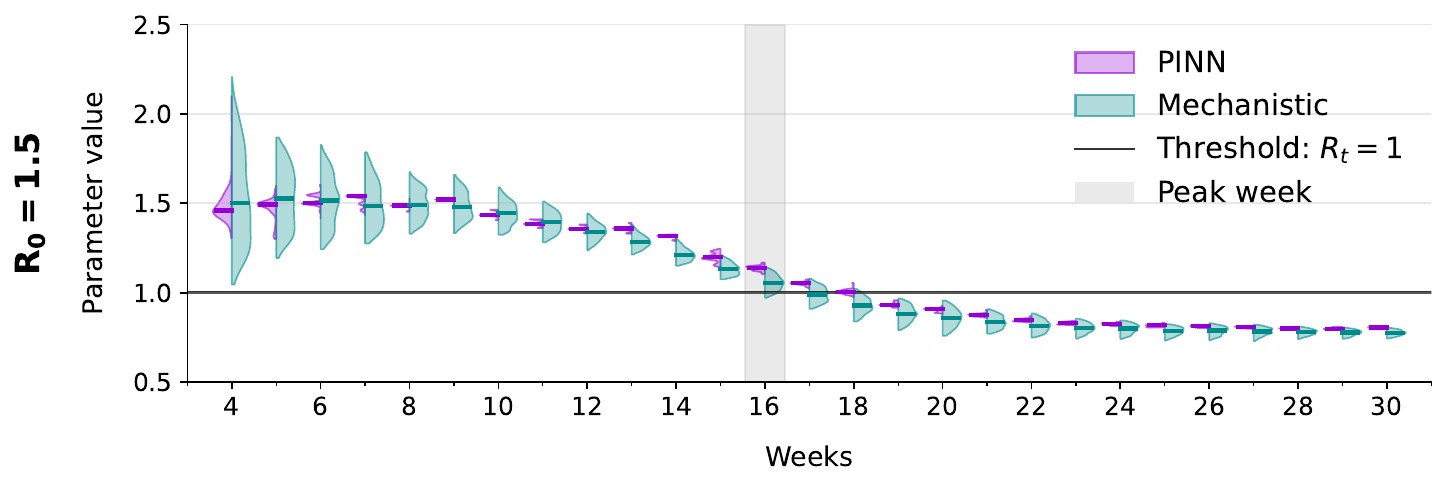}

    \vspace{0.5em}

    \includegraphics[width=0.9\textwidth]{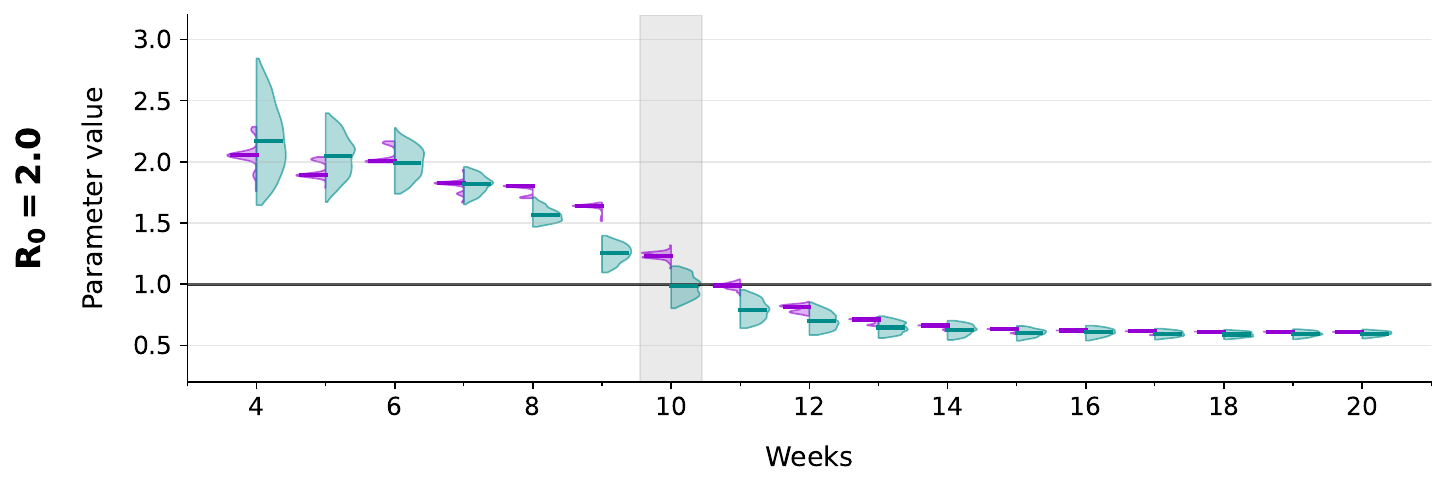}

    \caption{\textbf{Posterior distributions of $R_t$ across rounds in the two synthetic scenarios.} For each round, the split violin shows the posterior distribution of the effective reproduction number by the PINN-based framework with MCMC (left half, violet) and by the mechanistic framework via ABC-SMC (right half, teal). Each split violin shows the full shape of the $1\,000$ tail posterior samples for that window; horizontal ticks mark the per-window median for each model. In addition, the solid horizontal line marks the epidemic threshold $R_t = 1$, below which the outbreak is expected to decline, and the shaded vertical band marks the week of the peak in weekly new infections, week $16$ in the first scenario ($R_0=1.5$, top) and week $10$ in the second ($R_0=2.0$, bottom).} 
    \label{fig:Re_violin_synth}
\end{figure}

A similar behavior is observed in the estimates of the effective reproduction number $R_t$ across the rounds, illustrated in Figure~\ref{fig:Re_violin_synth}. In both scenarios, the posterior distributions of the \gls{pinn}-based framework are narrow in all rounds, while those of the mechanistic framework are wide in the first rounds and become progressively narrower as more observations are included, consistently with the width of the respective prediction intervals.
The medians of the two frameworks are close over most of the epidemic: in the first rounds they are close to the value of $R_0$ used to generate the data, as expected when almost the whole population is susceptible, and they decrease below 1 after the peak of the observed incidence. The largest differences are concentrated around the peak, where the \gls{pinn}-based framework estimates higher values of $R_t$ than the mechanistic one, so that the mechanistic framework crosses the threshold $R_t=1$ at the same time as, or slightly earlier than, the \gls{pinn}-based one, at or shortly after the peak of the observed incidence. 

The same analyses, repeated on synthetic incidence generated with a higher level of noise for both scenarios, are also reported in the Appendix \ref{sec: appendix_noise_synthetic} in \nameref{S1_Appendix}, in terms of overall, horizon-specific  performance (see also Figure~\ref{fig:comparison_horizons_synth_noise}), and of the effective reproduction number estimation (Figure~\ref{fig:Re_violin_synth_noise}). The conclusions are similar to those drawn here, although the performance of both frameworks worsens slightly, as expected. One exception is for the \gls{pinn}-based framework that surprisingly improves the Coverage at $90\%$ with respect to both the synthetic scenarios generated with a lower level of noise.

\subsection{COVID-19 forecasting}
\label{sec: COVID_forecasting_results}

To generalize the comparison between the two frameworks beyond the synthetic scenarios, we apply the same pipeline to a considerably more challenging real test case: the weekly death incidence of the first COVID-19 wave in London.
The difficulty of this scenario resides in the dynamics of COVID-19, which rapidly evolves in response to time-varying factors, such as the implementation of non-pharmaceutical interventions and changes in human behavior, that are not included in either framework. Furthermore, the observed quantity is the number of reported deaths, aggregated across ages, which requires estimating also the death reporting delay.

Following the rolling-window strategy, both frameworks produce, at each round, probabilistic forecasts from $1$ to $4$ weeks ahead, using all the past data available up to the current week. The first round starts from the data available up to the week of 2020-03-15, and all the forecasts are evaluated up to the last week of 2020-07-19. As a consequence, the number of rounds decreases with the horizon, from $18$ at horizon 1 to $15$ at horizon 4. 

As for the synthetic scenarios, we evaluate the forecasts with the same metrics and over the same horizons, both through a horizon-specific analysis and by aggregating the results over all horizons and rounds. The error-based metrics (i.e., \gls{mae}, \gls{rmse}, \gls{wis}, and IS$_{80}$) are normalized by the score of the baseline, re-implemented for this new task, so that values above $1$ indicate worse performance than the baseline.

\paragraph{Overall performance.}
Table~\ref{tab: COVID_aggregated} reports the metrics aggregated over all horizons and rounds. Both frameworks obtain error-based scores above the baseline, by about $2.5$--$2.9$ times for the \gls{pinn}-based framework and $3.6$--$5.0$ times for the mechanistic one. The \gls{pinn}-based framework has lower errors than the mechanistic one on all the error-based metrics (i.e., \gls{mae}, \gls{rmse}, \gls{wis}, and IS$_{80}$), by about $29$--$41\%$.
The coverages of the prediction intervals are below the nominal levels for both frameworks: the $50\%$ interval contains the observations in $8\%$ of the cases for the \gls{pinn}-based framework and in $18\%$ for the mechanistic one, and the $90\%$ interval in $39\%$ and $44\%$, respectively. The baseline instead reaches $62\%$ and $79\%$.

\begin{table}[htbp]
\centering
\begin{tabular}{lcccccc}
\toprule
Model & MAE & RMSE & WIS & IS$_{80}$ & Cov.\ 50\% & Cov.\ 90\% \\
\midrule
\textsc{PINN} & 2.532 & 2.926 & 2.629 & 2.725 & 0.076 & 0.394 \\
\textsc{Mechanistic} & 3.577 & 4.954 & 3.879 & 4.144 & 0.182 & 0.439 \\
\textit{Baseline} & 1.000 & 1.000 & 1.000 & 1.000 & 0.621 & 0.788 \\
\bottomrule
\end{tabular}
\caption{\textbf{Comparison of overall forecasting performance on the real COVID-19 scenario.} Forecasts of weekly deaths for the first COVID-19 wave in London (UK), aggregated over all horizons (from 1 to 4) and rounds. MAE, RMSE, WIS, and IS$_{80}$ are normalized by the baseline score.}
\label{tab: COVID_aggregated}
\end{table}

In summary, although both frameworks remain well below the nominal levels, the \gls{pinn}-based framework outperforms the mechanistic one in both interval- and point-forecast accuracy, while the mechanistic framework has higher coverages. This contrasts with the synthetic scenarios described above, where the mechanistic framework obtained the lowest errors and its intervals were too wide, whereas the under-coverage of the \gls{pinn}-based framework is consistent across synthetic and real data.

\paragraph{Performance by forecast horizon.}
The metrics for each horizon are reported in the Appendix \ref{sec: appendix_tableCOVID} in \nameref{S1_Appendix} (Table~\ref{tab:COVID_forecast_h}). At horizon 1, the two frameworks are close, with the mechanistic one having a slightly lower \gls{mae} ($4\%$) and the \gls{pinn}-based one lower \gls{rmse}, \gls{wis}, and IS$_{80}$ ($9$--$19\%$). From horizon 2 onwards, the \gls{pinn}-based framework has lower errors on all the metrics, by about $23$--$35\%$ at horizon 2 and $34$--$44\%$ at horizons 3 and 4. The errors of both frameworks increase with the horizon, but at different rates: between horizon 1 and horizon 4 the normalized \gls{mae} increases by about $34\%$ for the \gls{pinn}-based framework and by about $113\%$ for the mechanistic one, so that the difference between the two widens with the horizon. As for the overall performance reported in Table~\ref{tab: COVID_aggregated}, the coverages of the prediction intervals remain below the nominal levels at every horizon for both frameworks, with the mechanistic framework having higher or equal values than the \gls{pinn}-based one at all horizons and for both levels.

\begin{figure}[htbp]
    \centering
    \includegraphics[width=0.75\textwidth]{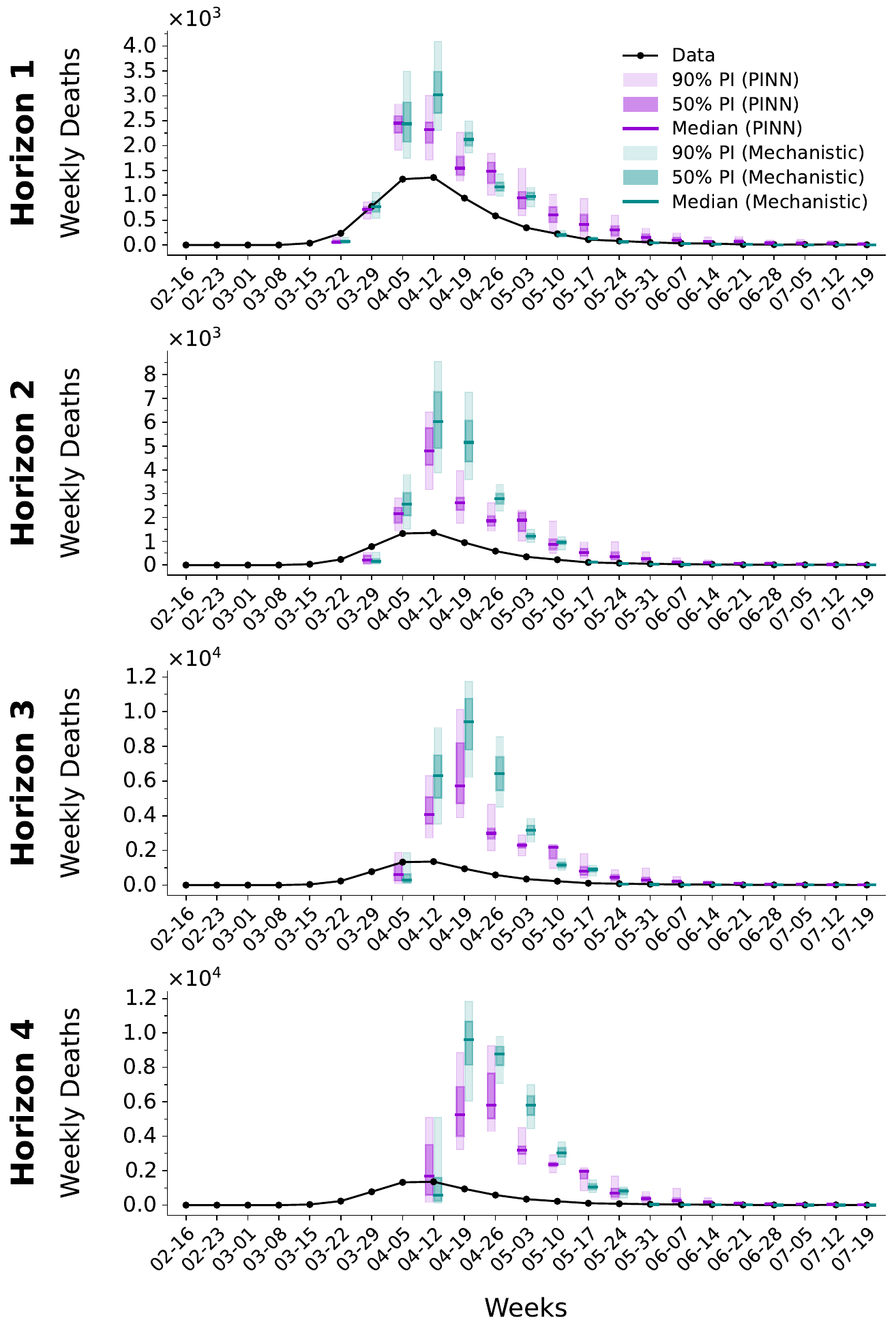}
    \caption{\textbf{Horizon-specific probabilistic forecasts of weekly COVID-19 deaths.} Forecasts of the \gls{pinn}-based and mechanistic frameworks for the first COVID-19 wave in London (UK), at fixed horizons of 1 to 4 weeks ahead (panels from top to bottom). In each panel, every box refers to the week on the x-axis and shows the forecast made for that week, as many weeks earlier as the horizon of the panel, using only the data available at that time. The lighter and darker boxes show the 90\% and 50\% prediction intervals, respectively, and the horizontal line inside each box shows the median. The observed weekly deaths are shown in black.}
    \label{fig:comparison_horizons}
\end{figure}

Figure~\ref{fig:comparison_horizons} shows the forecasts of the two frameworks at the four horizons, against the observed weekly deaths. As in the synthetic scenarios, each panel corresponds to a fixed horizon, and the boxes at a given week summarize the forecast that each framework, \gls{pinn}-based in violet and mechanistic in teal, had produced for that week, 1 to 4 weeks earlier, with the lighter box for the $90\%$ interval, the darker one for the $50\%$ interval, and the line for the median. The incidence of reported deaths is shown in black over the whole period from 2020-02-16, while the boxes start from the first week that can be forecast at each horizon, so that each panel contains $18$, $17$, $16$, and $15$ boxes per framework, as the number of rounds.

Excluding the first round, the two frameworks show a common pattern at all horizons, in the three phases of the wave. In the growth phase and around the peak, the observed deaths fall below the $90\%$ intervals of both frameworks, which overestimate them by a factor that increases with the horizon, from about $2$ times at horizon 1 to about $4$--$7$ times at horizons 3 and 4. In this phase, the \gls{pinn}-based framework produces lower forecasts than the mechanistic one, in both the medians and the intervals, and is therefore closer to the observations, which is consistent with its lower errors in the supplementary Table~\ref{tab:COVID_forecast_h}. After the peak, the pattern reverses: the \gls{pinn}-based framework predicts a slower decline of the deaths, with forecasts that remain above the observations and above those of the mechanistic framework, whose medians are closer. In the tail, from the end of May, the boxes of both frameworks are close to the observed deaths, which are almost zero, at all horizons.
The first round makes an exception: the \gls{pinn}-based medians are higher than the mechanistic ones at horizons 2 to 4, and therefore closer to the observed deaths, which are still low and underestimated by the forecasts of both frameworks. Notably, a few of the earliest forecasts, produced after observing only $5$ or $6$ weeks of data, are close to the observed values and contained in the $90\%$ intervals of both frameworks.  

The width of the intervals also depends on the phase of the epidemic: they are wide in the first rounds and around the peak --- reflecting the high uncertainty in predicting the right weekly death incidence curve during its most difficult phases, although not enough to contain the observations --- as the low coverages of the supplementary Table~\ref{tab:COVID_forecast_h} show. They shrink progressively in the decline. A similar behavior is observed in the estimates of the effective reproduction number $R_t$ (see Appendix \ref{sec: appendix_COVID_Re} in \nameref{S1_Appendix}, Figure~\ref{fig: violin_comparison}): the posterior distributions of both frameworks, particularly of the mechanistic one, are wide in the first rounds and become narrower in the following ones.

Repeating the analysis with a shorter look-back window, namely the last $5$ weeks of observations without the weighting strategy, improves the \gls{pinn}-based framework on all the error-based metrics (by about $5$--$16\%$) and increases its coverages from $0.076$ and $0.394$ to $0.379$ and $0.621$, while the mechanistic framework remains almost unchanged (see Appendix \ref{sec: appendix_COVID_n5} in \nameref{S1_Appendix}). With the shorter window, the coverages of the \gls{pinn}-based framework are also higher than those of the mechanistic one, so that the ordering observed with the full look-back window is reversed in this respect. 

In summary, in the case of the full look-back window, although the error-based metrics of both frameworks remain worse than those of the baseline, the \gls{pinn}-based framework obtains lower errors than the mechanistic one for almost all metrics and horizons, while the mechanistic framework has higher coverages, closer to the nominal levels. This contrasts with the synthetic scenarios, where both frameworks substantially outperform the baseline and the mechanistic framework obtains the lowest errors.

The same pattern can be observed in the forecasts of all the rounds, from the week of 2020-03-15 to the week of 2020-06-28, reported in Figure~\ref{fig:forecastig_comparison} in Appendix \ref{sec: appendix_COVID} in \nameref{S1_Appendix}.

\section{Discussion}
\label{sec: discussion}

We presented a systematic comparison between a \gls{pinn}-based framework and a stochastic mechanistic model. By relying on the same age-structured SEIRD model, shared setting choices, and an identical rolling-window forecasting pipeline, we aimed to compare the two modeling paradigms under controlled and comparable conditions. We evaluated their performance on two synthetic epidemic scenarios and on the first COVID-19 wave in London, considering both point forecast accuracy and the quality of probabilistic predictions over horizons ranging from one to four weeks. Overall, our findings indicate that neither approach consistently outperforms the other across all settings and evaluation criteria. Rather, their relative performance depends on the characteristics of the forecasting scenario, highlighting both the potential and the current limitations of \gls{pinn}-based approaches in epidemic forecasting.

The results obtained on synthetic data provide a controlled setting in which to evaluate the ability of the two frameworks to recover epidemic dynamics from noisy observations. In both scenarios, the two approaches successfully estimated the basic reproduction number and produced forecasts that consistently outperformed the reference baseline in terms of error-based metrics. The normalized error scores were generally well below the baseline, indicating that both approaches were able to extract meaningful information from the epidemic observations and provide substantially more accurate forecasts than a model relying only on past incidence variations. Nevertheless, the mechanistic model achieved better performance across the error-based metrics, including \gls{mae}, \gls{rmse}, \gls{wis}, and IS$_{80}$. This advantage was maintained when the observation noise was increased (see Appendix \ref{sec: appendix_noise_synthetic} in \nameref{S1_Appendix}), suggesting that, within the synthetic settings considered, the stochastic mechanistic framework was more robust in terms of predictive accuracy.

The two synthetic scenarios illustrate how epidemic dynamics influence forecasting performance. 
In the first scenario ($R_0=1.5$), characterized by slower epidemic growth and a lower incidence peak, both frameworks yield higher baseline-normalized error scores at every forecast horizon than in the second scenario ($R_0=2.0$).
Despite their structural differences, the frameworks exhibit the same pattern, suggesting that their relative performance depends partly on the epidemic trajectory. 

An important difference between the two frameworks emerged in the uncertainty associated with their predictions. In the synthetic scenarios, the \gls{pinn}-based framework consistently produced narrower prediction intervals, resulting in substantial under-coverage at both the $50\%$ and $90\%$ nominal levels. Conversely, the mechanistic model tended to produce wider intervals, with coverage generally closer to or above the nominal levels. 
This difference can be partly understood in terms of the distinct sources of uncertainty of the two frameworks. In the \gls{pinn}-based framework, the epidemic trajectories are deterministic function approximations of the estimated parameters, so the prediction intervals reflect only parameter uncertainty, which can be small when the parameters are well identified. In the mechanistic framework, each sampled parameter set also generates a stochastic realization of the epidemic process, so its predictive distributions include, in addition, the intrinsic stochasticity of the transmission and progression processes, which contributes to wider intervals~\cite{kharazmi2021identifiability,millevoi2024physics,qian2025physics}.
The difference in interval width should therefore not be interpreted exclusively as a difference in the quality of parameter inference, but also as a consequence of the different uncertainty representations adopted by the two frameworks.
The narrow \gls{pinn} intervals observed in the synthetic experiments may also be influenced by the observation model and the specific noise structure considered. In particular, the negative binomial noise used to generate the synthetic observations, together with the Poisson likelihood adopted for parameter inference, may affect the concentration of the \gls{mcmc} posterior distributions and contribute to the resulting narrow predictive intervals. This possibility deserves further investigation, especially considering that the \gls{pinn} intervals remain narrow even when stronger observation noise is introduced. 

Consistent with these findings, the age-specific analysis shows that the behavior observed with the aggregated incidence is reflected in all the age classes considered (see Appendix \ref{sec: appendix_sanity_synthetic} in \nameref{S1_Appendix}). This has to be mainly interpreted as an additional sanity check, showing that the age structure of the model is correctly implemented into the frameworks, in such a way that they reconstruct the distribution of 
new infections across age classes despite being calibrated only on aggregated incidence data. 

The results on the first COVID-19 wave in London reveal a different picture. In this setting, the \gls{pinn}-based framework achieved lower forecasting errors than the mechanistic model across almost all the evaluated forecast horizons and error-based metrics. However, neither framework outperformed the baseline in terms of these metrics, indicating that forecasting this real epidemic trajectory was considerably more challenging compared to synthetic scenario forecasting. 
First, real-world epidemic data are noisy and subject to underreporting and reporting delays, which can complicate model calibration and forecasting~\cite{roda2020difficult, starnini2021impact}.
Second, the minimal epidemiological setting adopted here does not explicitly incorporate information on interventions, changes in human mobility, or other time-varying factors that may have influenced epidemic transmission~\cite{perra2021non, nouvellet2021reduction,crawford2022impact}. 

Despite this limitation, the \gls{pinn}-based framework had lower forecasting errors than the mechanistic one, which performed better in the synthetic scenarios but deteriorated in the real one. 
One possible interpretation is that the neural network representation may provide some flexibility in capturing epidemic trajectories that deviate from the dynamics assumed by the underlying \glspl{ode} system. Conversely, the mechanistic model, despite explicitly accounting for stochasticity in the epidemic process, is more directly constrained by its mathematical structure, which could affect its forecasting performance when the observed data deviate from the assumed dynamics, especially when further assumptions and constraints are imposed to obtain the shared setting.
However, this interpretation should be regarded as a possible explanation rather than a definitive conclusion, since the real-world evaluation is restricted to a single epidemic wave.

The uncertainty results in the London case study further emphasize the differences between the synthetic and real settings. Unlike in the synthetic scenarios, the \gls{pinn}-based framework produces prediction intervals that are not uniformly narrower than those of the mechanistic model. Nevertheless, both frameworks show low empirical coverage, despite the relatively wide intervals produced in several forecasting rounds. However, this is primarily attributable to a displacement of the predicted trajectories with respect to the observations, rather than to insufficient interval width.

The estimates of the effective reproduction number further highlight differences between the two frameworks. In the synthetic scenarios, both frameworks similarly recover the temporal evolution of $R_t$ as the epidemic progresses. In the real case study, however, their posterior distributions differ markedly in the initial phase of the epidemic, in both their central estimates and their uncertainty, reaching similar values only in the final round (see Appendix \ref{sec: appendix_COVID_Re} in \nameref{S1_Appendix}). In addition, the two frameworks cross the threshold $R_t=1$ at different times. 
These results are consistent with the forecast results, where the two frameworks differ most in the initial phase and around the wave peak, which neither framework estimates accurately, confirming how difficult it is to capture the dynamics of an epidemic such as COVID-19 when no information on interventions or human behavior is available~\cite{gozzi2025comparative}.

An important aspect of the proposed \gls{pinn}-based framework is that the neural network is trained only once over a predefined temporal and parameter domain, and is subsequently reused across all forecasting rounds and scenarios without retraining. 
This design differs from approaches in which a neural network must be repeatedly trained as new observations become available, and represents a relevant practical advantage for sequential forecasting applications. Once trained, the network allows epidemic trajectories to be evaluated at different parameter values, while parameter inference is performed through \gls{mcmc} sampling. In this way, the computationally demanding training phase is separated from the repeated calibration and forecasting steps. The results presented here demonstrate that this strategy can be used in conjunction with a rolling-window forecasting pipeline, while preserving the structure of the underlying compartmental model. Nevertheless, the present comparison does not aim to establish a general computational advantage over stochastic mechanistic models, since the two frameworks rely on different inference procedures and computational operations, and were also implemented on different servers, making their computational costs not directly comparable.

Several limitations should be considered when interpreting our findings. First, the synthetic scenarios are generated using the same age-structured epidemic dynamics assumed by the two frameworks. Although this setting is useful for evaluating parameter recovery and forecasting performance under controlled conditions, it does not test the robustness of either approach to structural mis-specification. Second, the real-world evaluation is restricted to the first COVID-19 wave in London. Although this case study provides an application to observed epidemiological data, further analyses across different geographical settings, epidemic phases, and infectious diseases are needed to establish the broader validity of the findings. Third, both frameworks rely on a relatively constrained compartmental structure and on a set of fixed epidemiological assumptions. In particular, the transmission process and contact structure are not explicitly adapted to time-varying interventions or behavioral changes. Such extensions would be relevant to assess the performance of the two approaches in more complex epidemic settings. Fourth, the results are sensitive to design choices such as the length of the look-back window and the weighting strategy (see Appendix \ref{sec: appendix_COVID_n5} in \nameref{S1_Appendix}): the \gls{pinn}-based framework appears to benefit more than the mechanistic one from fully anchoring its calibration to the most recent observations, but since we explored only one alternative, the individual effects of the two choices cannot be isolated. As already mentioned, a further limitation concerns uncertainty quantification for the \gls{pinn}-based framework. Improving the representation of predictive uncertainty is therefore an important direction for future work. In particular, it would be valuable to understand what drives the narrow intervals of the \gls{pinn}-based framework, by accounting more comprehensively for the different sources of uncertainty (parameters, observation noise, process stochasticity, inference algorithm, and framework choices). Finally, the comparison deliberately adopts a minimal shared setting, rather than two independently optimized systems, to isolate the differences between the two approaches from those in model structure and forecasting procedure. At the same time, the two frameworks necessarily retain distinct characteristics, including their parameterization and inference algorithms. The results should therefore be interpreted as a comparison of two specific implementations of these modeling paradigms, rather than as a definitive ranking of \glspl{pinn} and mechanistic models in general. An interesting direction for future research would be to investigate hybrid approaches that combine the complementary strengths of the two frameworks. For instance, implementing an ensemble model or combining the flexibility of neural networks with stochastic mechanistic formulations. 

In conclusion, our results show that \gls{pinn}-based approaches represent a promising alternative for epidemic forecasting, but that their advantages are context-dependent. While the stochastic mechanistic model achieved more accurate forecasts and higher coverages in the synthetic scenarios, the \gls{pinn}-based framework performed better relative to the mechanistic model when applied to the first COVID-19 wave in London, although neither approach outperformed the baseline. At the same time, the differences in predictive uncertainty observed across the two settings highlight the need for further methodological developments in uncertainty quantification. More broadly, our findings emphasize that rigorous side-by-side comparisons under shared conditions are essential to identify the settings in which \gls{pinn}-based methods can provide a useful contribution to epidemic forecasting. 

\section*{CRediT authorship contribution statement}

\textbf{Martina Rama:} Conceptualization, Data curation, Formal analysis, Investigation, Methodology, Software, Validation, Visualization, Writing – original draft preparation, Writing – review and editing.
\textbf{Yuhan Li:} Conceptualization, Data curation, Formal analysis, Investigation, Methodology, Software, Validation, Writing – original draft preparation, Writing – review and editing.
\textbf{Bruno Lepri:} Conceptualization, Supervision, Writing – review and editing.
\textbf{Nicola Perra:} Conceptualization, Supervision, Methodology, Validation, Writing – review and editing.
\textbf{Michele Tizzoni:} Conceptualization, Methodology, Supervision, Validation, Writing – review and editing.

\section*{Declaration of competing interests}

The authors have declared that no competing interests exist.

\section*{Data and Code Availability Statement}

All data and code needed to reproduce the results are publicly available. The weekly COVID-19 deaths in London are taken from the Greater London Authority and Public Health England~\cite{glacovid2020} (\url{https://data.london.gov.uk/dataset/coronavirus-covid-19-weekly-update-e6yg0}); the age-specific population sizes from the Office for National Statistics, through Nomis~\cite{nomis2020} (\url{https://www.nomisweb.co.uk/datasets/pestnew}); the contact matrix for the United Kingdom from~\cite{prem2017projecting}; and the age-specific infection fatality ratios from~\cite{verity2020estimates}. All these data can be downloaded from the original sources, and the processed versions used in this work are also included in the repository below, together with the code to generate the synthetic data.

The code of the PINN-based framework, the pretrained network, the baseline, the posterior samples and forecasts of both frameworks, and the notebooks to reproduce all the figures and tables of the manuscript are available at \url{https://github.com/martina-rama/PINNvsMechanistic-models-for-epidemic-modeling-and-forecasting}. The code of the mechanistic framework is available at \url{https://github.com/Jadecoool/Comparing-physics-informed-neural-networks-and-mechanistic-models-for-epidemic-forecasting}, and its results, converted to the same format as those of the PINN-based framework, are included in the first repository.

\bibliography{bibtex}

\newpage
\begin{appendices}

\section*{Supporting Information}\label{S1_Appendix}

\section{Methods implementation details}
\label{sec: methods_details}

This section reports the methodological implementation details of the two forecasting frameworks developed for the COVID-19 case studies, including the synthetic scenarios, described in Sections~\ref{sec: PINN_model}--\ref{sec: mechanistic_model}. 

All simulations were coded in the Python programming language. The \gls{pinn}-based method and the validation results were coded using the libraries \texttt{PyTorch} version $2.6.0$~\cite{PyTorch2019}; in addition, for the \gls{mcmc} we take advantage of the \texttt{Pymcmcstat} package (version $1.9.1$)~\cite{Miles2019}. 
Furthermore, the mechanistic-based framework implemented the \gls{abcsmc} algorithm~\cite{toni2009approximate}, leveraging the Python library \texttt{pyabc}~\cite{schalte2022pyabc}.

This section first reports the details of the epidemiological structure shared by the two frameworks (Section~\ref{sec: appendix_epi}). It then follows the structure of the two frameworks. Sections~\ref{sec: pinn_details} and~\ref{sec: mcmc_setting} report the implementation details of the \gls{pinn}-based framework (violet box in Fig~\ref{fig:frameworks}), namely the architecture and training of the network (\textit{Step 1.}) and the \gls{mcmc} setting used for parameter inference (\textit{Step 2.}). Section~\ref{sec: mech_details} reports those of the mechanistic framework (teal box), namely the simulation of the stochastic model (\textit{Step 1.}) and its calibration via \gls{abcsmc} (\textit{Step 2.}). 

A summary of the parameters of both frameworks, explained in detail in the following Section~\ref{sec: mcmc_setting} and Section~\ref{sec: mech_details}, is given in Table~\ref{tab: parameters_summary}.

\begin{table}[htbp]
\centering
\footnotesize
\setlength{\tabcolsep}{3pt}
\newcolumntype{P}[1]{>{\raggedright\arraybackslash}p{#1}}
\begin{tabular}{P{0.08\textwidth}P{0.19\textwidth}P{0.21\textwidth}P{0.18\textwidth}P{0.23\textwidth}}
\toprule
\textbf{Param} & \textbf{Description} & \textbf{Value / bounds} & \textbf{PINN-based} & \textbf{Mechanistic-based} \\
\midrule
$\varepsilon$ & Latency rate & $1/3.7$ day$^{-1}$~\cite{kissler2020projecting,ma2020epidemiological,backer2020incubation,zhao2021estimating,xin2022estimating} & Fixed & Fixed \\
$\xi$ & Removal rate & $1/2.5$ day$^{-1}$~\cite{kissler2020projecting,ma2020epidemiological,backer2020incubation,zhao2021estimating,xin2022estimating} & Fixed & Fixed \\
$\text{IFR}_a$ & Infection fatality ratio & Age-specific~\cite{verity2020estimates} & Fixed & Fixed \\
$C$ & Contact matrix & UK~\cite{prem2017projecting} & Fixed & Fixed \\
$N_a$ & Population per age class & London~\cite{nomis2020} & Fixed & Fixed \\
\midrule
$R_0$ & Basic reproduction number & $[1.3, 3]$ & Estimated (\gls{mcmc}) & Estimated (\gls{abcsmc})\\
$1/\delta$ & Death reporting delay & $[7, 35]$ days & Real: estimated (\gls{mcmc}). Synthetic: fixed at $21$ days & Real: estimated (\gls{abcsmc}). Synthetic: set to $0$ \\
$\tau_0$ & Alignment time & $[0, 364]$ days ($[0, 150]$ in the first four weeks) & Estimated (\gls{mcmc}) & -- \\
$I_0$ & Initial number of infected individuals & $[1, 10\,000]$ & Fixed: one infectious individual per age class in training & Estimated (\gls{abcsmc})\\
$\Delta\tau$ & Alignment time & $[0, 90]$ days & -- & Estimated (\gls{abcsmc})\\
\midrule
$\beta$ & Transmission rate & Derived from $R_0$~(\ref{eq:R0}) & Derived & Derived \\
$R_t$ & Effective reproduction number & Derived from each posterior sample~(\ref{eq: Re}) & Derived & Derived \\
\bottomrule
\end{tabular}
\caption{Parameters of the two frameworks: fixed and shared (top), estimated (middle), and derived (bottom). In the synthetic scenarios, the observed quantity (newly exposed individuals) does not depend on $\delta$ nor on $\text{IFR}_a$.}
\label{tab: parameters_summary}
\end{table}

\subsection{Common structural details}
\label{sec: appendix_epi}

Both frameworks rely on the age-structured \gls{seird} system of \glspl{ode}~(\ref{eq: ODEs system}) introduced in Section~\ref{sec: common_structural_setting}. In both of them, the basic reproduction number $R_0$ is estimated at each round, and the effective reproduction number $R_t$ is derived from it.

The basic reproduction number is defined as the expected number of new infections produced by a single infected individual introduced into a population of fully susceptible, and, for the system~(\ref{eq: ODEs system}), can be defined as 
\begin{equation}
R_0 = \frac{\beta}{\xi} \cdot \rho(C),
\label{eq:R0}
\end{equation}
where $\rho(C)$ denotes the spectral radius (i.e., the dominant eigenvalue) of the contact matrix $C$, following the standard next-generation matrix approach for age-structured compartmental models~\cite{diekmann1990definition}. Analogously, the effective, time-dependent reproduction number $R_t$ (namely, the number of new infections caused by an infectious individual at any time point during the outbreak) can be obtained as:
\begin{equation}
R_t = \rho\big(K(t)\big), \qquad K(t)_{aa'} = \frac{\beta}{\xi} \, s_a(t) \, C_{aa'},
\label{eq: Re}
\end{equation}
which reduces to $R_0$ when the population is fully susceptible, i.e., $s_a(t) = 1$ for all $a$~\cite{nishiura2009effective}.
Under the assumption of time-constant transmission and recovery rates, the effective reproduction number can also be estimated based on the basic reproduction number as: $R_t = R_0\, s(t)$~\cite{nishiura2009effective}. In this work, we adopt this formulation extended to the age-structured setting by replacing $s(t)$ with its sum across age classes.

\subsection{PINN architecture and training}
\label{sec: pinn_details}
Physics-informed neural networks allow the use of comparatively small networks, minimizing the computational cost typically associated with deeper architectures~\cite{raissi2019physics, karniadakis2021physics}.
Hence, consistently with the design principles of SIR-INN~\cite{rama2026forecasting}, we adopt a simple feedforward neural network, while still reflecting the increased structural complexity of our age-structured epidemic system of reference ~(\ref{eq: ODEs system}). We therefore implement a \gls{pinn} consisting of five hidden layers, with an overall architecture of $[3, 100, 100, 100, 100, 100, 96]$, where the input layer corresponds to the three network inputs $t, R_0,$ and $1/\delta$, and the output layer corresponds to the $96$ normalized compartments ($16$ age classes $\times$ $6$ compartments). As activation functions, we use smooth differentiable functions: hyperbolic tangent for the hidden layers, and a sigmoid activation function for the output layer, to constrain the network output to $[0,1]^{96}$. These architecture choices are commonly adopted for \glspl{pinn}, as described in~\cite{cuomo2022scientific}.
 
As mentioned in Section~\ref{sec: PINN_modelapproximation}, to train our \gls{pinn} we rely on synthetic data for both the training set $(X_{tr},Y_{tr})$ and the collocation set $X_{coll}$. The training set enters the data term $\mathcal{L}_{data}$ of the \gls{pinn} loss function~(\ref{eq: loss_pinn}), while the collocation set enters the \gls{ode} term $\mathcal{L}_{ODE}$ through the residual function:
\begin{equation}
R(u,t;\lambda) := \frac{du(t;\lambda)}{dt} + F\big[u(t;\lambda);\lambda\big], \qquad t \in [t_0,T], \quad \lambda \in \mathbb{R}^2_{>0}.
\label{eq: residual}
\end{equation}
For the time component of $X_{tr}$, we use weekly time data over a window of $365$ days, i.e., $t_i^u = 0, 7, \dots, 364$. Regarding the epidemic parameters, we define a uniform grid of size $33 \times 33$ for $(R^u_0, 1/\delta^u)$, with $R^u_0 \in [1.3, 3]$ and $1/\delta^u \in [7, 35]$ (days), obtaining $1089$ parameter pairs, without any additional restriction on the grid. The output training data $Y_{tr}$ are generated via \textit{odeint} (from \texttt{SciPy} library~\cite{Virtanen2020}), collecting the numerical solution of the system ~(\ref{eq: ODEs system}) for the corresponding parameters, starting with the initial condition of one infectious individual in each age class and the rest of the population fully susceptible. All the components are already normalized with respect to the total population of each age class, $N_a$, consistently with the network output range $[0,1]^{96}$. The specific values adopted for the fixed parameters $\psi$ and the epidemiological justification for the ranges of $\lambda$, shared with the mechanistic model, are reported in Section~\ref{sec: common_structural_setting} (see also Table~\ref{tab: parameters_summary}).

We then construct the collocation set $X_{coll}$ independently from the training set $X_{tr}$. Specifically, the epidemic parameter pairs $(R^c_0, 1/\delta^c)$ are built by taking the midpoint values between adjacent nodes of the training grid, obtaining an independent uniform grid of size $32 \times 32$, i.e., $1024$ parameter pairs located between those of $X_{tr}$. Regarding the time component, we adopt a denser sampling in the same domain, with points spaced daily rather than weekly.

Following the notation introduced in Section~\ref{sec: PINN_modelapproximation}, the temporal-epidemic domain is thus $\mathcal{T} = [0, 364]$ days, $\mathcal{R} = [1.3, 3]$, and $\mathcal{D} = [7, 35]$ days. Hence, to summarize, we can define the training set and the collocation set as:
\begin{align}
X_{tr} &= \left\{ \big(t_i^u;\, R_{0 j}^u,\, 1/\delta_k^u\big) \right\}_{\substack{i=1,\dots,53 \\ j,k=1,\dots,33}} \subset \mathcal{T} \times \mathcal{R} \times \mathcal{D}, \\
Y_{tr} &= \left\{ u\big(t_i^u;\, R_{0 j}^u,\, 1/\delta_k^u\big) \right\}_{\substack{i=1,\dots,53 \\ j,k=1,\dots,33}} \subset [0,1]^{96}, \\
X_{coll} &= \left\{ \big(t_i^c;\, R_{0 j}^c,\, 1/\delta_k^c\big) \right\}_{\substack{i=1,\dots,365 \\ j,k=1,\dots,32}} \subset \mathcal{T} \times \mathcal{R} \times \mathcal{D}.
\end{align}
This setup is motivated by the aim of improving the ability to satisfy the governing equations~(\ref{eq: ODEs system}) and the generalization capabilities of the neural network. Furthermore, this choice does not affect the computational cost of network training because, while constructing $X_{tr}$ requires generating the corresponding labels $Y_{tr}$ via numerical integration of the system~(\ref{eq: ODEs system}), evaluating the residual loss on $X_{coll}$ does not require any labeled data, relying only on automatic differentiation of the network output~\cite{raissi2019physics}.
 
Regarding the weighting of the two loss terms in ~(\ref{eq: loss_pinn}), since the data loss $\mathcal{L}_{data}$ and the \gls{ode} loss $\mathcal{L}_{ODE}$ are on different scales, we set $\alpha = 100$, compensating for this imbalance.

We train our \gls{pinn} for $4\,000$ epochs using the Adamax optimization algorithm, with a learning rate of $0.001$. The neural network is trained with batches of $100$ training, validation, and collocation points, and a train size of $0.9$. The same trained network is used in all the scenarios, which allows us to generalize the framework and to save computational cost, since no retraining is needed across scenarios. In the synthetic cases, $1/\delta$ is provided as input at its fixed value ($21$ days), which is included in the training range $\mathcal{D}$.

Notably, by construction~\cite{rama2026forecasting}, the implemented \gls{pinn} learns to respect the physical constraint of population conservation~(\ref{eq: pop_conservation}), even though this is not explicitly enforced, either in the neural network architecture or in the training process.
 
\subsection{MCMC setting}
\label{sec: mcmc_setting}

As explained in Section~\ref{sec: PINN_inference&forecasting}, parameter estimation is performed via \gls{mcmc}. As in the SIR-INN framework~\cite{rama2026forecasting}, in addition to $\lambda = \{R_0, 1/\delta\}$, we estimate an alignment time $\tau_0 \in [0,364]$, which denotes the time at which we start to observe the data relative to the temporal domain over which $u_N$ is trained.

The observed quantity depends on the scenario: the weekly incidence of COVID-19 deaths, aggregated across age classes, in the real scenario, and the weekly incidence of new infections (i.e., exposed individuals) in the synthetic ones (see Section~\ref{sec: epidemiological_data}). Since the neural network outputs normalized compartments, in both cases we define a function $G$ that extracts from the network output the corresponding estimated weekly incidence. To this end, we fix a weekly time grid $\tau_j = \tau_0 + 7j$, $j = 0, 1, \dots$, depending on an unknown alignment time $\tau_0 \in [0,364]$.

In the real scenario, $G$ is based on the output of the network referred to the observed deaths in age class $a$, namely $u_N^{(D^{obs}_a)}$:
\begin{equation}
G_{\text{real}}(j; \tau_0, R_0, 1/\delta) = \sum_{a=1}^{16} \Big[u_N^{(d^{obs}_a)}(\tau_j; R_0, 1/\delta) - u_N^{(d^{obs}_a)}(\tau_{j-1}; R_0, 1/\delta)\Big] \cdot N_{a}.
\end{equation}
The difference between two consecutive weeks is rescaled by the age-specific population size $N_{a}$, coherently with the normalized network outputs, and then summed over age classes, to obtain values expressed as counts, consistently with the reference data and with the choice of the likelihood.

In the synthetic scenarios, the incidence of newly exposed individuals equals the weekly decrease of the susceptible compartment, so that
\begin{equation}
G_{\text{syn}}(j; \tau_0, R_0) = \sum_{a=1}^{16} \Big[u_N^{(s_a)}(\tau_{j-1}; R_0, 1/\delta^{*}) - u_N^{(s_a)}(\tau_{j}; R_0, 1/\delta^{*})\Big] \cdot N_a,
\end{equation}
where $1/\delta^{*}$ days is a fixed value of the reporting delay.

Parameter estimation consists of estimating $\tau_0$, $R_0$ and, only in the real COVID-19 scenario, $1/\delta$, in $G$ from the observed weekly data, over a rolling window of $M$ weeks, whose specific values are reported in Section~\ref{sec: forecasting_setting}. As already mentioned, we assume a Poisson distribution, reflecting the count nature of the epidemiological data, with mean given by $G$, the \gls{pinn} estimated weekly incidence, and perform parameter inference by minimizing the corresponding negative Poisson log-likelihood. 
The contribution of each weekly observation to the log-likelihood is weighted as explained in Section~\ref{sec: forecasting_setting}, normalized by their sum, so that more recent weeks have a greater influence on the inference.
We assume uniform priors for all parameters $\tau_0, R_0,$ and $1/\delta$, with lower and upper bounds coinciding with those adopted for the corresponding training grid (Section~\ref{sec: pinn_details}), and the chain is initialized at the midpoint of each range. As already mentioned, for the synthetic scenarios, we fixed $1/\delta^{*} = 21$ days to be the midpoint of the \gls{pinn} training set range of the deaths reporting delay. In this way, the network is always evaluated within the domain in which it was trained; instead, the mechanistic framework sets the reporting delay to zero.
Furthermore, only for the first four weeks, when the observed weekly death incidence could be uniformly zero, we restrict the upper bound of $\tau_0$ from $364$ to $150$. Without this constraint, the null incidence values observed at the initial phase of the epidemic could be wrongly attributed to its late declining phase, resulting in an incorrect parameter inference. This forces the pre-trained \gls{pinn} to be constrained to the correct initial phase, without starting to observe future values instead of the correct initial ones. These bound choices are made to perform the inference with minimal prior assumptions, as the structure of our framework allows us to do so~\cite{rama2026forecasting}. In addition, values of $\tau_0$ for which the network does not cover the entire look-back window are rejected, by setting the objective function to a very large constant ($10^{6}$).

For each observation window, the \gls{mcmc} sampling is performed for $10\,000$ iterations, discarding the first $9\,000$ as burn-in. From the remaining $1\,000$ posterior samples (i.e., $\hat\tau_0, \hat R_0, 1/\hat\delta$), we derive summary statistics and probabilistic forecasts, as described in Section~\ref{sec: PINN_inference&forecasting}. 
In addition, for each posterior set $\big(\hat\tau_0, \hat R_0, 1/\hat\delta\big)$ we derive a corresponding sample of the effective reproduction number $\hat R_t$. Specifically, we evaluate the pre-trained \gls{pinn} $u_N$ at the current week of the observation window (according to $\hat\tau_j$), with parameters $\big(\hat R_0, 1/\hat\delta\big)$, and we compute the normalized sum of the susceptibles $\hat s = \sum_{a=1}^{16} \Big[u_N^{(s_a)}(\hat\tau_j; \hat{R}_0, 1/\hat{\delta})\, \cdot N_a \Big]/ N_{\text{tot}}$. In the synthetic scenarios, the posterior sets reduce to $\big(\hat\tau_0, \hat R_0\big)$, and the network is evaluated with $1/\delta^{*} = 21$ days fixed. Then, according to Section~\ref{sec: common_structural_setting} (Eq.~(\ref{eq: Re}), we derive the $1\,000$ posterior samples of $R_t$ for each observation window as $\hat R_t = \hat R_0 \cdot \hat s$, at no additional inference cost.

The final step, the forecast of the future epidemic states, reduces to two operations, as in~\cite{rama2026forecasting}: evaluating the pre-trained \gls{pinn} $u_N$ at the estimated parameters, with their uncertainty, over the whole time window, and then translating time zero to the estimated alignment time $\hat\tau_0$ and selecting the desired future time window. In this way, we obtain probabilistic forecasts, in the form of prediction quantiles, over the period of interest (i.e., the future time window of observations).

Finally, to assess the reproducibility of our results and mitigate the potential variability of \gls{mcmc} independent runs, due to the stochastic nature of the sampling procedure, for each observation window we repeat the sampling procedure over $100$ independent chains. We then keep the chain achieving the lowest fitting \gls{wis}, evaluating the pre-trained \gls{pinn} on the past forecasting weeks. Specifically, the \gls{wis} metric is weighted: its values over the weeks of the look-back window are averaged using the same weights adopted for the inference (see Eq.~(\ref{eq: weights}) in Section~\ref{sec: forecasting_setting}), so that the fit to the most recent weeks counts more in the selection.
We adopt \gls{wis} as the selection criterion since, unlike point-forecast metrics such as \gls{mae} or \gls{wmape}, it jointly captures both the accuracy of the median forecast and the calibration of the associated prediction intervals, providing a more comprehensive measure of fit quality; this is consistent with its role among the metrics adopted for forecast evaluation presented in this work (Section~\ref{sec: metrics}).

\subsection{Mechanistic model implementation details}\label{sec: mech_details}

For age class $a$, the stochastic model is a discrete-time counterpart of system~(\ref{eq: ODEs system}), with the compartments expressed in numbers of individuals instead of proportions. It is defined by the following set of stochastic equations:
\begin{equation}
\begin{aligned}
S_a(t+\Delta t) &= S_a(t) - \text{Bin}\big(S_a(t), \Lambda_a(t)\big) \\
E_a(t+\Delta t) &= E_a(t) + \text{Bin}\big(S_a(t), \Lambda_a(t)\big) - \text{Bin}\big(E_a(t), \varepsilon\big) \\
I_a(t+\Delta t) &= I_a(t) + \text{Bin}\big(E_a(t), \varepsilon\big) - \text{Bin}\big(I_a(t), \xi\big) \\
R_a(t+\Delta t) &= R_a(t) + \text{Bin}\big(I_a(t), \xi\big)\,(1-\text{IFR}_a) \\
D_a(t+\Delta t) &= D_a(t) + \text{Bin}\big(I_a(t), \xi\big)\,\text{IFR}_a \\
D^{\text{obs}}_a(t+\Delta t) &= D_a\big(t+\Delta t-1/\delta\big)
\end{aligned}
\label{eq: stochastic_system}
\end{equation}
where the force of infection acting on age class $a$ is
\begin{equation}
\Lambda_a(t) = \beta \sum_{a'=1}^{16} C_{aa'} \frac{I_{a'}(t)}{N_{a'}}.
\label{eq:force_of_infection}
\end{equation}
Here $C$ is the contact matrix, and $C_{aa'}$ indicates the number of contacts of an individual in age group $a$ with the individuals in age group $a'$. $I_{a'}$ is the number of infectious individuals in age class $a'$, and $\beta$ is the transmission rate, computed from the basic reproduction number $R_0$ according to Eq.~(\ref{eq:R0}). 

For the stochastic model, we denote with capital letters the numbers of individuals in each compartment, while the lowercase letters of system~(\ref{eq: ODEs system}) denote the corresponding proportions with respect to the population of the age class. For simplicity and for the goal of comparison between the two frameworks, we keep the same notation for the constants ($\beta$, $\varepsilon$, $\xi$, $C$, $\text{IFR}_a$, $\delta$), even though the two formulations are not strictly identical. 

The initialization of the stochastic models is as follows. At the beginning of the simulation, $I_0$ individuals are infected. They are assigned to the exposed and infectious compartments proportionally to the mean duration of each stage, $\varepsilon^{-1}/(\varepsilon^{-1}+\xi^{-1})$ and $\xi^{-1}/(\varepsilon^{-1}+\xi^{-1})$, respectively, and then distributed across age classes proportionally to the population size $N_a$. The initial number of exposed individuals in age class $a$ is thus
\begin{equation}
E_a(0) = I_0 \cdot \frac{\varepsilon^{-1}}{\varepsilon^{-1}+\xi^{-1}} \cdot \frac{N_a}{N_{\text{tot}}},
\label{eq: E0}
\end{equation}
and that of infectious individuals is
\begin{equation}
I_a(0) = I_0 \cdot \frac{\xi^{-1}}{\varepsilon^{-1}+\xi^{-1}} \cdot \frac{N_a}{N_{\text{tot}}}.
\label{eq: I0}
\end{equation}
All the other individuals are assumed susceptible at the beginning of the epidemic, so that $S_0 := S(0) = N_{\text{tot}} - I_0$, distributed across age classes in the same way:
\begin{equation}
S_a(0) = S_0 \cdot \frac{N_a}{N_{\text{tot}}}.
\end{equation}
The remaining compartments ($R_a$, $D_a$, $D^{\text{obs}}_a$) are empty at $t=0$.

In addition to $\lambda = \{R_0, 1/\delta\}$, the mechanistic framework estimates two parameters that specify the state of the epidemic when the observations begin, and that play a role analogous to that of the alignment time $\tau_0$ in the \gls{pinn}-based framework (Sections~\ref{sec: PINN_inference&forecasting} and~\ref{sec: mcmc_setting}). These parameters are the initial number of infected individuals $I_0$ and the alignment time $\Delta\tau$.
Specifically, in the real-data setting, each simulation is seeded at $t_{\text{start}} - \Delta\tau$ with$\Delta\tau \sim \mathcal{U}\{0,\dots,90\}$ days and $t_{\text{start}}$ is the start date of the data for calibration. The simulation is run until an end date $t_{\text{end}}$ of the real data, i.e. over a horizon of $T + \Delta\tau$ days, where $T = t_{\text{end}}-t_{\text{start}}$ days. The trajectories are cut from the calibration start date $t_{\text{start}}$ to matched to lenght of the real data for calibration. The parameter $\Delta\tau$ thus sets the seeding date relative to the observations and, together with $I_0$, plays a role analogous to that of the alignment time $\tau_0$ in the \gls{pinn}-based framework. In the synthetic setting $\Delta\tau$ is fixed to 0.

As described in Section~\ref{sec: calibration&forecast_mech}, we use the \gls{abcsmc} algorithm to calibrate the compartmental model~(\ref{eq: stochastic_system}). 
The goal is to estimate the posterior distribution for models' parameters given the input of their prior distribution. The algorithm consists of $N_g$ generations (i.e., iterations). The first generation is based on a rejection algorithm step where the tolerance is set to a high value. In the second generation, the tolerance is decreased, parameters are sampled from those accepted in the previous step and perturbed via a kernel to avoid converging on local minima of the phase space. The process is repeated for $N_g$ generations of $N_p$ particles (i.e, samples) each. Then, the set of accepted parameters in the last generation is used as the empirical posterior distribution. In our simulations, we set $N_g=10$ and $N_p = 1\,000$.

We assume uniform priors for all the unknown parameters, with the same bounds adopted for the \gls{pinn}-based framework where the parameters coincide (Section~\ref{sec: common_structural_setting}):
\begin{itemize}
\item the basic reproduction number $R_0 \in [1.3, 3]$;
\item the death reporting delay $1/\delta$, over the integers in $[7, 35]$ days, estimated only in the real COVID-19 scenario;
\item the initial number of infected individuals $I_0$, over the integers in $[1, 10\,000]$;
\item the alignment time $\Delta\tau \in [0, 90]$ days.
\end{itemize}
In the synthetic scenarios, the observed quantity is the weekly incidence of newly exposed individuals rather than the reported deaths, and it does not depend on the notification delay. So $1/\delta$ is not estimated and it is set to zero, as discussed in Section~\ref{sec: common_structural_setting}.

Regarding the distance between simulated and observed data over the look-back window, we opted for the log-transformation of the weighted \gls{wmape} (see Section~\ref{sec: metrics} for the metric formulation and Eq.~(\ref{eq: weights}) in Section~\ref{sec: forecasting_setting} for the weighting criterion). We adopt this version in all the scenarios, as it improves the results on the real COVID-19 data, and using the same distance across scenarios supports the generalizability of the framework. Observed zeros, which are problematic for the \gls{wmape}, are treated as follows: at the beginning of the epidemic, the distance and the weights are computed starting from the first non-zero observation; in the tail of the epidemic, the computation stops before the first zero.

\subsection{Baseline model}
\label{sec: baseline}

As a reference for the evaluation, we also consider a baseline model, which does not rely on any epidemiological knowledge and is not part of the comparison between the two frameworks. We adopt the baseline of~\cite{gozzi2025comparative}, generated with the same method in all the scenarios. At each round, let $t_c$ be the current week and $y_1, \dots, y_{t_c}$ the observed weekly values. We compute the weekly increments $d_i = y_i - y_{i-1}$, $i = 2, \dots, t_c$, over all the look-back window of the incidence data, without the weighting strategy adopted by the two frameworks, and we symmetrize them by including their opposites, $\Omega = \{d_i\} \cup \{-d_i\}$. We then generate $N_{\text{traj}} = 10\,000$ trajectories: for each of them, four increments (one per forecast week) are selected from $\Omega$ with replacement, and the forecast at horizon $h$ is
\begin{equation}
f_h = y_{t_c} + \sum_{i=1}^{h} \hat d_i,
\end{equation}
where $\hat d_i$ is the $i$-th increment selected. The prediction quantiles are computed from the empirical distribution of the trajectories. Since the increments are symmetrized, the median forecast remains approximately equal to the last observed value. As weekly counts cannot be negative, the trajectories are truncated at zero. The error-based metrics of the two frameworks are normalized by the score of this baseline (Section~\ref{sec: metrics}), so that values below $1$ indicate better performance than the baseline.

\clearpage
\newpage
\section{Full forecasting comparison results - synthetic scenarios}
\label{sec: appendix_full_synthetic}
\begin{figure}[htbp]
\centering
\includegraphics[width=0.9\textwidth]{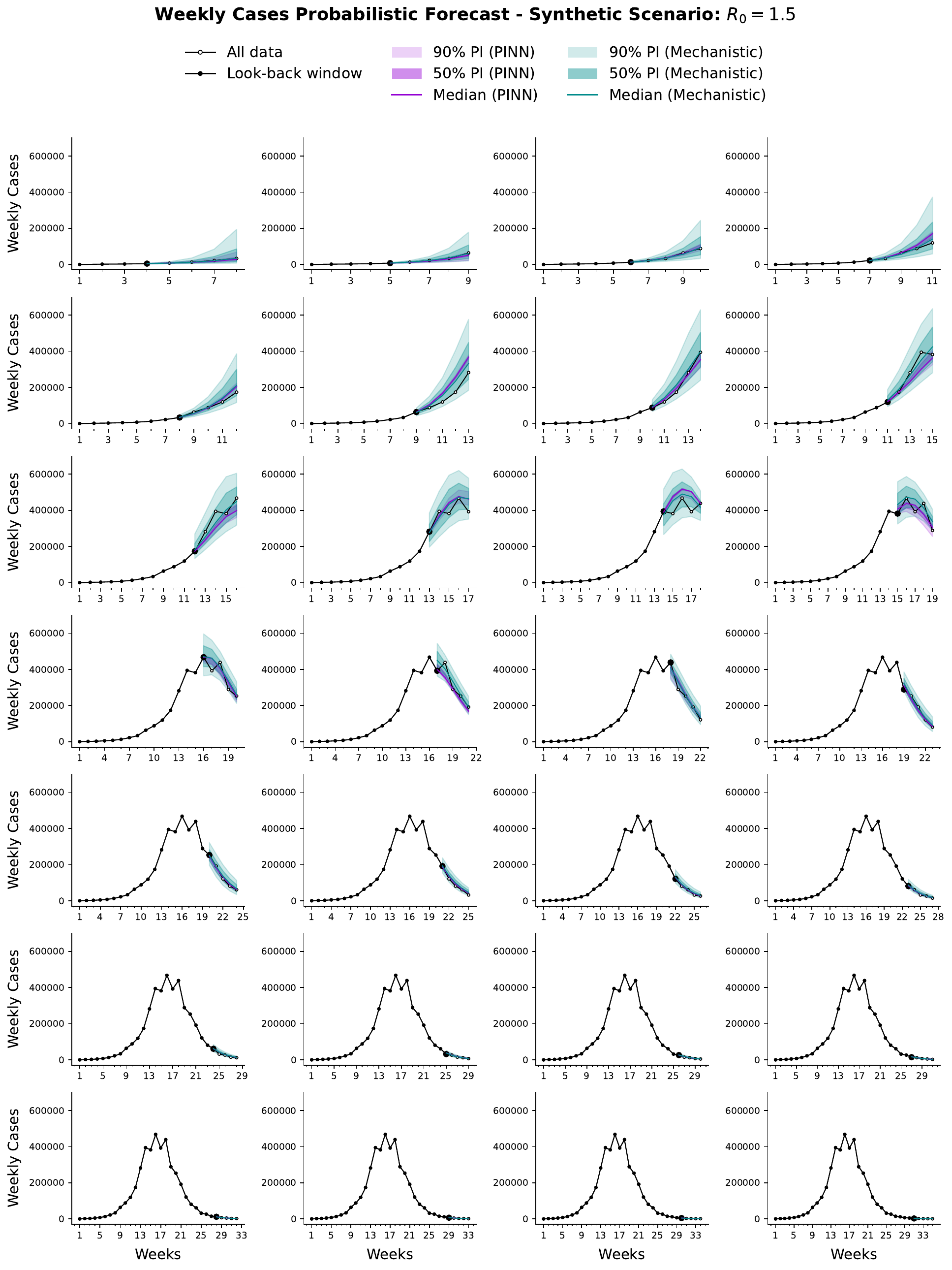}
\caption{Comparison of \gls{pinn} and mechanistic forecasts of weekly new infections on the noisy synthetic data with $R_0 = 1.5$, across 28 forecast rounds (i.e., weeks). In each panel, black dots denote the observed weekly new infections used for model calibration up to the current week, while white dots denote the subsequent four weeks to forecast. Shaded bands show the 50\% and 90\% prediction intervals of each model over the 1- to 4-week-ahead forecast horizons, and solid lines indicate the corresponding medians.}
\label{fig:forecastig_comparison_synt1}
\end{figure}

\begin{figure}[htbp]
\centering
\includegraphics[width=0.9\textwidth]{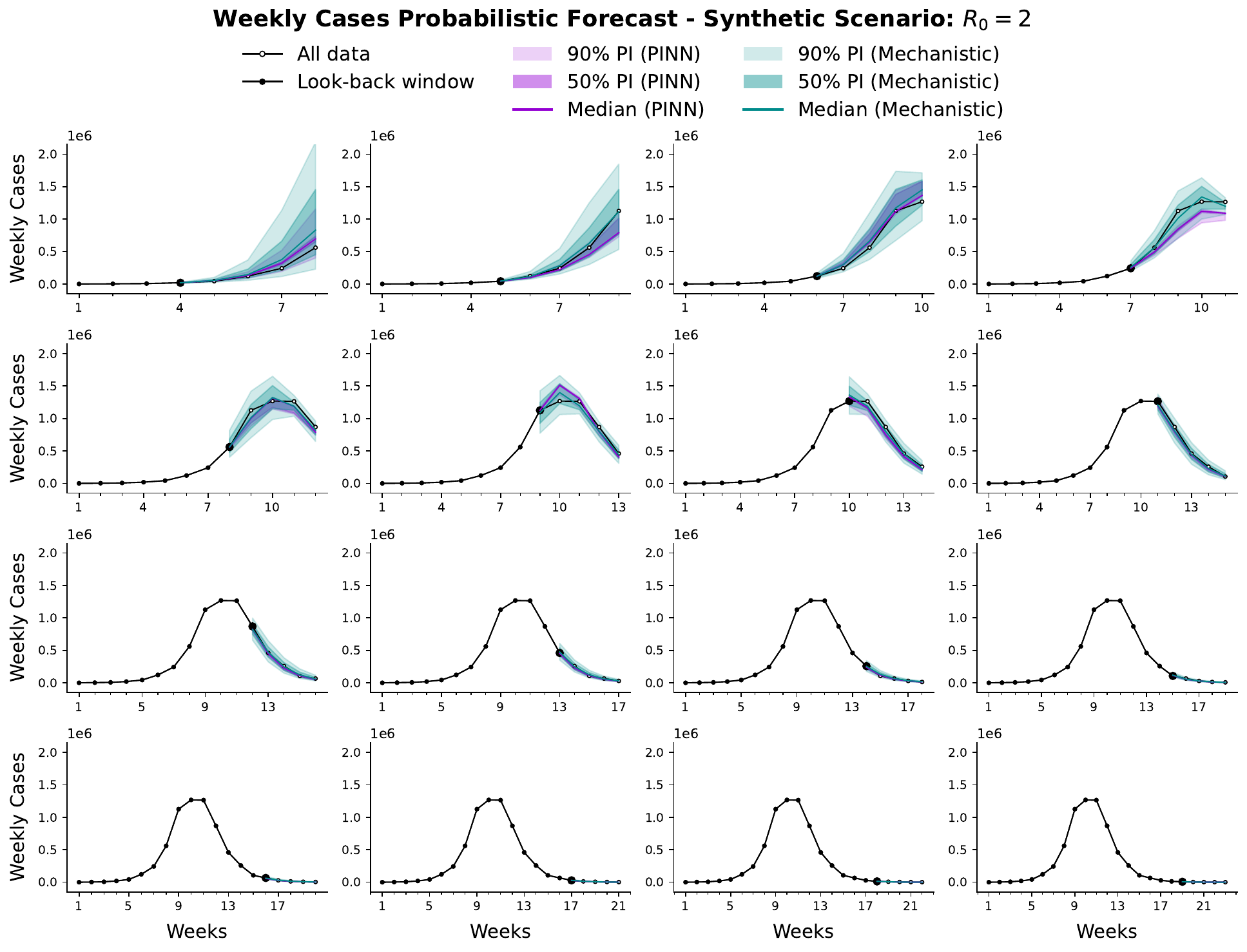}
\caption{Comparison of \gls{pinn} and mechanistic forecasts of weekly new infections on the noisy synthetic data with $R_0 = 2.0$, across 16 forecast rounds (i.e., weeks). In each panel, black dots denote the observed weekly new infections used for model calibration up to the current week, while white dots denote the subsequent four weeks to forecast. Shaded bands show the 50\% and 90\% prediction intervals of each model over the 1- to 4-week-ahead forecast horizons, and solid lines indicate the corresponding medians.}
\label{fig:forecastig_comparison_synt2}
\end{figure}

\clearpage
\newpage
\section{Age-classes sanity-check}
\label{sec: appendix_sanity_synthetic}

\begin{figure}[htbp]
\centering
\includegraphics[width=\textwidth]{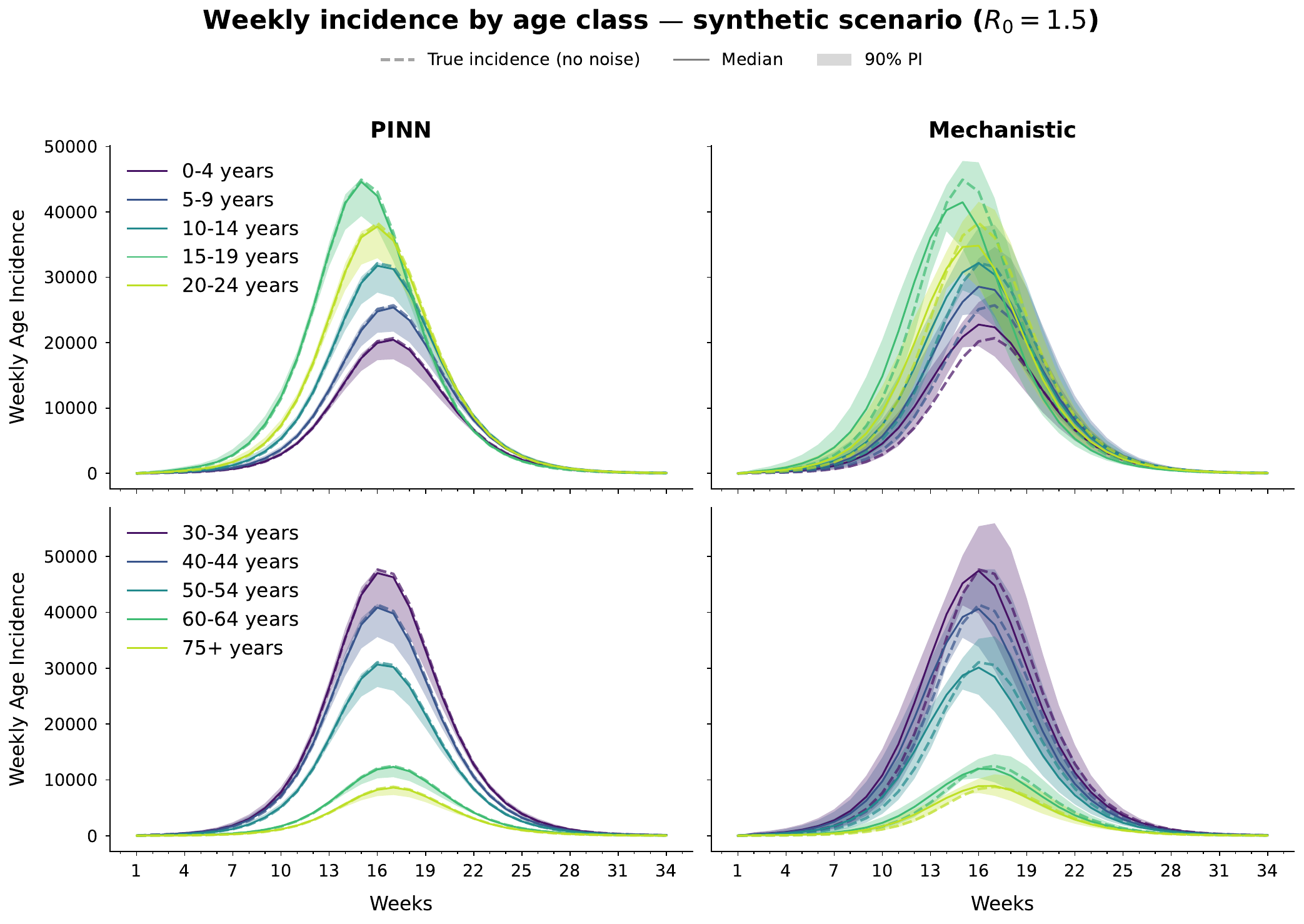}
\caption{\textbf{Retrospective fit of the age-specific weekly incidence on the first synthetic scenario ($R_0=1.5$).} Weekly new infections by age class estimated by the \gls{pinn}-based (left) and mechanistic (right) frameworks over the $34$ weeks of the simulation, for the age classes 0--4, 5--9, 10--14, 15--19, and 20--24 years (top row) and 30--34, 40--44, 50--54, 60--64, and 75+ years (bottom row). Lines show the median predictions of the models and shaded bands the $90\%$ prediction interval of the estimates, and the dashed lines show the true incidence of the simulation, before the age-classes aggregation and noise addition.}
\label{fig:fitting_ages_synth1}
\end{figure}

\begin{figure}[htbp]
\centering
\includegraphics[width=\textwidth]{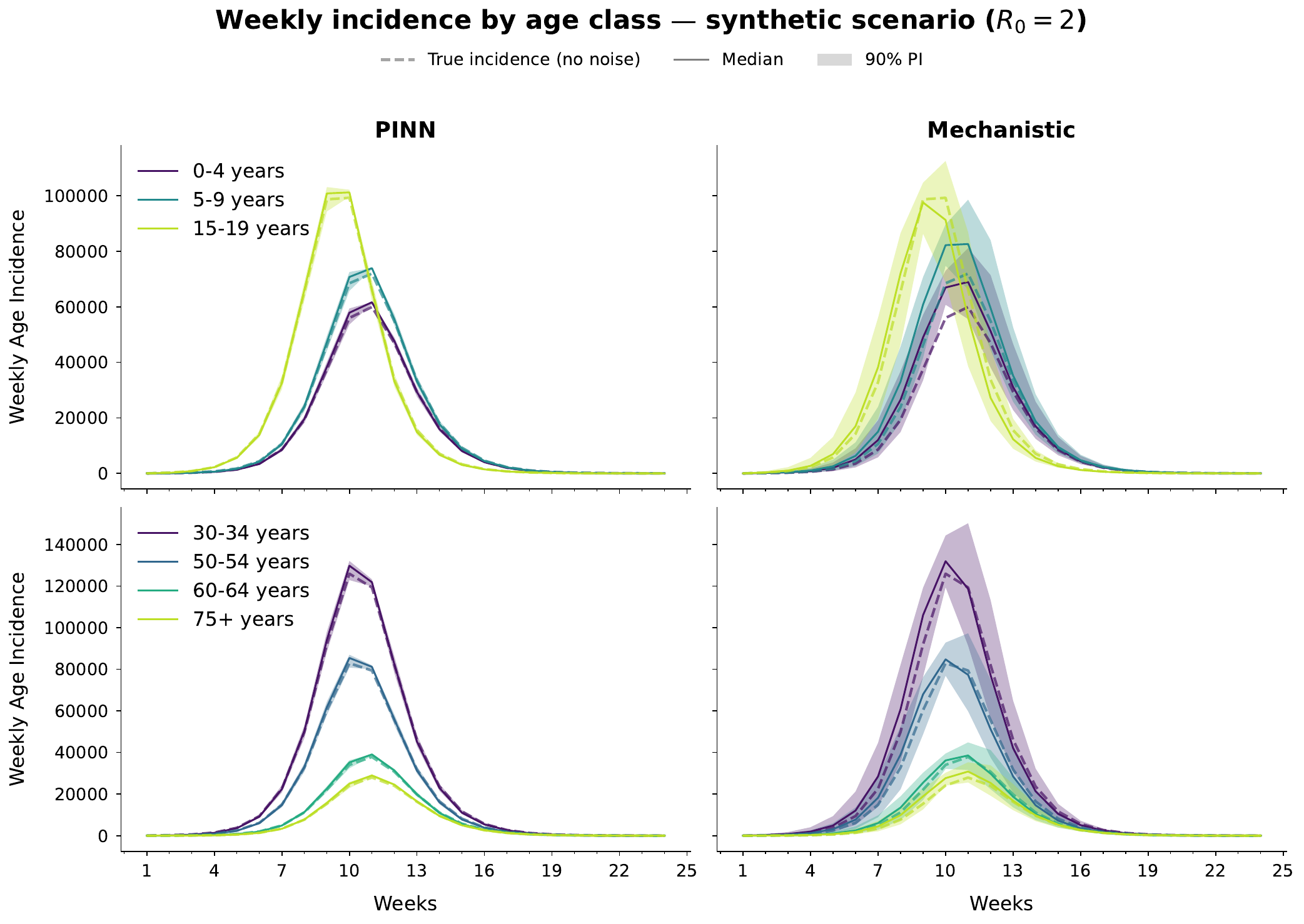}
\caption{\textbf{Retrospective fit of the age-specific weekly incidence on the second synthetic scenario ($R_0=2.0$).} Weekly new infections by age class estimated by the \gls{pinn}-based (left) and mechanistic (right) frameworks over the $24$ weeks of the simulation, for the age classes 0--4, 5--9, and 15--19 years (top row) and 30--34, 50--54, 60--64, and 75+ years (bottom row). Lines show the median predictions of the models and shaded bands the $90\%$ prediction interval of the estimates, and the dashed lines show the true incidence of the simulation, before the agage-class aggregation and noise addition.}
\label{fig:fitting_ages_synth2}
\end{figure}

As a further check on the two synthetic scenarios, in Fig~\ref{fig:fitting_ages_synth1} and Fig~\ref{fig:fitting_ages_synth2}, for the first and second scenario, respectively, we compared the age-specific weekly infections incidence reconstructed by the two frameworks with the true age-specific incidence of the simulation, taken before the aggregation across age classes and before the addition of the noise (see Section~\ref{sec: epidemiological_data}). Since the frameworks are calibrated only on the noisy incidence aggregated across ages, the age distribution of the new infections is not an input of the parameter inference but an output of the age-structured model, determined by the contact matrix and the population sizes together with the few estimated parameters. This comparison is therefore a sanity check, which is possible only because the true age-specific incidence is known in the synthetic scenarios: it verifies that the age structure is accurately reconstructed from the aggregated data. Since the synthetic data are generated by the same age-structured dynamics assumed by the two frameworks, a good recovery is expected, and the check should be read as a verification of the implementation and of the identifiability of the age profile, and not as a test of robustness to differences between the process generating the data and the model assumed by the frameworks.

For readability, we show $10$ (including the $7$ of the second scenario) of the $16$ age classes, since several classes produce very similar trajectories. The median incidence of the \gls{pinn}-based framework closely follows the true incidence in all the shown age classes and in both scenarios. The mechanistic medians are instead slightly displaced from the true curve: they underestimate the peak in some age classes of the first scenario (e.g., 10--14 and 20--24 years), and overestimate it in the youngest ones (0--4 and 5--9 years) in both scenarios. In addition, in these youngest classes, the $90\%$ interval of the mechanistic framework does not contain all the points of the true incidence.

The width of the intervals differs between the two frameworks. The $90\%$ intervals of the \gls{pinn}-based framework are always narrower than those of the mechanistic one, as shown by the forecast results (Section~\ref{sec: synth_forecasting_results}), and in the second scenario they are almost null. The shape of the intervals also differs between the two frameworks. The $90\%$ intervals of the mechanistic framework are approximately symmetric around its median. Those of the \gls{pinn}-based framework, which are visible only in the first scenario, are instead markedly asymmetric: the median, which practically coincides with the true incidence, lies close to the upper limit of the interval, so that the interval extends almost only below it. This behavior is also reflected in the retrospective fit of the aggregated incidence (Fig~\ref{fig: synth_retro_R0}, left column).

Overall, both frameworks recover the age-specific epidemic curves from the aggregated incidence. The \gls{pinn}-based framework is more accurate in the median but has very narrow intervals, while the mechanistic one shows larger deviations of the median, mainly in the youngest age classes, and wider intervals that more likely better contain the age-specific weekly new infections incidence with noise, close to a more realistic scenario.

\clearpage
\newpage
\section{Forecasting weekly new infections incidence under stronger noise}
\label{sec: appendix_noise_synthetic}

\begin{figure}[htbp]
    \centering
    \resizebox{\textwidth}{!}{
        \includegraphics[height=4.7cm]{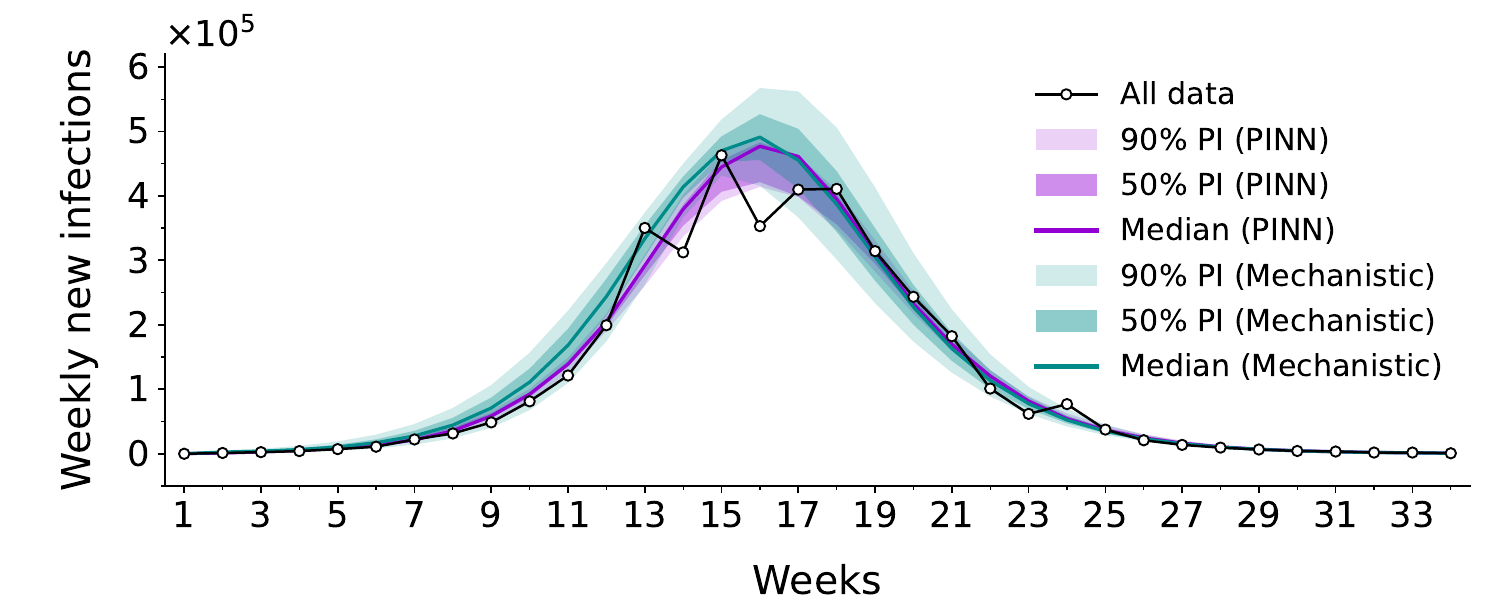}
        \hspace{0.3cm}%
        \includegraphics[height=4.7cm]{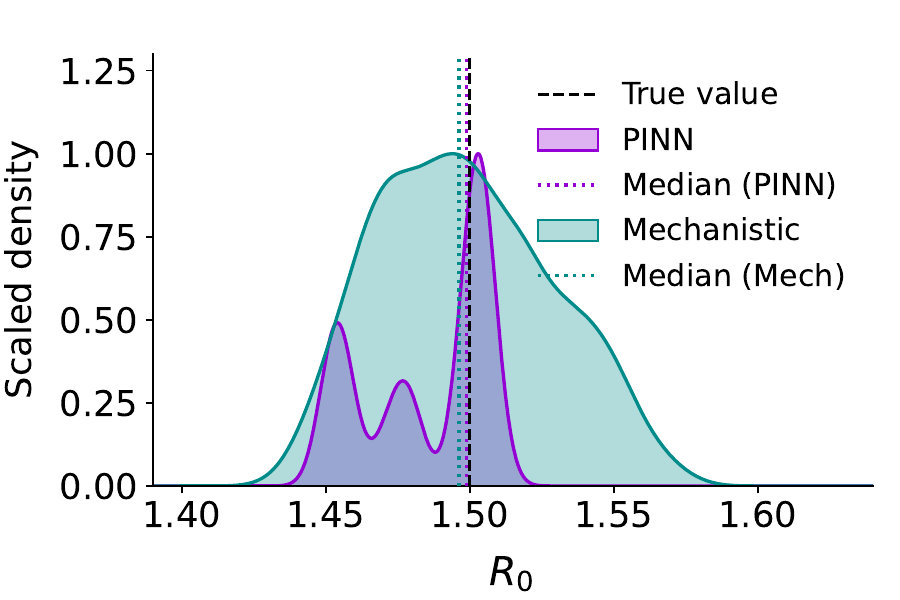}
        }

    \vspace{0.5em}

    \resizebox{\textwidth}{!}{
        \includegraphics[height=4.7cm]{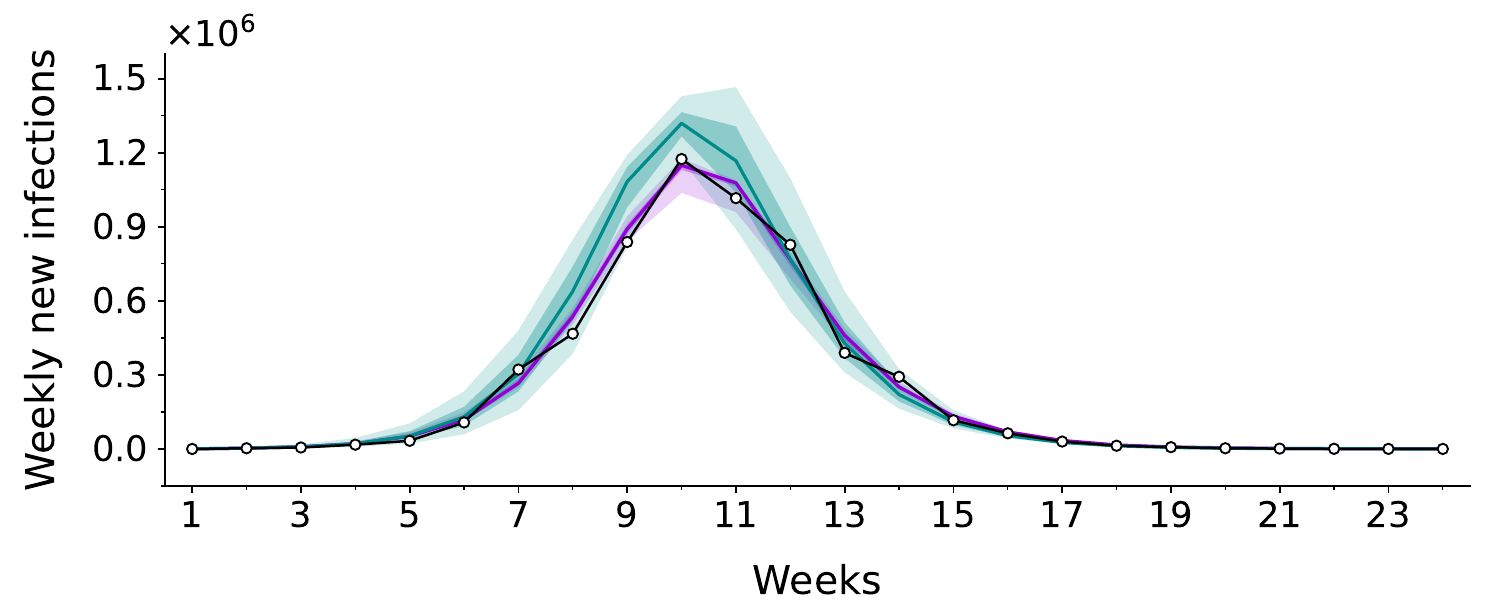}
        \hspace{0.3cm}
        \includegraphics[height=4.7cm]{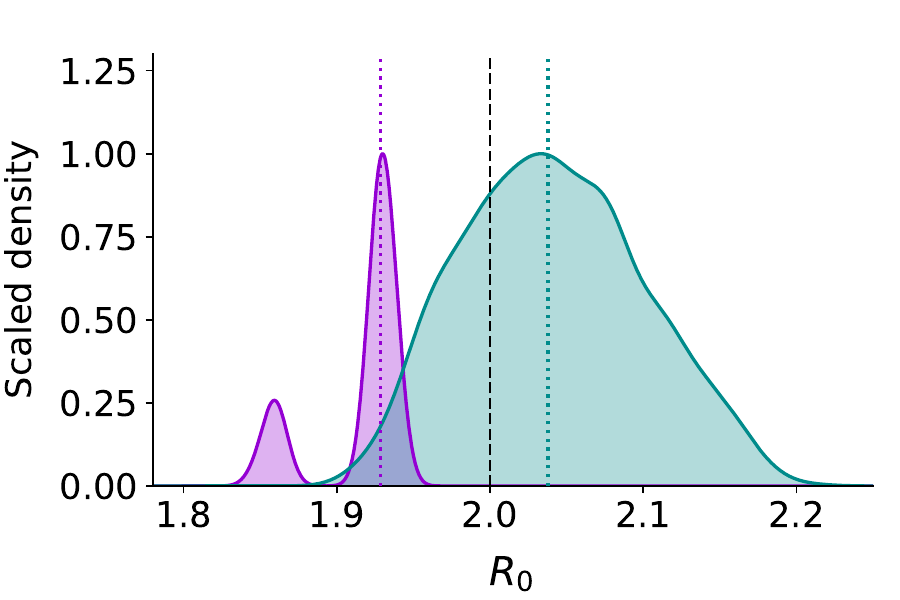}
        }
        \vspace{0.4em}
    \caption{\textbf{Retrospective fit and $R_0$ posteriors in the two synthetic scenarios under stronger noise.} Top row: first scenario ($R_0=1.5$), with data up to week $34$; bottom row: second scenario ($R_0=2.0$), with data up to week $24$. Both are generated from the synthetic data with those $R_0$ and stronger negative binomial noise ($r=50$). Left panels: retrospective fit of the weekly new infections obtained by the \gls{pinn}-based (violet) and mechanistic-based (teal) frameworks, calibrated on all the noisy observations up to the last week (white dots). Shaded areas show the $90\%$ (lighter) and $50\%$ (darker) prediction intervals, and lines the medians. Right panels: posterior distributions of $R_0$ estimated by the two frameworks at the same week, with densities scaled to a maximum of $1$. Dotted lines mark the posterior medians and the dashed black line the true value of $R_0$ used to generate the data.}
    \label{fig: synth_retro_R0_noise}
\end{figure}

\begin{table}[htbp]
\centering
\begin{tabular}{llcccccc}
\toprule
Scenario & Model & MAE & RMSE & WIS & IS$_{80}$ & Cov.\ 50\% & Cov.\ 90\% \\
\midrule
\multirow{3}{*}{\makecell[l]{Scenario 1\\ $R_0=1.5$ \\ stronger noise}}
& \textsc{PINN} & 0.325 & 0.465 & 0.345 & 0.360 & 0.204 & 0.593 \\
& \textsc{Mechanistic} & \textbf{0.298} & \textbf{0.379} & \textbf{0.270} & \textbf{0.245} & 0.676 & \textbf{0.935} \\
& \textit{Baseline} & 1.000 & 1.000 & 1.000 & 1.000 & \textbf{0.352} & 0.611 \\
\midrule
\multirow{3}{*}{\makecell[l]{Scenario 2\\ $R_0=2.0$ \\ stronger noise}}
&\textsc{PINN} & 0.215 & 0.294 & 0.212 & 0.233 & 0.132 & 0.515 \\
&\textsc{Mechanistic} & \textbf{0.192} & \textbf{0.233} & \textbf{0.159} & \textbf{0.130} & 0.706 & \textbf{0.956} \\
&\textit{Baseline} & 1.000 & 1.000 & 1.000 & 1.000 & \textbf{0.397} & 0.588 \\
\bottomrule
\end{tabular}
\caption{\textbf{Comparison of overall forecasting performance on the two synthetic scenarios under stronger noise.} Forecasts of weekly new infections, generated from the synthetic data with $R_0=1.5$ and stronger negative binomial noise ($r=50$), aggregated over all horizons (from 1 to 4) and rounds. MAE, RMSE, WIS, and IS$_{80}$ are normalized by the baseline score of the same scenario. Bold indicates the best performance among \textsc{PINN}, \textsc{Mechanistic}, and \textit{Baseline} within each scenario; for coverage, the value closest to the nominal level.}
\label{tab:results_pooled_synth3&4}
\end{table}

\begin{table}[htbp]
\centering
\footnotesize
\setlength{\tabcolsep}{2pt}   
\resizebox{\textwidth}{!}{%
\begin{tabular}{cl cccccc | cccccc}
\toprule
 & & \multicolumn{6}{c|}{Scenario 1: $R_0=1.5$, stronger noise} & \multicolumn{6}{c}{Scenario 2: $R_0=2.0$, stronger noise} \\
\cmidrule(lr){3-8} \cmidrule(lr){9-14}
Horizon & Model & MAE & RMSE & WIS & IS$_{80}$ & $C_{50\%}$ & $C_{90\%}$ & MAE & RMSE & WIS & IS$_{80}$ & $C_{50\%}$ & $C_{90\%}$ \\
\midrule
H1 & \textsc{PINN}        & 0.555 & 0.774 & 0.651 & 0.744 & 0.185 & 0.519 & \textbf{0.365} & 0.458 & 0.398 & 0.423 & 0.118 & 0.353 \\
   & \textsc{Mechanistic} & \textbf{0.487} & \textbf{0.656} & \textbf{0.458} & \textbf{0.403} & \textbf{0.667} & \textbf{0.926} & 0.381 & \textbf{0.421} & \textbf{0.313} & \textbf{0.233} & 0.647 & \textbf{0.941} \\
   & \textit{Baseline}    & 1.000 & 1.000 & 1.000 & 1.000 & 0.296 & 0.667 & 1.000 & 1.000 & 1.000 & 1.000 & \textbf{0.412} & 0.706 \\
\midrule
H2 & \textsc{PINN}        & 0.373 & 0.480 & 0.395 & 0.416 & 0.148 & 0.519 & 0.292 & 0.446 & 0.300 & 0.347 & 0.118 & 0.647 \\
   & \textsc{Mechanistic} & \textbf{0.359} & \textbf{0.471} & \textbf{0.326} & \textbf{0.311} & \textbf{0.667} & \textbf{0.926} & \textbf{0.238} & \textbf{0.317} & \textbf{0.194} & \textbf{0.145} & 0.765 & \textbf{1.000} \\
   & \textit{Baseline}    & 1.000 & 1.000 & 1.000 & 1.000 & 0.333 & 0.630 & 1.000 & 1.000 & 1.000 & 1.000 & \textbf{0.353} & 0.588 \\
\midrule
H3 & \textsc{PINN}        & \textbf{0.267} & 0.491 & 0.274 & 0.283 & 0.333 & 0.704 & 0.182 & 0.265 & 0.186 & 0.212 & 0.176 & 0.588 \\
   & \textsc{Mechanistic} & 0.271 & \textbf{0.377} & \textbf{0.243} & \textbf{0.222} & 0.704 & \textbf{0.926} & \textbf{0.173} & \textbf{0.222} & \textbf{0.135} & \textbf{0.109} & 0.765 & \textbf{0.941} \\
   & \textit{Baseline}    & 1.000 & 1.000 & 1.000 & 1.000 & \textbf{0.444} & 0.593 & 1.000 & 1.000 & 1.000 & 1.000 & \textbf{0.412} & 0.529 \\
\midrule
H4 & \textsc{PINN}        & 0.261 & 0.369 & 0.269 & 0.265 & 0.148 & 0.630 & 0.146 & 0.200 & 0.126 & 0.132 & 0.118 & 0.471 \\
   & \textsc{Mechanistic} & \textbf{0.219} & \textbf{0.277} & \textbf{0.196} & \textbf{0.178} & \textbf{0.667} & \textbf{0.963} & \textbf{0.118} & \textbf{0.161} & \textbf{0.110} & \textbf{0.109} & 0.647 & \textbf{0.941} \\
   & \textit{Baseline}    & 1.000 & 1.000 & 1.000 & 1.000 & 0.333 & 0.556 & 1.000 & 1.000 & 1.000 & 1.000 & \textbf{0.412} & 0.529 \\
\bottomrule
\end{tabular}%
}
\caption{\textbf{Comparison of horizon-specific forecasting performance on the two synthetic scenarios under stronger noise.} Forecasts of weekly new infections on the synthetic data with $R_0=1.5$ and $R_0=2.0$ and stronger negative binomial noise ($r=50$), aggregated over all rounds (27 rounds in the first scenario and 17 in the second) for each horizon, from 1 to 4. MAE, RMSE, WIS, and IS$_{80}$ are normalized by the baseline score of the same scenario and horizon. Bold indicates the best performance among \textsc{PINN}, \textsc{Mechanistic}, and \textit{Baseline} within each scenario and horizon; for coverage, the value closest to the nominal level.}
\label{tab:results_multihorizon_synth_noise}
\end{table}

\begin{figure}[htbp]
    \centering
    \resizebox{\textwidth}{!}{%
        \includegraphics[height=10cm]{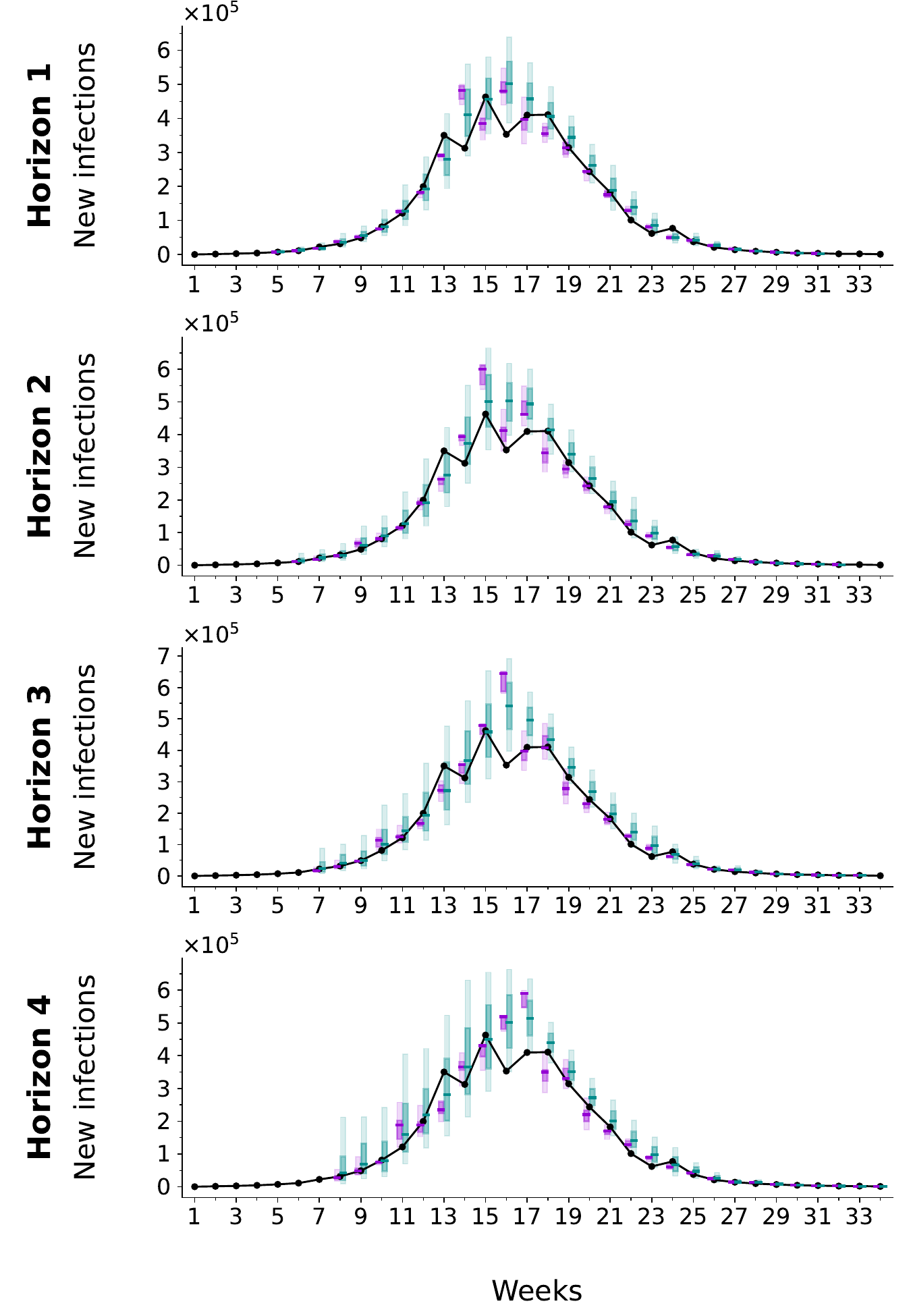}%
        \hspace{0.3cm}%
        \includegraphics[height=10cm]{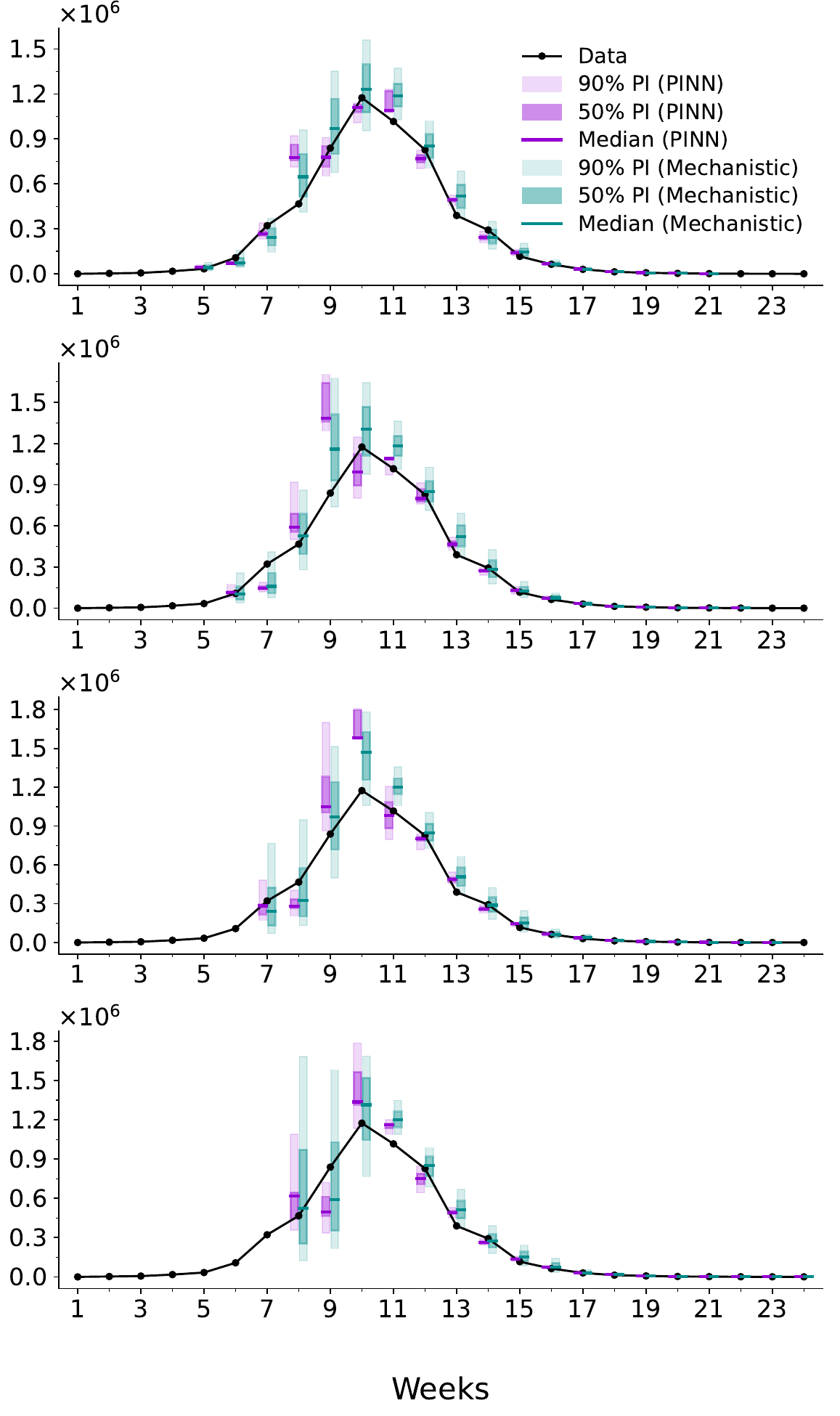}%
    }
   \caption{\textbf{Horizon-specific probabilistic forecasts on the two synthetic scenarios under stronger noise.} Forecasts from the \gls{pinn}-based (in violet) and mechanistic (in teal) of the weekly new infections on the noisy synthetic data with stronger negative binomial noise ($r=50$) with $R_0=1.5$ (on the left) and $R_0=2.0$ (on the right), at fixed horizons of 1 to 4 weeks ahead (panels from top to bottom). In each panel, every box refers to the week on the x-axis and shows the forecast made, for that week, as many weeks earlier as the horizon of the panel, using only the data available at that time. The lighter and darker boxes show the 90\% and 50\% prediction intervals, respectively, and the horizontal line inside each box shows the median. The observed weekly new infections are shown in black.}    
   \label{fig:comparison_horizons_synth_noise}
\end{figure}

\begin{figure}[htbp]
    \centering
    \includegraphics[width=\textwidth]{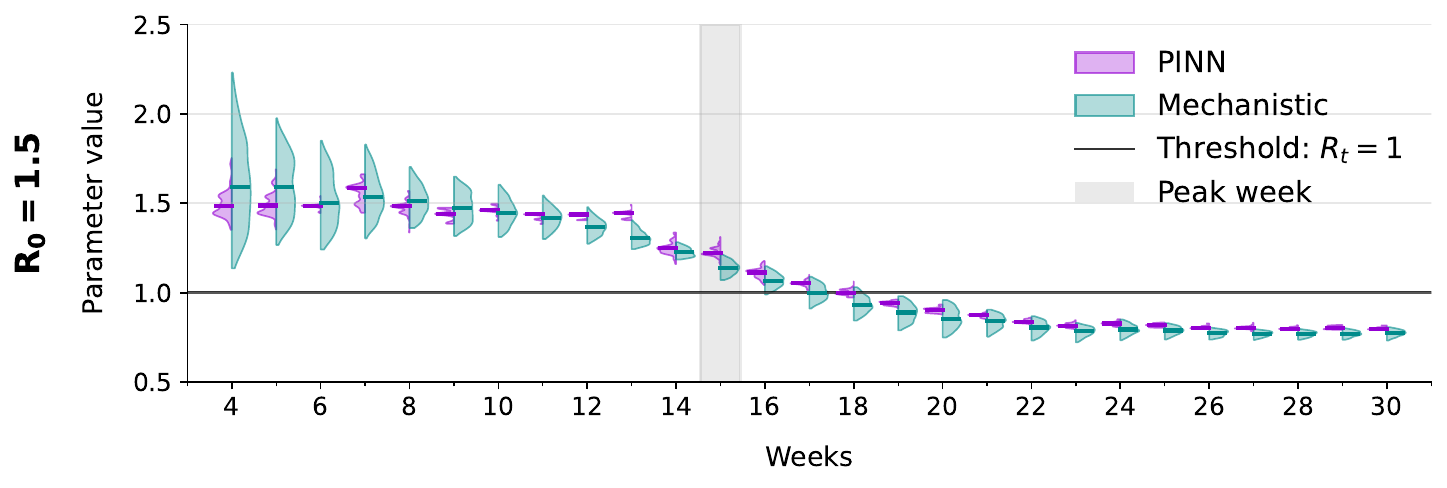}

    \vspace{0.5em}

    \includegraphics[width=\textwidth]{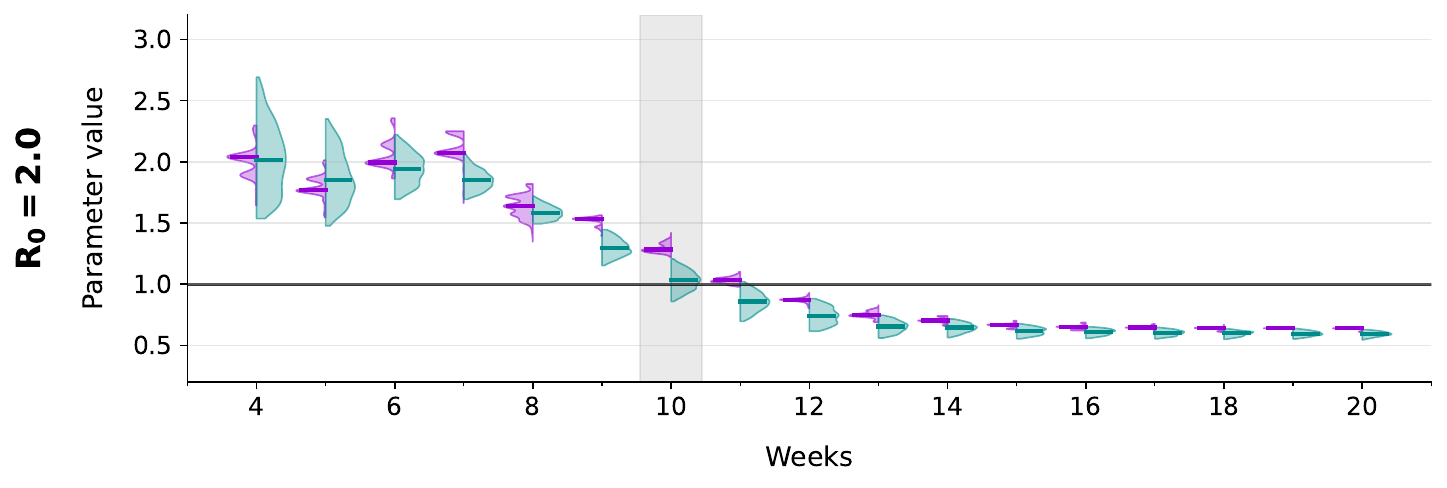}

    \caption{\textbf{Posterior distributions of $R_t$ across rounds in the two synthetic scenarios under stronger noise.} For each round, the split violin shows the posterior distribution of the effective reproduction number estimated by the PINN-based framework with MCMC (left half, violet) and by the mechanistic framework via ABC-SMC (right half, teal). Each split violin shows the full shape of the $1\,000$ tail posterior samples for that window; horizontal ticks mark the per-window median for each model. In addition, the solid horizontal line marks the epidemic threshold $R_t = 1$, below which the outbreak is expected to decline, and the shaded vertical band marks the week of the peak in weekly new infections generated with stronger negative binomial noise ($r=50$), week $16$ in the first scenario ($R_0=1.5$, top) and week $10$ in the second ($R_0=2.0$, bottom).} 
    \label{fig:Re_violin_synth_noise}
\end{figure}

\clearpage
\newpage
\section{Additional results on COVID-19 forecasting}
\label{sec: appendix_COVID}

\begin{figure}[htbp]
\centering
\includegraphics[width=1\textwidth]{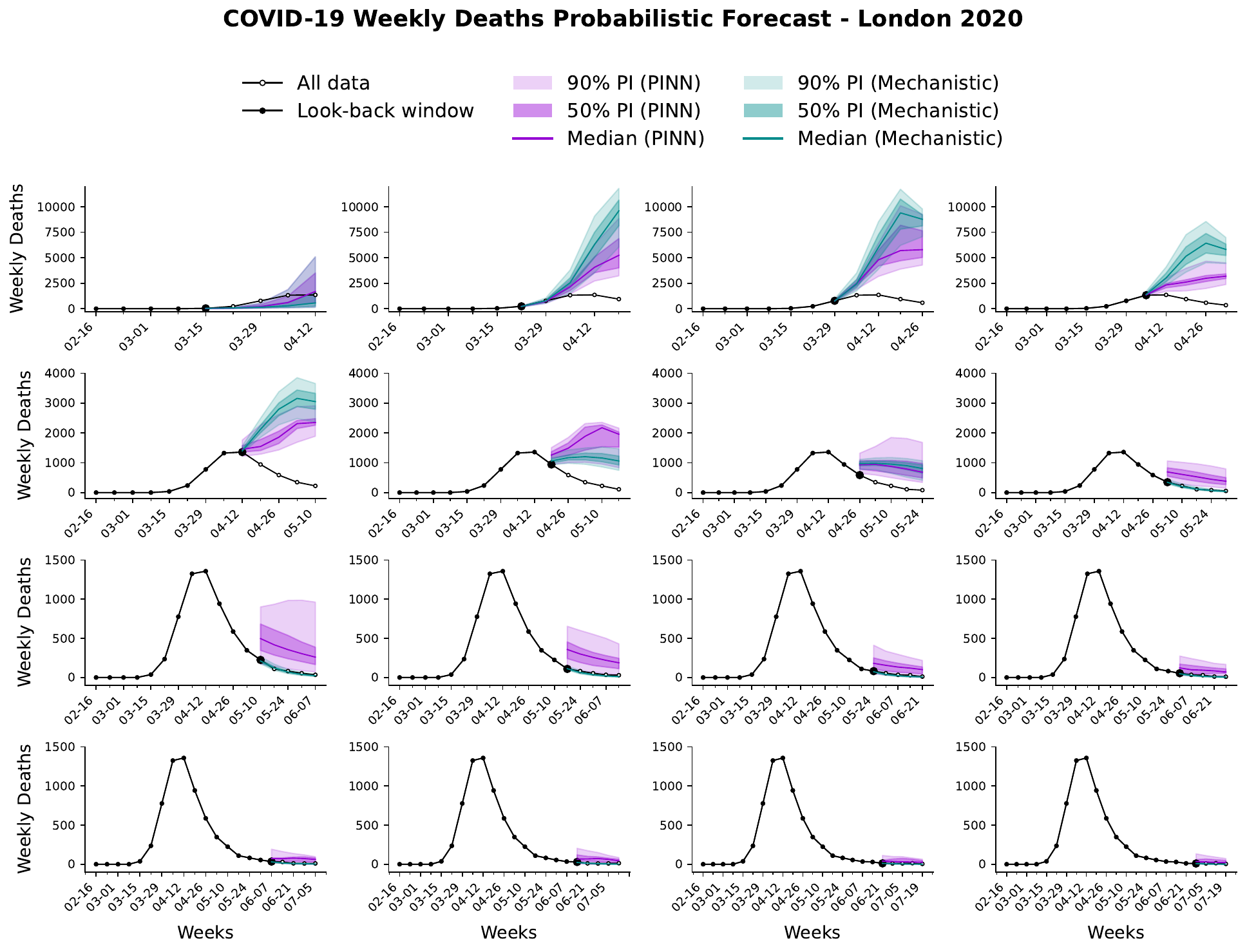}
\caption{Comparison of \gls{pinn} and mechanistic forecasts of weekly COVID-19 deaths in London across sixteen forecast rounds (i.e., weeks). In each panel, black dots denote the observed weekly deaths used for model calibration up to the current week, while white dots denote the subsequent four weeks to forecast. Shaded bands show the 50\% and 90\% prediction intervals of each model over the 1- to 4-week-ahead forecast horizons, and solid lines indicate the corresponding medians.}
\label{fig:forecastig_comparison}
\end{figure}

\clearpage
\newpage
\section{Horizon-specific forecasting performance on the COVID-19 scenario}
\label{sec: appendix_tableCOVID}

\begin{table}[htbp]
\centering
\begin{tabular}{clcccccc}
\toprule
Horizon & Model & MAE & RMSE & WIS & IS$_{80}$ & Cov.\ 50\% & Cov.\ 90\% \\
\midrule
H1 (18 rounds) & \textsc{PINN} & 2.124 & 2.002 & 1.926 & 1.689 & 0.056 & 0.500 \\
 & \textsc{Mechanistic} & 2.032 & 2.461 & 2.112 & 2.033 & 0.278 & 0.500 \\
 & \textit{Baseline} & 1.000 & 1.000 & 1.000 & 1.000 & 0.667 & 0.833 \\
\midrule
H2 (17 rounds) & \textsc{PINN} & 2.211 & 2.458 & 2.186 & 2.373 & 0.059 & 0.353 \\
 & \textsc{Mechanistic} & 2.880 & 3.757 & 3.036 & 3.399 & 0.235 & 0.412 \\
 & \textit{Baseline} & 1.000 & 1.000 & 1.000 & 1.000 & 0.588 & 0.706 \\
\midrule
H3 (16 rounds) & \textsc{PINN} & 2.640 & 2.888 & 2.684 & 2.674 & 0.125 & 0.375 \\
 & \textsc{Mechanistic} & 4.034 & 5.125 & 4.220 & 4.296 & 0.125 & 0.438 \\
 & \textit{Baseline} & 1.000 & 1.000 & 1.000 & 1.000 & 0.625 & 0.812 \\
\midrule
H4 (15 rounds) & \textsc{PINN} & 2.854 & 3.336 & 3.254 & 3.588 & 0.067 & 0.333 \\
 & \textsc{Mechanistic} & 4.338 & 5.706 & 5.021 & 5.652 & 0.067 & 0.400 \\
 & \textit{Baseline} & 1.000 & 1.000 & 1.000 & 1.000 & 0.600 & 0.800 \\
\bottomrule
\end{tabular}
\caption{\textbf{Comparison of horizon-specific forecasting performance on the real COVID-19 scenario.} Forecasts of weekly deaths, for the first COVID-19 wave in London (UK), aggregated over all rounds for each horizon (from 1 to 4). MAE, RMSE, WIS, and IS$_{80}$ are normalized by the baseline score.}
\label{tab:COVID_forecast_h}
\end{table}

\clearpage
\newpage
\section{Effective reproduction number inference on the COVID-19 scenario}
\label{sec: appendix_COVID_Re}

\begin{figure}[htbp]
    \centering
    \includegraphics[width=\textwidth]{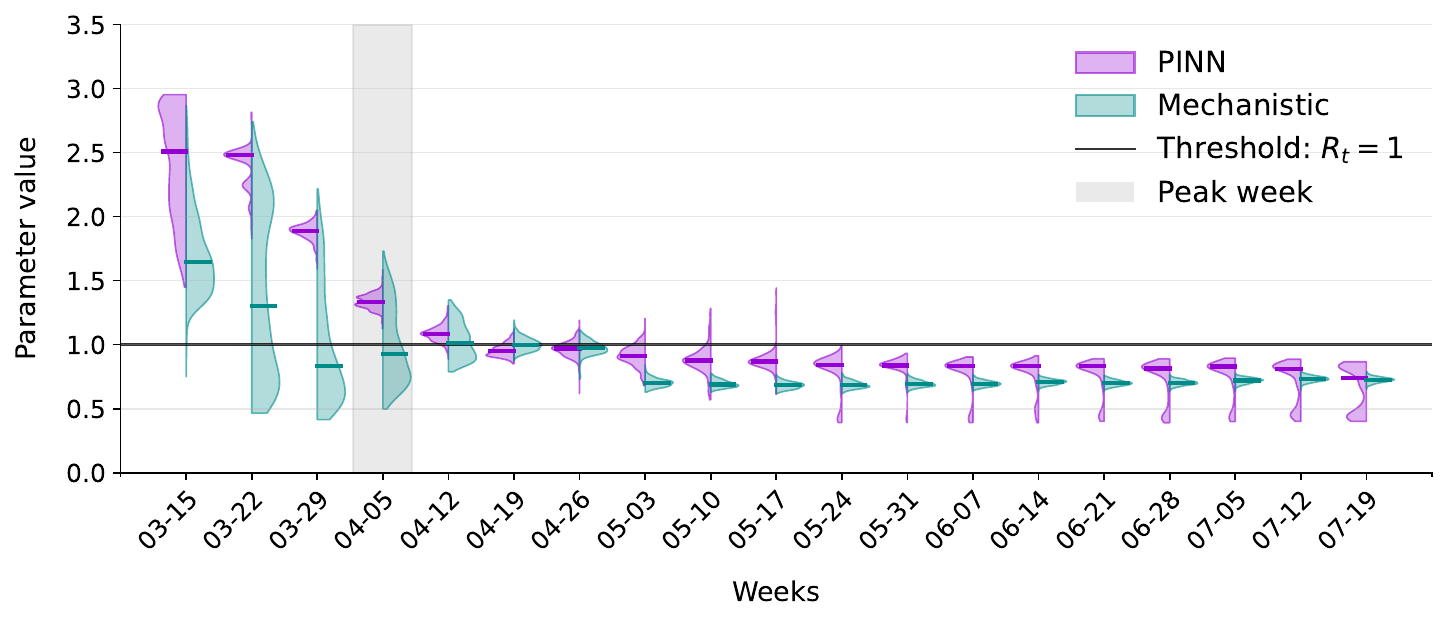}
    \caption{\textbf{Posterior distributions of $R_t$ between the two frameworks across weeks.} For each round, the split violin shows the posterior distribution of the effective reproduction number estimated by the PINN-based framework with MCMC (left half, violet) and by the mechanistic framework via ABC-SMC (right half, teal). Each split violin shows the full shape of the $1\,000$ tail posterior samples for that window; horizontal ticks mark the per-window median for each model. In addition, the solid horizontal line marks the epidemic threshold $R_t = 1$, below which the outbreak is expected to decline, and the shaded vertical band marks the week of the peak in reported weekly new cases.
    }
    \label{fig: violin_comparison}
\end{figure}

Fig~\ref{fig: violin_comparison} shows the posterior distributions of the effective reproduction number $R_t$ estimated by the two frameworks at each round, as explained in Section~\ref{sec: common_structural_setting}, Eq.~(\ref{eq: Re}), together with their medians and the threshold $R_t = 1$. In the first rounds, the estimates of the \gls{pinn}-based framework are higher than those of the mechanistic one: at week 2020-03-15 the medians are about 2.5 and 1.6, respectively, and the \gls{pinn}-based distribution accumulates close to the upper bound of the prior range of $R_0$ ($3$), which $R_t$ approaches since almost the whole population is still susceptible. The mechanistic distributions are instead wider in these rounds. The medians of both frameworks decrease as the epidemic progresses, but at different speeds: the mechanistic ones fall below $1$ at the rounds of 2020-03-29 and 2020-04-05, and are close to $1$ in the following rounds, while the \gls{pinn}-based ones decrease more gradually and remain above $1$ until the round of 2020-04-12. 
Specifically, the two frameworks differ in the timing at which the estimated $R_t$ crosses the threshold $R_t=1$. The mechanistic medians fall below $1$ already at the rounds of 2020-03-29 and 2020-04-05, which is consistent with the peak of the new cases preceding by some weeks the peak of the reported deaths (see the shaded band), whereas the \gls{pinn}-based medians cross the threshold later, at the round of 2020-04-19. After crossing the threshold, however, the \gls{pinn}-based medians remain below $1$ at all the following rounds, while the mechanistic ones return close to $1$ around the peak of the reported deaths (rounds from 2020-04-12 to 2020-04-26) before decreasing again. 
From the round of 2020-04-26, the medians of both frameworks are at or below $1$, consistent with the decline of the observed deaths after their peak in the week of 2020-04-12.
In the last round of 2020-07-19 the two medians converge, to 0.74 (\gls{pinn}) and 0.73 (mechanistic), in agreement with the convergence of the medians inference of the basic reproduction number, $R_0$: to 1.64 for the \gls{pinn}-based framework and 1.63 for the mechanistic-based one.

Furthermore, the width of the posterior distributions decreases with the rounds for both frameworks, but more markedly for the mechanistic one, whose distributions are narrow from the round of 2020-04-12, while the \gls{pinn}-based ones maintain a long lower tail until the last rounds. This trend, with higher uncertainty in the first phases of the epidemic and overconfidence in the tail, is also reflected in the width of the prediction intervals of the forecasts shown in Fig~\ref{fig:comparison_horizons}.

\clearpage
\newpage
\section{COVID-19 forecasting with shorter look-back window}
\label{sec: appendix_COVID_n5}

\begin{table}[htbp]
\centering
\begin{tabular}{llcccccc}
\toprule
Look-back window & Model & MAE & RMSE & WIS & IS$_{80}$ & Cov.\ 50\% & Cov.\ 90\% \\
\midrule
\multirow{2}{*}{\makecell[l]{5 weeks}}
& \textsc{PINN} & 2.257 & 2.778 & 2.206 & 2.313 & 0.379 & 0.621 \\
& \textsc{Mechanistic} & 3.507 & 5.071 & 3.841 & 4.121 & 0.167 & 0.364 \\
\midrule
\multirow{2}{*}{\makecell[l]{All past data}}
& \textsc{PINN} & 2.532 & 2.926 & 2.629 & 2.725 & 0.076 & 0.394 \\
& \textsc{Mechanistic} & 3.577 & 4.954 & 3.879 & 4.144 & 0.182 & 0.439 \\
\midrule
\multirow{1}{*}{\makecell[l]{--}}
& \textit{Baseline} & 1.000 & 1.000 & 1.000 & 1.000 & 0.621 & 0.788 \\
\bottomrule
\end{tabular}
\caption{\textbf{Comparison of overall forecasting performance on the real COVID-19 scenario with different look-back windows.} Forecasts of weekly deaths for the first COVID-19 wave in London (UK), aggregated over all horizons (from 1 to 4) and rounds. The first two rows report the results obtained with a look-back window of the last $M = 5$ weeks of observations, without the weighting strategy, while the other two rows report the results of the main analysis (Table~\ref{tab: COVID_aggregated}), where all the past data available are used, with the weighting strategy. Baseline is reported in the last row, since its implementation does not depend on the look-back window length. MAE, RMSE, WIS, and IS$_{80}$ are normalized by the baseline score.}
\label{tab: COVID_aggregated_n5}
\end{table}

\begin{table}[htbp]
\centering
\begin{tabular}{clcccccc}
\toprule
Horizon & Model & MAE & RMSE & WIS & IS$_{80}$ & Cov.\ 50\% & Cov.\ 90\% \\
\midrule
H1 (18 rounds) & \textsc{PINN} & 1.765 & 1.970 & 1.431 & 1.213 & 0.389 & 0.667 \\
 & \textsc{Mechanistic} & 1.868 & 2.447 & 1.920 & 1.803 & 0.278 & 0.500 \\
 & \textit{Baseline} & 1.000 & 1.000 & 1.000 & 1.000 & 0.667 & 0.833 \\
\midrule
H2 (17 rounds) & \textsc{PINN} & 1.864 & 2.247 & 1.781 & 1.967 & 0.412 & 0.588 \\
 & \textsc{Mechanistic} & 2.851 & 3.942 & 2.985 & 3.315 & 0.176 & 0.353 \\
 & \textit{Baseline} & 1.000 & 1.000 & 1.000 & 1.000 & 0.588 & 0.706 \\
\midrule
H3 (16 rounds) & \textsc{PINN} & 2.390 & 2.692 & 2.241 & 2.268 & 0.312 & 0.625 \\
 & \textsc{Mechanistic} & 3.999 & 5.296 & 4.226 & 4.334 & 0.062 & 0.312 \\
 & \textit{Baseline} & 1.000 & 1.000 & 1.000 & 1.000 & 0.625 & 0.812 \\
\midrule
H4 (15 rounds) & \textsc{PINN} & 2.647 & 3.236 & 2.870 & 3.195 & 0.400 & 0.600 \\
 & \textsc{Mechanistic} & 4.242 & 5.763 & 5.019 & 5.717 & 0.133 & 0.267 \\
 & \textit{Baseline} & 1.000 & 1.000 & 1.000 & 1.000 & 0.600 & 0.800 \\
\bottomrule
\end{tabular}
\caption{\textbf{Comparison of horizon-specific forecasting performance on the real COVID-19 scenario with a look-back window of length $5$.} Forecasts of weekly deaths, for the first COVID-19 wave in London (UK), aggregated over all rounds for each horizon (from 1 to 4). Unlike in the main analysis, where all the past data available are used (see Table~\ref{tab: COVID_aggregated} for comparison), at each round both frameworks are calibrated on the last $M = 5$ weeks of observations only, without the weighting strategy. MAE, RMSE, WIS, and IS$_{80}$ are normalized by the baseline score.}
\label{tab:COVID_forecast_h_n5}
\end{table}

\end{appendices}

\end{document}